\documentclass[11pt,a4paper,longbibliography]{article}
\usepackage{jcappub}
\usepackage{amsmath,amssymb,mathtools}
\usepackage{graphicx}
\usepackage{xcolor}
\usepackage{hyperref}
\usepackage{booktabs}
\usepackage{multirow}
\usepackage{xspace}
\usepackage{fontawesome5}
\usepackage{pdflscape}
\usepackage{orcidlink}

\newcommand{\orcid}[1]{\,\orcidlink{#1}}

\usepackage[acronym, toc, nonumberlist]{glossaries}

\glsdisablehyper
\setacronymstyle{long-short}

\newacronym{ns}{NS}{neutron star}
\newacronym{bh}{BH}{black hole}
\newacronym{bbh}{BBH}{binary black hole}
\newacronym{bns}{BNS}{binary neutron star}
\newacronym{nsbh}{NSBH}{neutron star black hole}

\newacronym{eos}{EoS}{equation of state}
\newacronym{gw}{GW}{gravitational wave}
\newacronym{gr}{GR}{general relativity}
\newacronym{snr}{SNR}{signal-to-noise ratio}

\newacronym{lisa}{LISA}{Laser Interferometer Space Antenna}
\newacronym{ligo}{LIGO}{Laser Interferometer Gravitational-Wave Observatory}
\newacronym{kagra}{KAGRA}{KAmioka GRavitational wave detector}
\newacronym{eob}{EOB}{effective one-body}
\newacronym{em}{EM}{electromagnetic}
\newacronym{lcdm}{$\Lambda$CDM}{$\Lambda$ cold dark matter}
\newacronym{pl}{PL}{power law}
\newacronym{plg}{PLG}{power law and Gaussian}
\newacronym{kde}{KDE}{kernel density estimate}
\newacronym{de}{DE}{dark energy}
\newacronym{cdf}{CDF}{cumulative density function}

\newacronym{lvk}{LVK}{LIGO-Virgo-KAGRA}
\newacronym{ego}{EGO}{European gravitational observatory}
\newacronym{asd}{ASD}{amplitude spectral density}
\newacronym{psd}{PSD}{power spectral density}
\newacronym{mcmc}{MCMC}{Markov chain Monte Carlo}
\newacronym{hlv}{HLV}{Hanford Livingston Virgo}
\newacronym{pe}{PE}{parameter estimation}
\newacronym{cbc}{CBC}{compact binary coalescence}
\newacronym{aligo}{aLIGO}{advanced LIGO}
\newacronym{far}{FAR}{false alarm rate}
\newacronym{emri}{EMRI}{extreme mass-ratio inspiral}
\newacronym{agn}{AGN}{active galactic nucleus}

\newacronym{cl}{CL}{confidence level}

\newacronym{pn}{PN}{post-Newtonian}
\newacronym{nr}{NR}{numerical relativity}
\newacronym{ppisn}{PPISN}{pulsational pair-instability supernova}
\newacronym{pisn}{PISN}{pair-instability supernova}
\newacronym{et}{ET}{Einstein Telescope}
\newacronym{ce}{CE}{Cosmic Explorer}

\newacronym{cmb}{CMB}{cosmic microwave background}
\newacronym{lss}{LSS}{large scale structure}
\newacronym{isco}{ISCO}{innermost stable circular orbit}

\newacronym{oi}{Oi}{observation run $i$}
\newacronym{gwtci}{GWTC-i}{gravitational wave transient catalog $i$}

\newacronym{2g}{2G}{second generation}
\newacronym{3g}{3G}{third generation}

\newacronym{nf}{NF}{normalizing flow}
\newacronym{ml}{ML}{machine learning}
\newacronym{lfi}{LFI}{likelihood-free inference}
\newacronym{nn}{NN}{neural network}
\newacronym{dingo}{DINGO}{deep inference for gravitational wave observations}
\newacronym{gpu}{GPU}{graphics processing unit}
\newacronym{hba}{HBA}{hierarchical Bayesian analysis}

\newacronym{kl}{KL}{Kullback-Leibler}
\newacronym{js}{JS}{Jensen-Shannon}
\newacronym{ks}{KS}{Kolmogorov--Smirnov}

\newacronym{ifar}{IFAR}{inverse false-alarm rate}

\newacronym{dm}{DM}{dark matter}

\newacronym{gp}{GP}{Gaussian process}
\newacronym{grf}{GRF}{Gaussian random field}
\newacronym{fft}{FFT}{Fast Fourier Transform}

\newacronym{mltp}{MLTP}{multi-peak truncated power law}
\newacronym{ppd}{PPD}{posterior predictive distribution}
\newacronym{tpl}{TPL}{truncated power law}
\newacronym{ad}{AD}{automatic differentiation}
\newacronym{hmc}{HMC}{Hamiltonian Monte Carlo}
\newacronym{nuts}{NUTS}{No-U-Turn Sampler}
\newacronym{xas}{XAS}{aligned-spin}

\newacronym{gp1d}{GP-1D}{one-dimensional Gaussian process}

\newcommand{\hyperredshift}{\Lambda_{z}}
\newcommand{\hypercosmology}{\Lambda_{\rm c}}
\newcommand{\hypermass}{\Lambda_{\rm m}}
\newcommand{\lognorm}{\mathcal{N}}

\newcommand{\deffrom}{\coloneqq}

\newcommand{\dd}[1]{\mathrm{d}#1}

\newcommand{\Omegam}{\Omega_{{\rm m}}}

\newcommand{\hu}{\,{\rm km \,s^{-1} \, Mpc^{-1}}} %
\newcommand{\Mpc}{\,{\rm Mpc}} %
\newcommand{\Hubble}{H_0}

\newcommand{\msun}{M_\odot}

\newcommand{\icarogw}{\texttt{icarogw}\xspace}
\newcommand{\gwcosmo}{\texttt{gwcosmo}\xspace}
\newcommand{\cosmopyro}{\texttt{CosmoPyro}\xspace}

\newcommand{\md}{m_{\rm d}}
\newcommand{\ms}{m_{\rm s}}
\newcommand{\mmin}{m_{\rm min}}
\newcommand{\mmax}{m_{\rm max}}

\newcommand{\msone}{m_{1,{\rm s}}}
\newcommand{\mstwo}{m_{2,{\rm s}}}

\newcommand{\mdone}{m_{1,{\rm d}}}
\newcommand{\mdtwo}{m_{2,{\rm d}}}

\newcommand{\massratio}{q}

\newcommand{\logMtot}{\mathcal{S}}
\newcommand{\minuslogq}{\delta}

\newcommand{\alphapl}{\alpha}
\newcommand{\deltam}{\delta_m}
\newcommand{\deltamtwo}{\delta_{m,2}}
\newcommand{\betaq}{\beta}
\newcommand{\betazero}{\beta_0}
\newcommand{\betaone}{\beta_1}
\newcommand{\msref}{m_{1,{\rm ref}}}
\newcommand{\lambdag}{\lambda_{g}}
\newcommand{\lambdaglow}{\lambda_{g,{\rm low}}}
\newcommand{\muglow}{\mu_{g,{\rm low}}}
\newcommand{\sigmaglow}{\sigma_{g,{\rm low}}}
\newcommand{\mughigh}{\mu_{g,{\rm high}}}
\newcommand{\sigmaghigh}{\sigma_{g,{\rm high}}}

\newcommand{\sigmalowfrac}{\sigma_{\rm f,  low}}
\newcommand{\sigmahighfrac}{\sigma_{\rm f,  high}}

\newcommand{\alpharef}{\alpha_{\mathrm{ref}}}
\newcommand{\betaref}{\beta_{\mathrm{ref}}}

\newcommand{\Ssmooth}{S}

\newcommand{\data}{d}

\newcommand{\datagwsub}[1]{\ensuremath{\data_{#1}}}

\newcommand{\hyper}{\Lambda}
\newcommand{\param}{\lambda}
\newcommand{\params}{\param}

\newcommand{\nobs}{N_{\rm obs}}

\newcommand{\ninjtot}{N_{\rm inj, tot}}
\newcommand{\ninjsel}{N_{\rm inj, sel}}
\newcommand{\injparams}{\params_{\rm inj}}

\newcommand{\dVdz}{\frac{\mathrm{d}V}{\mathrm{d}z}}
\newcommand{\gaussianfield}{G_F}
\newcommand{\gaussianfieldwhitened}{G_{F,{\rm w}}}

\newcommand{\numpyro}{\texttt{NumPyro}\xspace}
\newcommand{\jax}{\texttt{jax}\xspace}

\newcommand{\powerspectrumamplitude}{\mathcal{A}_{\rm PS}}
\newcommand{\powerspectrumcutoff}{k_{\rm cut}}

\newcommand{\mltp}{\textsc{Multi Peak}\xspace}

\newcommand{\fullpopthreepeak}[0]{\textsc{FullPop 3 Peaks}\xspace}

\newcommand{\mumassratio}{\mu_{\massratio}}
\newcommand{\sigmamassratio}{\sigma_{\massratio}}

\newcommand{\modelheading}[1]{%
  \vspace{0.5cm}
  \noindent
  \textbf{#1}.
}

\newcommand{\numsim}{1000}
\newcommand{\hgponed}{\ensuremath{0.66}}
\newcommand{\hgponedlow}{\ensuremath{0.20}}
\newcommand{\hgponedhigh}{\ensuremath{0.17}}
\newcommand{\hgptwod}{\ensuremath{0.57}}
\newcommand{\hgptwodlow}{\ensuremath{0.15}}
\newcommand{\hgptwodhigh}{\ensuremath{0.20}}
\newcommand{\alphazerogponed}{\ensuremath{-2.3}}
\newcommand{\alphazerogponedlow}{\ensuremath{0.5}}
\newcommand{\alphazerogponedhigh}{\ensuremath{0.4}}
\newcommand{\powerspectrumamplitudegponed}{\ensuremath{9.6}}
\newcommand{\powerspectrumamplitudegponedlow}{\ensuremath{3.7}}
\newcommand{\powerspectrumamplitudegponedhigh}{\ensuremath{6.2}}
\newcommand{\powerspectrumcutoffgponed}{\ensuremath{434}}
\newcommand{\powerspectrumcutoffgponedlow}{\ensuremath{110}}
\newcommand{\powerspectrumcutoffgponedhigh}{\ensuremath{142}}
\newcommand{\alphazerogptwod}{\ensuremath{-2.7}}
\newcommand{\alphazerogptwodlow}{\ensuremath{0.8}}
\newcommand{\alphazerogptwodhigh}{\ensuremath{0.6}}
\newcommand{\powerspectrumamplitudegptwod}{\ensuremath{65}}
\newcommand{\powerspectrumamplitudegptwodlow}{\ensuremath{24}}
\newcommand{\powerspectrumamplitudegptwodhigh}{\ensuremath{40}}
\newcommand{\powerspectrumcutoffgptwod}{\ensuremath{40}}
\newcommand{\powerspectrumcutoffgptwodlow}{\ensuremath{12}}
\newcommand{\powerspectrumcutoffgptwodhigh}{\ensuremath{13}}
\newcommand{\hlvkmltp}{\ensuremath{0.69}}
\newcommand{\hlvkmltplow}{\ensuremath{0.16}}
\newcommand{\hlvkmltphigh}{\ensuremath{0.18}}
\newcommand{\hlvkfullpopthreepeaks}{\ensuremath{0.74}}
\newcommand{\hlvkfullpopthreepeakslow}{\ensuremath{0.13}}
\newcommand{\hlvkfullpopthreepeakshigh}{\ensuremath{0.15}}

\newcommand{\gwtcfivehgponed}{h=\hgponed^{+\hgponedhigh}_{-\hgponedlow}}
\newcommand{\gwtcfivehgptwod}{h=\hgptwod^{+\hgptwodhigh}_{-\hgptwodlow}}
\newcommand{\alphazerogwtcfivegponed}{\alpharef=\alphazerogponed^{+\alphazerogponedhigh}_{-\alphazerogponedlow}}
\newcommand{\alphazerogwtcfivegptwod}{\alpharef=\alphazerogptwod^{+\alphazerogptwodhigh}_{-\alphazerogptwodlow}}
\newcommand{\powerspectrumamplitudegwtcfivegponed}{\powerspectrumamplitude=\powerspectrumamplitudegponed^{+\powerspectrumamplitudegponedhigh}_{-\powerspectrumamplitudegponedlow}}
\newcommand{\powerspectrumamplitudegwtcfivegptwod}{\powerspectrumamplitude=\powerspectrumamplitudegptwod^{+\powerspectrumamplitudegptwodhigh}_{-\powerspectrumamplitudegptwodlow}}
\newcommand{\powerspectrumcutoffgwtcfivegponed}{\powerspectrumcutoff=\powerspectrumcutoffgponed^{+\powerspectrumcutoffgponedhigh}_{-\powerspectrumcutoffgponedlow}}
\newcommand{\powerspectrumcutoffgwtcfivegptwod}{\powerspectrumcutoff=\powerspectrumcutoffgptwod^{+\powerspectrumcutoffgptwodhigh}_{-\powerspectrumcutoffgptwodlow}}
\newcommand{\gwtcfivehlvkmltp}{h=\hlvkmltp^{+\hlvkmltphigh}_{-\hlvkmltplow}}
\newcommand{\gwtcfivehlvkfullpop}{h=\hlvkfullpopthreepeaks^{+\hlvkfullpopthreepeakshigh}_{-\hlvkfullpopthreepeakslow}}

\newcommand{\gwtcfivegponedpeakone}{10}
\newcommand{\gwtcfivegponedpeaktwo}{18}
\newcommand{\gwtcfivegponedpeakthree}{34}
\newcommand{\gwtcfivegptwodpeakone}{10}
\newcommand{\gwtcfivegptwodpeaktwo}{20}

\newcommand{\hlvkmltpgwtcfivewide}{\ensuremath{0.69}}
\newcommand{\hlvkmltpgwtcfivewidelow}{\ensuremath{0.17}}
\newcommand{\hlvkmltpgwtcfivewidehigh}{\ensuremath{0.19}}
\newcommand{\hgwtcfivemltpcosmopyro}{\ensuremath{0.71}}
\newcommand{\hgwtcfivemltpcosmopyrolow}{\ensuremath{0.17}}
\newcommand{\hgwtcfivemltpcosmopyrohigh}{\ensuremath{0.19}}

\newcommand{\hmeasurelvkmltpgwtcfivewide}{h=\hlvkmltpgwtcfivewide^{+\hlvkmltpgwtcfivewidehigh}_{-\hlvkmltpgwtcfivewidelow}}
\newcommand{\hmeasurecosmopyromltpgwtcfivewide}{h=\hgwtcfivemltpcosmopyro^{+\hgwtcfivemltpcosmopyrohigh}_{-\hgwtcfivemltpcosmopyrolow}}

\newcommand{\hsimgponed}{\ensuremath{0.58}}
\newcommand{\hsimgponedlow}{\ensuremath{0.09}}
\newcommand{\hsimgponedhigh}{\ensuremath{0.09}}
\newcommand{\hsimgptwod}{\ensuremath{0.58}}
\newcommand{\hsimgptwodlow}{\ensuremath{0.10}}
\newcommand{\hsimgptwodhigh}{\ensuremath{0.11}}

\newcommand{\powerspectrumamplitudesimgponed}{\ensuremath{7.2}}
\newcommand{\powerspectrumamplitudesimgponedlow}{\ensuremath{2.5}}
\newcommand{\powerspectrumamplitudesimgponedhigh}{\ensuremath{5.4}}
\newcommand{\powerspectrumcutoffsimgponed}{\ensuremath{312}}
\newcommand{\powerspectrumcutoffsimgponedlow}{\ensuremath{74}}
\newcommand{\powerspectrumcutoffsimgponedhigh}{\ensuremath{93}}

\newcommand{\powerspectrumamplitudesimgptwod}{\ensuremath{49}}
\newcommand{\powerspectrumamplitudesimgptwodlow}{\ensuremath{18}}
\newcommand{\powerspectrumamplitudesimgptwodhigh}{\ensuremath{37}}
\newcommand{\powerspectrumcutoffsimgptwod}{\ensuremath{29}}
\newcommand{\powerspectrumcutoffsimgptwodlow}{\ensuremath{8}}
\newcommand{\powerspectrumcutoffsimgptwodhigh}{\ensuremath{10}}

\newcommand{\simhgponed}{h=\hsimgponed^{+\hsimgponedhigh}_{-\hsimgponedlow}}
\newcommand{\simhgptwod}{h=\hsimgptwod^{+\hsimgptwodhigh}_{-\hsimgptwodlow}}

\newcommand{\simpowerspectrumamplitudegponed}{\powerspectrumamplitude=\powerspectrumamplitudesimgponed^{+\powerspectrumamplitudesimgponedhigh}_{-\powerspectrumamplitudesimgponedlow}}
\newcommand{\simpowerspectrumcutoffgponed}{\powerspectrumcutoff=\powerspectrumcutoffsimgponed^{+\powerspectrumcutoffsimgponedhigh}_{-\powerspectrumcutoffsimgponedlow}}

\newcommand{\simpowerspectrumamplitudegptwod}{\powerspectrumamplitude=\powerspectrumamplitudesimgptwod^{+\powerspectrumamplitudesimgptwodhigh}_{-\powerspectrumamplitudesimgptwodlow}}
\newcommand{\simpowerspectrumcutoffgptwod}{\powerspectrumcutoff=\powerspectrumcutoffsimgptwod^{+\powerspectrumcutoffsimgptwodhigh}_{-\powerspectrumcutoffsimgptwodlow}}

\title{CosmoPyro: Gradients for Gravitational-Wave Cosmology}

\author{Konstantin Leyde$^a$\orcid{0000-0001-7661-2810},}
\author{Elena Colangeli$^b$\orcid{0009-0009-9783-3407}}
\affiliation[a]{Center for Computational Astrophysics, Flatiron Institute, 162 5th Ave, New York, NY 10010}
\affiliation[b]{Institute of Cosmology and Gravitation, University of Portsmouth, \\
Burnaby Road, Portsmouth PO1 3FX, United Kingdom}

\emailAdd{kleyde@flatironinstitute.org}
\emailAdd{elena.colangeli@port.ac.uk}

\abstract{
    Gravitational-wave (GW) observations of stellar-mass compact binary coalescences directly measure the source luminosity distance. Combined with the source redshift, these measurements constrain the current expansion rate of the Universe, the Hubble constant, $H_0$, or $h=H_0 / [100 \,{\rm km \,s^{-1} \, Mpc^{-1}}]$.
    For most GW signals no electromagnetic redshift measurement is expected, but the GW signal itself depends on the redshifted (detector-frame) masses.
    Assuming a source-frame mass distribution therefore enables a redshift estimate for each source. Combining the redshift estimates with the distance measurements provides a weak constraint on $H_0$ for each individual source that tightens with the number of sources in the catalog.
    However, the shape of the source-frame mass distribution is not known a priori, and previous work has relied on parametric models (piecewise power-laws with Gaussian components), and one-dimensional Gaussian processes.
    Here, we introduce \href{https://github.com/konstantinleyde/cosmopyro}{\textcolor{black}{\cosmopyro}}, a fully differentiable hierarchical Bayesian inference code that models the mass distribution using either one- or two-dimensional Gaussian processes.
    With the latest GW transient catalog (GWTC-5) we find $\gwtcfivehgponed$ and $\gwtcfivehgptwod$ (median with $1\sigma$ uncertainty), for the one- and two-dimensional case, respectively.
    Despite the noticeably different inferred mass distributions, both models yield $h$ values consistent with the latest LVK measurements within $1 \sigma$.
    While our main results marginalize over the Gaussian-process power-spectrum hyperparameters, the measurement is also robust against fixing these hyperparameters over a range comparable to their measured uncertainty.
}

\begin{document}
\maketitle

\section{Introduction}
\label{sec:intro}

The observation of \glspl{gw} from \glspl{cbc} provides a direct measurement of the luminosity distance to the source~\cite{Schutz:1986gp}.
When combined with redshift information, \gls{gw} sources can constrain any cosmological parameters that enter the luminosity distance-redshift relation, most notably the Hubble constant $\Hubble$, the current expansion rate of the Universe.
However, the redshift only enters the \gls{gw} signal through the redshifted source-frame masses, yielding the detector-frame mass $\md = \ms(1+z)$, with $\ms$ the source-frame mass.

In the absence of an identified \gls{em} counterpart, the ``spectral siren'' method exploits this relation to infer the Hubble constant $\Hubble$ (see~\cite{Taylor:2011fs, Taylor:2012db, Farr:2019twy, Mastrogiovanni:2021wsd, Ezquiaga:2022zkx} for an introduction).
Importantly, the source-frame mass distribution has features that, when translated to the detector-frame mass, evolve as a function of the luminosity distance.
The Hubble constant, $\Hubble$, controls how fast these features shift.
The spectral siren method is forecasted to achieve percent-level uncertainty for $h\deffrom \Hubble / [100\hu]$ with third-generation \gls{gw} detectors \cite{You:2020wju, Ezquiaga:2022zkx, Tagliazucchi:2026dpr}.
Identifying potential host galaxies from the GW sky localization can further inform the redshift of a given GW event \cite{Schutz:1986gp}. Thus, the spectral siren $h$ measurement can be improved \cite{DelPozzo:2011vcw, LIGOScientific:2018gmd, Gray:2019ksv, DES:2020nay, Gray:2021sew, Palmese:2021mjm, Finke:2021aom, Borghi:2023opd, Turski:2023lxq, Mastrogiovanni:2023emh, DESI:2023fij, Alfradique:2023giv, Gray:2023wgj, Gair:2022zsa, Dalang:2024gfk, Beirnaert:2025wcx, Cross-Parkin:2025xwf, Tagliazucchi:2025ofb, Naveed:2025kgk, Borghi:2025pav, Li:2025hrh, Turski:2025flk, McMahon:2026nhi, Andrade-Oliveira:2026jjm}, with simulated studies finding percent-level precision already in O5-like, albeit optimistic, observing scenarios \cite{Alfradique:2023giv, Borghi:2025pav}.

The precision of spectral siren measurements depends on the accuracy and flexibility of the assumed mass model~\cite{Ezquiaga:2022zkx, MaganaHernandez:2024uty, Mali:2024wpq}.
Broadly speaking, the modeling of the mass distribution can be categorized into parametric and non-parametric (sometimes referred to as weakly-modeled) approaches.
Parametric approaches model the mass distribution of the heavier component, $\msone$, as a power law, e.g.~$p(\msone | \hyper) \propto \msone^{-\alpha}$, or a broken power law, with additional over- or under-densities, whereas examples of non-parametric approaches include Gaussian processes or binned histograms (examples are discussed below).
Rather than assuming a specific functional form, non-parametric models impose smoothness through correlations of $p(\msone | \hyper)$ at nearby masses.
Both approaches have advantages and disadvantages.

Overly simple parametric models may introduce systematic biases if the true mass distribution deviates from the assumed functional form \cite{Mastrogiovanni:2021wsd}, and tend to extrapolate in regions where the data have low constraining power.
Non-parametric models have larger uncertainties in the estimated mass distribution, since fewer data points inform a given point of the probability distribution.
A balance between flexibility and constraining power is therefore important.

A number of works have modeled the mass distribution in a weakly parametric way for cosmology. The authors in \cite{Farah:2024xub} explore a Gaussian process on the (logarithm of the) heavier mass component, modeling the lighter component mass with a power law, while \cite{MaganaHernandez:2024uty} uses a one-dimensional binned Gaussian process for GWTC-3 \cite{LIGOScientific:2021djp} (the third data release from the LIGO-Virgo-KAGRA collaboration, following the third observing run, O3), but also models the two-dimensional mass distribution in $\msone$ and $\massratio$, the ratio of the two component masses, $\massratio\deffrom \mstwo/\msone$.
More recently, \cite{Tagliazucchi:2026gxn} uses a semi-parametric model for the heavier mass component, including splines to improve the $H_0$ constraints from GWTC-4 \cite{LIGOScientific:2025slb} (the fourth data release from the LIGO-Virgo-KAGRA collaboration, following the first part of the fourth observing run, O4a) spectral sirens to a relative $H_0$ uncertainty of 21\%, while the secondary mass follows a simpler model.
Other examples in GW cosmology for non-parametric models include \cite{Pierra:2025hoc} where the Hubble rate, $H(z)$, is modeled non-parametrically.

Upon fixing the cosmological parameters, various works have explored non-parametric mass models \cite{Tiwari:2020vym, Edelman:2021zkw, Edelman:2022ydv, KAGRA:2021duu, Sadiq:2023zee, Callister:2023tgi, Heinzel:2023hlb, Ray:2023upk, Ray:2024hos, Heinzel:2024hva, LIGOScientific:2025pvj, Alvarez-Lopez:2025ltt, Flanagan:2026ayy}.
We particularly highlight the \textsc{PixelPop} code \cite{Heinzel:2024jlc} since it models various combinations of the full distribution of mass ratio, redshift and effective spin through nearest-neighbor correlated bins.
More generally, see \cite{Edwards:2023sak, Talbot:2024yqw, Wouters:2025zju, Demasi:2026ltw} for applications of gradient-based inference in GW science.

In this work we present \cosmopyro{}, a fully differentiable spectral siren\footnote{While the code also allows for a dark siren analysis, i.e.~using other tracers of the \gls{cbc} distribution, such as galaxies \cite{Gray:2023wgj, Mastrogiovanni:2023emh} or hydrogen maps \cite{Dupletsa:2026uqs}, we restrict this present work to the spectral siren case. } code built on \jax~\cite{jax2018github} and \texttt{NumPyro}~\cite{Phan:2019elc, bingham2019pyro}.
By leveraging \gls{ad}, \cosmopyro{} enables the use of gradient-based samplers such as \gls{hmc} and \gls{nuts}~\cite{Hoffman:2011ukg}, which can efficiently explore the high-dimensional parameter spaces that arise in non-parametric mass models.
Our new contributions are two \gls{gp} mass distributions of increasing flexibility:
\begin{enumerate}
    \item A one-dimensional model (GP-1D, Sec.~\ref{subsec: mass distributions}) for $p(\msone | \hyper)$, with a conditional running power law for the mass ratio;
    \item A two-dimensional model (GP-2D, Sec.~\ref{subsec: mass distributions}) that jointly describes the distribution in log total source-frame mass $\logMtot \deffrom \log(\msone+\mstwo)$ and minus-log mass ratio $\minuslogq \deffrom - \log \massratio$.\footnote{Throughout, we use $\log$ to denote the natural logarithm. }
\end{enumerate}
We apply GP models in one and two dimensions to GWTC-5 data \cite{LIGOScientific:2026sit, LIGOScientific:2026wfs, LIGOScientific:2026ifv}, and subsequently to a dataset with O5-like sensitivity of the \gls{lvk} detectors \cite{det1-aligo2015,det2-aLIGO:2020wna,det3-Tse:2019wcy,det4-VIRGO:2014yos,det5-Virgo:2019juy}.
For simulated data, we show that the two flexible models recover the injected Hubble constant at the $1\sigma$ level.
On an A100 or H100 GPU, a production-level GP-2D analysis of the $\numsim$-event catalog takes approximately 15 to 25~hours to run.
The code is publicly available at \faGithub\,\href{https://github.com/konstantinleyde/cosmopyro}{\textcolor{black}{\cosmopyro}}.

This paper is organized as follows.
Sec.~\ref{sec:methodology} gives a brief review of the hierarchical Bayesian framework for spectral sirens and details the GP mass models.
Sec.~\ref{sec:data} describes the GWTC-5 data and our simulated dataset.
Following this structure, Sec.~\ref{sec:results} presents our results, first on GWTC-5 and then on simulated O5-like data.
We conclude in Sec.~\ref{sec:conclusions}.
The appendices specify the parametric mass model, the Gaussian process, the priors, and further technical details, as well as a number of validation and robustness tests.

Readers interested in the main results are referred to Fig.~\ref{fig: hubble constant comparison gwtc5}, which shows the inferred $h$ from GWTC-5, and Fig.~\ref{fig: reconstructed mass distribution comparison gp1d vs gp2d gwtc5}, which shows the corresponding reconstructed source-frame distribution.

\section{Methodology}
\label{sec:methodology}

We follow the standard hierarchical Bayesian framework for \gls{gw} population inference~\cite{Mandel:2018mve, Thrane:2018qnx, Vitale:2020aaz}.

\subsection{Hierarchical likelihood}
\label{sec:hierarchical_likelihood}

Given a catalog of $\nobs$ \gls{gw} events with data $\{\datagwsub{i}\}_{i=1}^{\nobs}$, the likelihood of the population hyperparameters $\hyper$ is \cite{Mandel:2018mve, Thrane:2018qnx, Vitale:2020aaz}
\begin{equation}
    \label{eq:hyperlikelihood}
    p\!\left(\{\datagwsub{i}\} \mid \hyper\right)
    \propto
    \prod_{i=1}^{\nobs}
    \frac{
        \int p(\datagwsub{i} \mid \params) \, p(\params \mid \hyper) \, \dd{\params}
    }{
        \int p_{\rm det}(\params) \, p(\params \mid \hyper) \, \dd{\params}
    },
\end{equation}
where $\params$ denotes the single-event parameters (for this particular problem, masses and redshift, but this can potentially also include sky position and spins), $p(\params \mid \hyper)$ is the population model, and $p_{\rm det}(\params)$ is the detection probability.
The detection probability depends on the detector sensitivities, which can vary day by day. Since we later consider a cumulative GW catalog extending over several observing runs, we compute a time-weighted average of the detection probabilities.
In practice, both the numerator and denominator integrals are typically evaluated via Monte Carlo sums (although for exceptions on the detection probability see \cite{Talbot:2020oeu, Gerosa:2020pgy, Callister:2024qyq, Lorenzo-Medina:2024opt}, and on single-event posterior samples see \cite{Mancarella:2025uat, Hussain:2025llf, Leyde:2026hvm}).

The approximation for the integral in the numerator for event $i$ through single-event posterior samples $\{\params_i^{(k)}\}_{k=1}^{n_i}$, obtained under a reference prior $\pi_{\rm ref}$ (where the index $k$ runs over all samples for this particular event, $k\in \{ 1,2,\ldots, n_i\}$) is
\begin{equation}
    \label{eq:numerator}
    \int p(\datagwsub{i} \mid \params) \, p(\params \mid \hyper) \, \dd{\params}
    \approx
    \frac{1}{n_i} \sum_{k=1}^{n_i}
    \frac{p(\params_i^{(k)} \mid \hyper)}{\pi_{\rm ref}(\params_i^{(k)})}\,,
\end{equation}
where the approximation refers to the finite number of posterior samples, and we have omitted the evidence $p(\datagwsub{i})$, which is independent of $\hyper$. 
The denominator (selection effect) is estimated using a set of simulated injections $\{\injparams^{(j)}\}_{j=1}^{\ninjtot}$ whose associated data have been passed through the detection pipeline, with
\begin{equation}
    \label{eq:selection}
    \int p_{\rm det}(\params) \, p(\params \mid \hyper) \, \dd{\params}
    \approx
    \frac{1}{\ninjtot} \sum_{j=1}^{\ninjsel}
    \frac{p(\injparams^{(j)} \mid \hyper)}{\pi_{\rm inj}(\injparams^{(j)})}\,,
\end{equation}
where the sum runs over the detected injections (a total of $\ninjsel$) and $\pi_{\rm inj}$ is the prior from which the injected signals are drawn.
While this prior can be unphysical, one has to ensure it covers a sufficiently large region of masses and distances compatible with the population models to be explored.

\subsection{Population model}
\label{sec:population_model}

We choose a population model that factorizes as
\begin{equation}
    \label{eq:pop_model}
    p(\params \mid \hyper) = p(\msone, \massratio \mid \hypermass) \, p(z \mid \hyperredshift, \hypercosmology) \,,
\end{equation}
where the mass and redshift distributions are modeled independently. The hyperparameters governing their respective distributions are denoted by $\hyperredshift$ (redshift), $\hypercosmology$ (cosmology) and $\hypermass$ (source-frame mass).
In the following section, we begin by summarizing the mass distribution we explore in this work, followed by the description of the redshift model.

\subsubsection{Source-frame mass distributions}
\label{subsec: mass distributions}

For both the one- and two-dimensional mass distribution explored here, we rely on Gaussian random fields, denoted as $\gaussianfield$.
These fields naturally enforce smoothness with a characteristic scale that can be related to the fields' two-point correlator.
Since we assume the correlations to be independent of the location in mass space, it is easier to work in the Fourier domain and use the \textit{power spectrum}, the Fourier transform of the two-point correlator.
The power spectrum controls the correlation lengthscale and the amplitude of the variations.
Thus, there are two important hyperparameters that are closely related to these: $\powerspectrumamplitude$ and $\powerspectrumcutoff$, see Eq.~\eqref{eq: power spectrum 1d} and Eq.~\eqref{eq: power spectrum 2d}.
The parameter $\powerspectrumamplitude$ determines the overall amplitude of the power spectrum, and $\powerspectrumcutoff$ sets the maximum inverse correlation length. A higher value of $\powerspectrumcutoff$ thus implies that the mass distribution is less smooth. 

To relate the Gaussian field to the mass distribution, we then set $\log p({\rm mass}) \sim \gaussianfield$.
For the technical details of how we generate random draws of the Gaussian random field, see App.~\ref{app:gaussian process}.

\begin{figure}[!tp]
    \includegraphics[width=\linewidth]{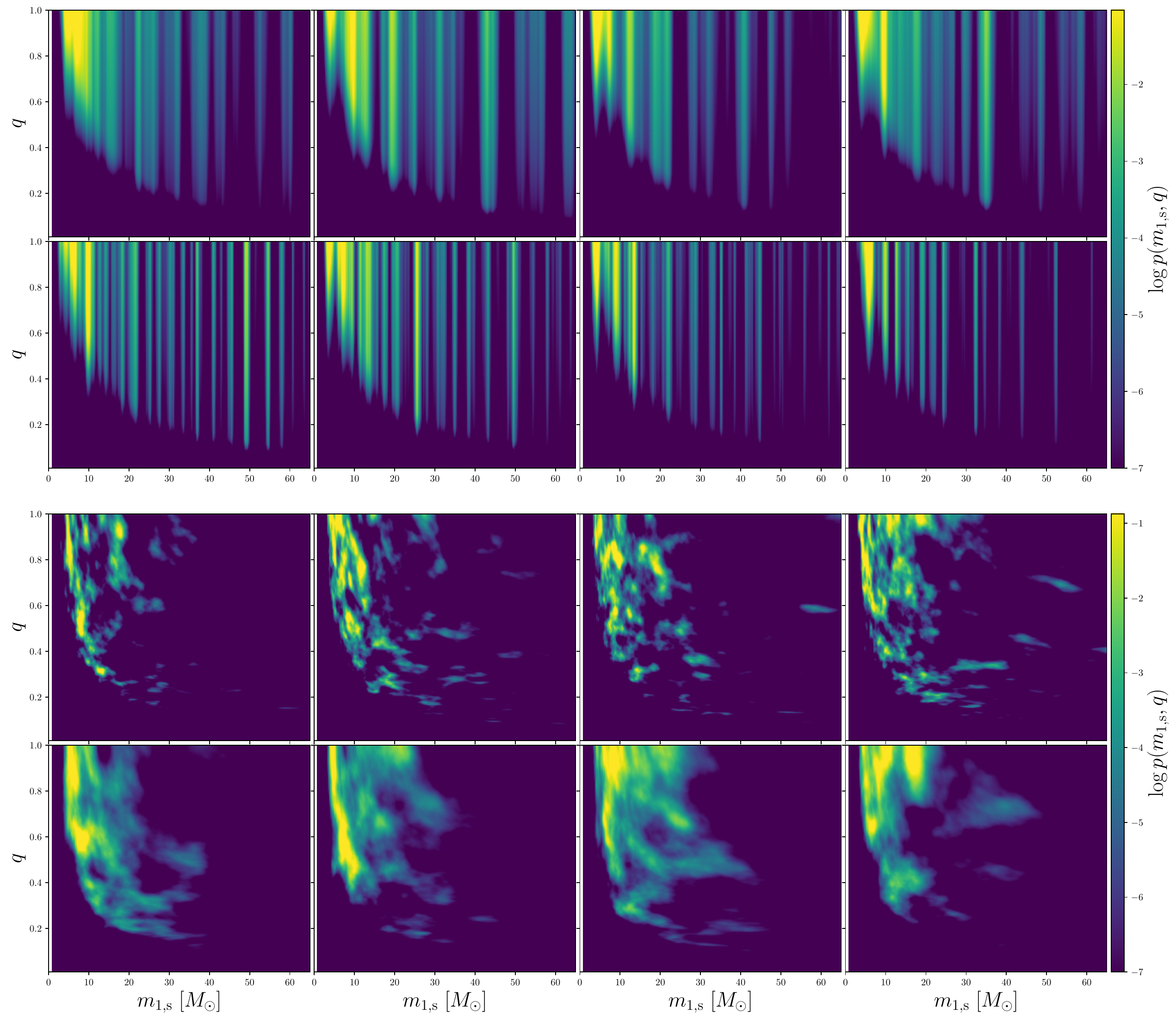}
    \caption{%
        Four prior draws from the GP-1D (two top rows) and GP-2D (two bottom rows) mass models.
        Each panel is an independent draw from the prior.
        Within the two blocks, the two rows illustrate the impact of the amplitude ($\powerspectrumamplitude$) and cutoff frequency ($\powerspectrumcutoff$) of the power spectrum.
        For a higher cutoff the distribution can exhibit sharper features, while a lower cutoff leads to smoother distributions.
        Tab.~\ref{tab:gp_parameters_prior_figure} summarizes corresponding mass parameter choices.
    }
    \label{fig:prior_draws}
\end{figure}

Readers who are primarily interested in the qualitative behavior of the mass distribution, rather than the technical details, may wish to proceed directly to Fig.~\ref{fig:prior_draws}, which shows representative prior draws for varying the aforementioned hyperparameters $\powerspectrumamplitude$ and $\powerspectrumcutoff$.

\modelheading{GP-1D (One-dimensional Gaussian process)}
We relate the one-dimensional Gaussian field to the mass distribution of the heavier component, $\msone$, via
\begin{equation}
\label{eq: link gaussian process mass distribution 1d}
    \log p(\msone | \hypermass, \text{GP-1D})
    =
    \gaussianfield
    +
    \log w(\msone)
    \,,
\end{equation}
where $w$ is a smooth window function given by
\begin{equation}
\label{eq: 1D window function}
    w(\msone)
    =
    \frac{1}{1+\exp\!\left[-\dfrac{\msone-\mmin}{\sigma_{\rm low}}\right]}
    \,
    \frac{1}{1+\exp\!\left[\dfrac{\msone-\mmax}{\sigma_{\rm high}}\right]} \,.
\end{equation}
We choose this specific functional form to avoid a vanishing probability, which is a nuisance for gradient computation and problematic for Hamiltonian Monte Carlo sampling.
The smoothing scales for the low- and high-mass cutoffs are taken to be proportional to the corresponding cutoff masses,
\begin{equation}
\label{eq:def sigma fractional}
    \sigma_{\rm low} = \sigmalowfrac\,\mmin,
    \qquad
    \sigma_{\rm high} = \sigmahighfrac\,\mmax,
\end{equation}
where $\sigmalowfrac$ and $\sigmahighfrac$ are dimensionless fractional smoothing parameters that are inferred jointly with the other mass parameters.
Since the probabilities at low and high mass differ by several orders of magnitude, we multiply by a power-law prior, so that the overall mass distribution becomes
\begin{equation}
\label{eq: Gp-1D all}
    p(\msone | \hypermass)
    \propto
    p(\msone | \hypermass, \text{GP-1D}) p(\msone | \hypermass, \text{PL})\,,
\end{equation}
with the power-law component given by $p(\msone | \hypermass, \text{PL}) \propto \msone^{\alpharef}$.
We infer this ``baseline power-law trend'' jointly with the Gaussian field.
The proportionality above becomes an equality upon adding a normalization constant $\lognorm(\hypermass)$ to $\log p(\msone | \hypermass)$; it is independent of $\msone$ and we leave it implicit throughout.

The mass ratio follows a power law with an exponent that runs with primary mass:
\begin{equation}
    \label{eq:mltp_q}
    p(\massratio \mid \msone, \hypermass) \propto \massratio^{\betaq(\msone)} \, \Ssmooth(\mstwo; \mmin, \deltamtwo)\,,
\end{equation}
where
\begin{equation}
    \label{eq:beta_running}
    \betaq(\msone) = \betazero + \betaone \log \frac{\msone}{\msref} \,,
\end{equation}
$\mstwo = \msone \, \massratio$, and $\deltamtwo$ controls the smoothing of the secondary mass at the low-mass boundary.
The parameters $\beta_0$ and $\beta_1$ are both dimensionless. 
The factor $\Ssmooth$ is a smooth approximation to the low-mass tapering function used in \gls{lvk} population analyses \cite{Talbot:2018cva, LIGOScientific:2026ctl}. We avoid the LVK smoothing window itself, since it vanishes below the lower mass boundary and can therefore be inconvenient in logarithmic calculations and gradient-based inference; the approximation we adopt instead is described in App.~\ref{app:mltp}.
Analogously to the LVK, this window has a free parameter that controls the width of the smoothing, which we denote by $\deltamtwo$.
Note that in the GP models the low- and high-mass cutoffs of the primary mass are controlled by the window function of Eq.~\eqref{eq: 1D window function}, i.e.~by $\sigmalowfrac$ and $\sigmahighfrac$; the smoothing scale $\deltam$ of the parametric \mltp model (App.~\ref{app:mltp}) does not enter here.

To illustrate the flexibility of the model, random draws of the one-dimensional field are plotted in the top panels of Fig.~\ref{fig:prior_draws}. The figure shows four prior draws along the horizontal axis for two choices of power-spectrum hyperparameters (along the vertical axis), with the corresponding parameters listed in Tab.~\ref{tab:gp_parameters_prior_figure}.
A smaller cutoff $\powerspectrumcutoff$ produces a distribution that varies more smoothly, while a larger cutoff allows for sharper features.

For computational efficiency, we use \glspl{fft}, which implicitly assume periodic signals. We thus generate the field on a sufficiently large domain and apply a window with a minimum and maximum mass to break this periodicity.

To complement the one-dimensional Gaussian process, we formulate in the following a flexible mass distribution in $\msone$ and the mass ratio $\massratio$.

\modelheading{GP-2D (Two-dimensional Gaussian process)}
We formulate the correlations of the source-frame mass distribution in the space of
\begin{equation}
\label{eq: def logMtot and minuslogq}
    \logMtot \deffrom \log (\msone + \mstwo) \,,
    \quad\quad
    \minuslogq \deffrom - \log \massratio \,.
\end{equation}
The Gaussian field then determines
\begin{equation}
\label{eq: link gaussian process mass distribution 2d}
    \log p(\logMtot, \minuslogq) =
    \gaussianfield + \log w(\logMtot, \minuslogq) \,,
\end{equation}
where $\gaussianfield$ is a Gaussian field and $w(\logMtot, \minuslogq)$ is a window function that enforces the smooth minimum and maximum mass cutoff in $\msone$ and $\mstwo$.
For each component, it is the same as Eq.~\eqref{eq: 1D window function}, where $\mmin$, $\mmax$, $\sigmalowfrac$, and $\sigmahighfrac$ are all inferred along with the Gaussian field.
As in the one-dimensional case, the additive normalization $\lognorm(\hypermass)$ is left implicit and fixed only once the baseline power-law trend of Eq.~\eqref{eq:prior_2D} below has been included.
To make the Gaussian-field contribution symmetric at unit mass ratio, and therefore give it zero slope along the $\minuslogq$ direction at $\massratio = 1$, we construct the whitened Gaussian-field components such that
\begin{equation}
    \gaussianfield(\logMtot, \minuslogq) = \gaussianfield(\logMtot, -\minuslogq)\,.
\end{equation}
To enforce this exchange symmetry, the white-noise field is mirrored in the $\minuslogq$ direction before applying the power spectrum and inverse \gls{fft}.
Only the physical half ($\minuslogq \ge 0$) is retained after the transform.

Similarly to the one-dimensional case, we use a baseline power-law trend on the mass distribution that is of the form
\begin{equation}
    \label{eq:prior_2D}
    p(\msone, \massratio) \propto \msone ^ {\alpharef} \, \massratio ^ {\betaref} \,,
\end{equation}
where $\alpharef$ and $\betaref$ are hyperparameters.
Only $\alpharef$ is inferred along with the Gaussian field, and we fix $\betaref=1$ throughout. While this aids the inference of the mass distribution, the impact (for reasonable values) of $\betaref$ on the Hubble constant is negligible.

The Jacobian relating the two-dimensional Gaussian process in $(\logMtot, \minuslogq)$ to the source-frame mass distribution in $(\msone, \massratio)$ is
\begin{equation}
    \label{eq:jacobian_sd}
    \left|\frac{\partial(\logMtot, \minuslogq)}{\partial(\msone, \massratio)}\right|
    = \frac{1}{\msone \, \massratio}\,.
\end{equation}
For evaluation, we factorize both the GP-1D and the GP-2D distribution into two one-dimensional distributions as
\begin{equation}
    p(\msone, \massratio) = p(\msone | \hypermass) \, p(\massratio | \msone, \hypermass)\,.
\end{equation}
For the GP-2D mass model, the Gaussian field itself is constructed uniformly in the $\logMtot$ and $\minuslogq$ coordinates, and $p(\logMtot, \minuslogq)$ is interpolated via cubic polynomials to the $(\msone, \massratio)$ coordinates.
We infer the whitened Gaussian-field components since they have a simpler structure facilitating sampling.

See the two lower panels of Fig.~\ref{fig:prior_draws} for a visualization of the GP-2D prior draws in the $(\msone, \massratio)$ coordinates.
We show four prior draws for two choices of the power-spectrum parameters $\powerspectrumamplitude$ and $\powerspectrumcutoff$.
App.~\ref{app: meta choices} explores the impact of these choices on the inferred source-frame mass distribution and the Hubble constant.

When compared to the GP-2D model, the GP-1D mass model imposes a more rigid structure on the mass distribution since it assumes $p(\massratio|\msone, \hypermass)$ to be a power law at all primary masses, with $\msone$ only setting the (slowly running) exponent of Eq.~\eqref{eq:beta_running} and the lower mass-ratio bound, while the GP-2D mass model can capture more complex correlations.

We also implement a parametric \mltp-like model (App.~\ref{app:mltp}), almost identical to the \mltp model used in \gls{lvk} population analyses~\cite{LIGOScientific:2021aug} but with a slightly different low-mass smoothing; a comparison against the \gls{lvk} GWTC-5 \mltp spectral siren result can be found in App.~\ref{app: validation against lvk gwtc-5 result}.

\begin{table}
\centering
\begin{tabular}{@{}c@{\hspace{1em}}c@{}}
\begin{minipage}[t]{0.48\textwidth}
\vspace{0pt}
\centering
\begin{tabular}{lll}
\toprule
\multicolumn{3}{c}{\textbf{GP-1D}} \\
\midrule
Parameter & Value & Reference \\
\midrule
$m_{\min}~[\msun]$ & 4.0 & Eq.~\eqref{eq: 1D window function} \\
$m_{\max}~[\msun]$ & 80.0 & Eq.~\eqref{eq: 1D window function} \\
$\alpharef$ & $-2.0$ & Eq.~\eqref{eq: Gp-1D all} \\
$\sigma_{\rm f, low}$ & 0.05 & Eq.~\eqref{eq:def sigma fractional} \\
$\sigma_{\rm f, high}$ & 0.05 & Eq.~\eqref{eq:def sigma fractional} \\
$\beta_0$ & 1.0 & Eq.~\eqref{eq:beta_running} \\
$\beta_1$ & 0.0 & Eq.~\eqref{eq:beta_running} \\
$m_{\mathrm{ref}}~[\msun]$ & 10.0 & Eq.~\eqref{eq:beta_running} \\
$\deltamtwo~[\msun]$ & 3.0 & Eq.~\eqref{eq:mltp_q} \\
$\powerspectrumamplitude$ & $10.0$ or $20.0$ & Eq.~\eqref{eq: power spectrum 1d} \\
$\powerspectrumcutoff$ & $400.0$ or $800.0$ & Eq.~\eqref{eq: power spectrum 1d} \\
\bottomrule
\end{tabular}
\end{minipage}
&
\begin{minipage}[t]{0.48\textwidth}
\vspace{0pt}
\centering
\begin{tabular}{lll}
\toprule
\multicolumn{3}{c}{\textbf{GP-2D} } \\
\midrule
Parameter & Value & Reference \\
\midrule
$m_{\min}~[\msun]$ & 4.0 & Eq.~\eqref{eq: 1D window function} \\
$m_{\max}~[\msun]$ & 120.0 & Eq.~\eqref{eq: 1D window function} \\
$\alpharef$ & $-3.0$ & Eq.~\eqref{eq:prior_2D} \\
$\sigma_{\rm f, low}$ & 0.05 & Eq.~\eqref{eq:def sigma fractional} \\
$\sigma_{\rm f, high}$ & 0.05 & Eq.~\eqref{eq:def sigma fractional} \\
$\betaref$ & 1.0 & Eq.~\eqref{eq:prior_2D} \\
$\powerspectrumamplitude$ & $120.0$ or $60.0$ & Eq.~\eqref{eq: power spectrum 2d} \\
$\powerspectrumcutoff$ & $80.0$ or $40.0$ & Eq.~\eqref{eq: power spectrum 2d} \\
\bottomrule
\end{tabular}
\end{minipage}
\end{tabular}

\caption{Reference parameters for prior draws (cf.~Fig.~\ref{fig:prior_draws}) used to generate the Gaussian-process mass distributions for the one- and two-dimensional models.
For the power-spectrum parameters, the first (second) number refers to the top (bottom) row of each of the mass models of Fig.~\ref{fig:prior_draws}. }
\label{tab:gp_parameters_prior_figure}
\end{table}

\subsubsection{Redshift distribution}

The redshift distribution is taken to be uniform in comoving volume and source-frame time,
\begin{equation}
    p(z \mid \hyperredshift, \hypercosmology) \propto \frac{\dVdz(z; \hypercosmology)}{1+z} \, \psi(z; \hyperredshift)\,,
\end{equation}
modulated by the Madau--Dickinson star formation rate~\cite{Madau:2014bja} with parameters $\hyperredshift = \{\gamma, z_p, \kappa\}$
\begin{equation}
\label{eq: def madau dickinson}
    \psi(z; \hyperredshift) = \frac{(1+z)^\gamma}{1 + \left(\frac{1+z}{1+z_p}\right)^{\gamma+\kappa}}\,.
\end{equation}
This is a standard modeling assumption in GW cosmology, see for instance \cite{LIGOScientific:2021aug, LIGOScientific:2025jau, LIGOScientific:2026uyd}.

\section{Data}
\label{sec:data}

\subsection{GWTC-5}

For the first part of the analysis, we use \gls{bbh} events from the GWTC-5 catalog~\cite{LIGOScientific:2026sit, LIGOScientific:2026wfs, LIGOScientific:2026ifv}.
Following the LVK work on constraints on the cosmic expansion history \cite{LIGOScientific:2025jau, LIGOScientific:2026uyd}, we use the results of the search pipelines \texttt{gstlal} \cite{Sachdev:2019vvd}, \texttt{pycbc} \cite{Usman:2015kfa}, and \texttt{mbta} \cite{Aubin:2020goo}.
We follow the LVK prescription in \cite{LIGOScientific:2026uyd} for event selection and require all events to have an \gls{ifar} of at least $4~{\rm yr}$.
This results in a catalog of 231 BBH mergers.
The LVK single-event posterior samples were obtained with the \texttt{bilby} parameter estimation pipeline~\cite{Ashton:2018jfp} and are available at \cite{ligo_scientific_collaboration_and_virgo_2022_6513631, ligo_scientific_collaboration_and_virgo_2021_5546663, ligo_scientific_collaboration_and_virgo_2025_16053484, ligo_scientific_collaboration_and_virgo_2026_20348006}, using between $10{,}000$ and $20{,}000$ posterior samples per event (depending on availability), with the \texttt{IMRPhenomXPHM}, \texttt{IMRPhenomXPHM-SpinTaylor}~\cite{Pratten:2020ceb}, and \texttt{NRSur7dq4}~\cite{Varma:2019csw} waveform models.
The selection function is estimated from the corresponding injection campaign \cite{Essick:2025zed}.
Also following the LVK prescription, we classify O1 and O2 injections as detected if their \gls{snr} exceeds 10, while injections from O3, O4a and O4b are classified using the IFAR threshold defined above.
While this selection is not exactly the same as the one used for real events, it is a good approximation for the purpose of estimating the selection function \cite{Essick:2023toz}.
This selection leads to a total of $\sim 1.5\times 10^6$ detected simulated GW signals.

\begin{table}[t]
\centering
\renewcommand{\arraystretch}{1.3}
\setlength{\tabcolsep}{4.5pt}
\begin{tabular}{lccccccccccc}
\hline
Param. &
$\alphapl$ &
$\mmin$ &
$\mmax$ &
$\deltam$ &
$\lambdag$ &
$\lambdaglow$ &
$\muglow$ &
$\sigmaglow$ &
$\mughigh$ &
$\sigmaghigh$ &
$\betaq$ \\
\hline
Units &
-- &
$\msun$ &
$\msun$ &
$\msun$ &
-- &
-- &
$\msun$ &
$\msun$ &
$\msun$ &
$\msun$ &
-- \\
Value &
2.61 &
5.24 &
69.44 &
3.05 &
0.44 &
0.91 &
9.49 &
0.45 &
31.76 &
1.40 &
0.85 \\
\hline
\end{tabular}
\renewcommand{\arraystretch}{1.0}
\caption{Parameters of the \mltp mass distribution model used to generate the simulated O5-like population.}
\label{tab:mass_model_parameters}
\end{table}

\subsection{Simulated O5-like data}
\label{sec: data simulated_data}

We generate a simulated catalog of $\numsim$ \gls{bbh} events, the conservative expectation of BBH detections during the fifth observing run (O5) \cite{KAGRA:2013rdx, Kiendrebeogo:2023hzf}. 
This synthetic catalog allows us to validate our analysis pipeline and determine the optimal settings for recovering the simulated mass distribution and Hubble constant.
The events are drawn from a \mltp source-frame mass distribution (as defined in App.~\ref{app:mltp}, parameters given in Tab.~\ref{tab:mass_model_parameters}), and redshifts are distributed according to a Madau--Dickinson redshift distribution with parameters $\gamma = 2.7$, $\kappa=3.0$ and $z_{\rm p}=2.0$ (cf.~Eq.~\eqref{eq: def madau dickinson}).
We assume a flat $\Lambda$CDM cosmology with parameters $h=0.67$ and $\Omegam=0.319$ to convert redshifts into luminosity distances.
We use a simplified \gls{snr} computation and parameter estimation, outlined in App.~\ref{app: mock pe}, mimicking capabilities at an O5 sensitivity configuration.
For the single-event parameter estimation we assume a prior that is uniform in detector-frame masses and luminosity distance, and use $15{,}000$ samples per event.

As selection criterion we impose a matched-filter \gls{snr} greater than $12$ (cf.~Eq.~\eqref{eq: scattered snr}), and produce $7.6\times 10^6$ detected injections.
The furthest detection of the resulting catalog is at a true distance of $10{,}000\,$Mpc, with the peak of detected sources at $2{,}000\,$Mpc.

\section{Results}
\label{sec:results}

In the following, we present the inferred Hubble constant and reconstructed mass distributions with the GP-1D and GP-2D models.
We first apply these models to the 231 BBH events of GWTC-5.
We then assess the faithfulness of the reconstruction for an O5-like dataset with $\numsim$ simulated GW events.
Throughout, we assume a flat $\Lambda$CDM Universe, fixing $\Omegam=0.3$.\footnote{The $\Omegam$ value does not exactly match the simulated GW catalog of Sec.~\ref{sec:results_simulated}, but this mismatch is negligible for current observations, see \cite{LIGOScientific:2025jau}. }

\subsection{Sampling}

The likelihood of Eq.~\eqref{eq:hyperlikelihood} is implemented in \jax~\cite{jax2018github}, which enables efficient automatic differentiation.
We draw samples of the posterior distribution through the NUTS sampler~\cite{Neal:2011mrf, Hoffman:2011ukg, Betancourt:2017ebh}, available in the \numpyro{} package \cite{Phan:2019elc}.
The GP-1D (GP-2D) mass model has $\mathcal{O}(1{,}500)$ ($\mathcal{O}(25{,}000)$) inferred Gaussian-field parameters, cf.~Tab.~\ref{tab:binning_details}.
Apart from these parameters, we adapt a dense (non-diagonal) mass matrix for all remaining parameters.
The chain undergoes $2{,}000$ warm-up steps, and we subsequently draw $3{,}000$--$20{,}000$ posterior samples, depending on the run, using either an A100 or an H100 GPU.
While the \mltp mass model runs typically take $\mathcal{O}(1)$\,h for GWTC-5, the one- and two-dimensional Gaussian-process models take $5$--$25$\,h for production results.
The priors are outlined in App.~\ref{app: priors}, and see App.~\ref{app: implementation details} for implementation details such as the binning resolution. 

Since this is a high-dimensional inference problem, we assess the robustness of sampler convergence by running selected analyses multiple times, varying the random seed, binning resolution and prior ranges.
We find indistinguishable inferred mass distributions and $h$ posteriors, with the exception of the power-spectrum parameters that can influence the reconstructed mass distributions, see App.~\ref{app: meta choices}. 
Hence, the main results of the following section marginalize over these parameters.

\subsection{GWTC-5}
\label{sec:results_gwtc5}

\begin{figure}
    \centering
    \includegraphics[width=0.8\linewidth]{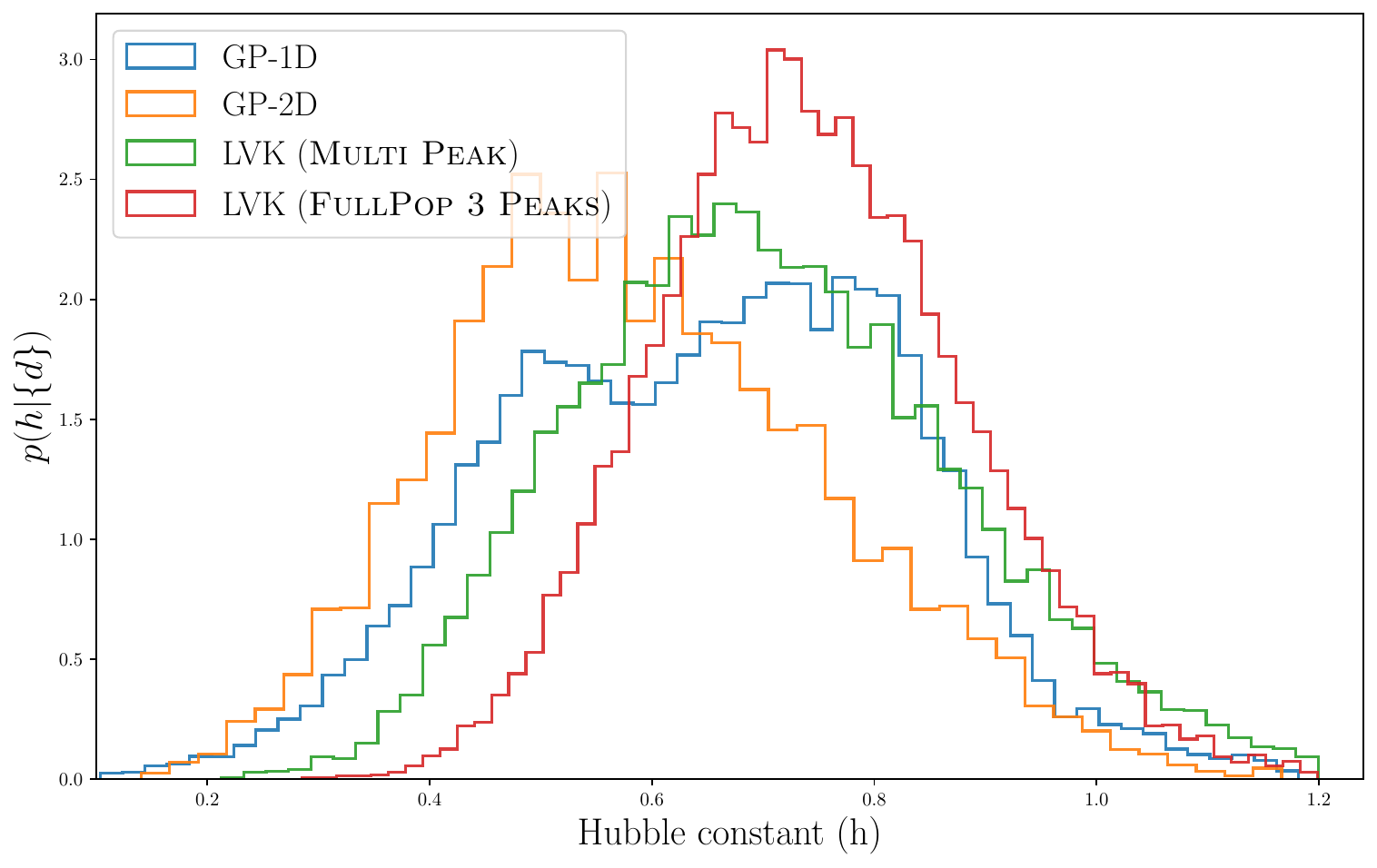}
    \caption{Hubble constant posterior using GWTC-5 with the two different GP mass models.
    The corresponding reconstructed mass distributions are shown in Fig.~\ref{fig: reconstructed mass distribution comparison gp1d vs gp2d gwtc5}.
    For comparison the LVK results for the \mltp and \fullpopthreepeak models are also shown.
    }
    \label{fig: hubble constant comparison gwtc5}
\end{figure}

\begin{figure}
    \centering
    \includegraphics[width=0.87\linewidth]{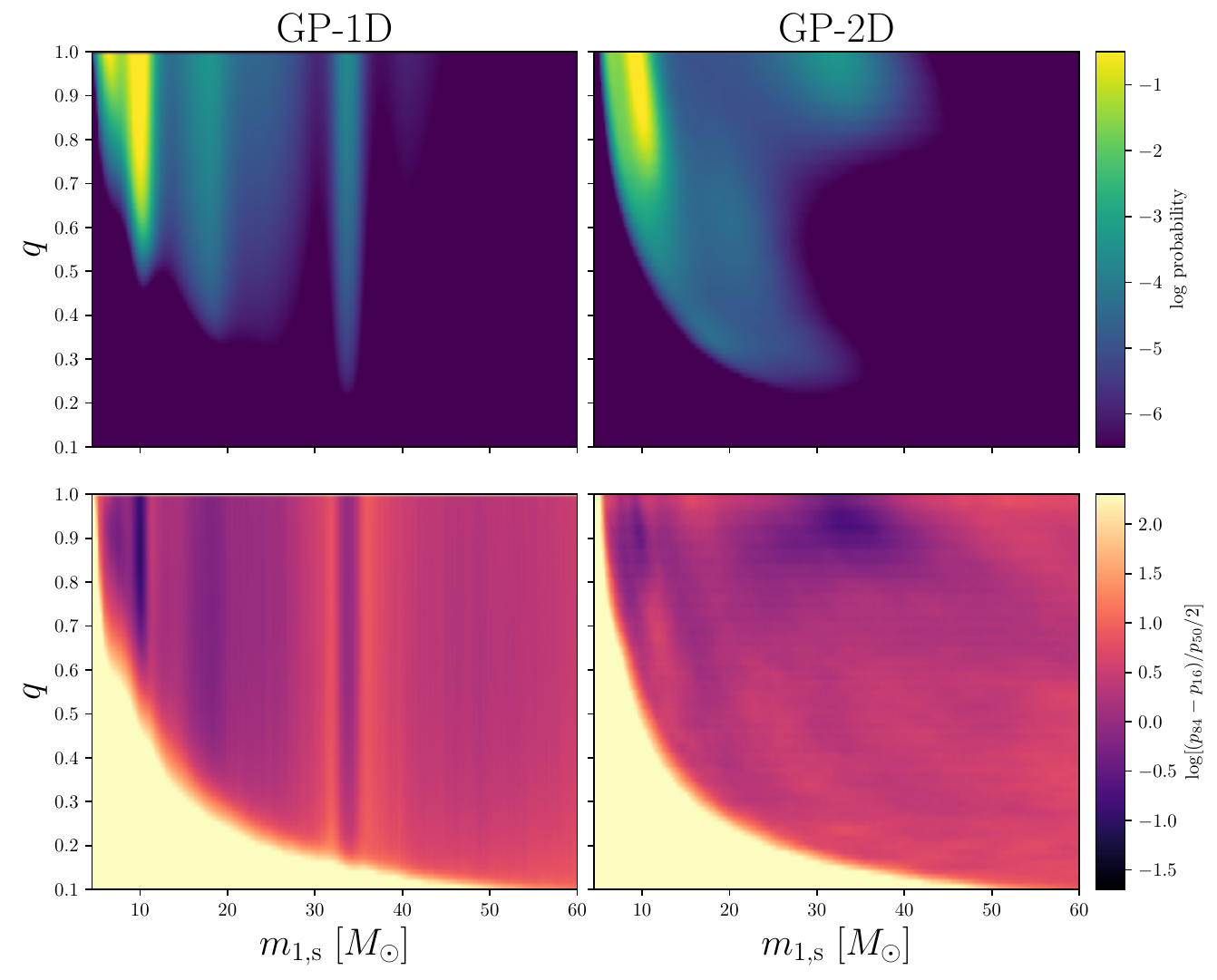}
    \includegraphics[width=0.87\linewidth]{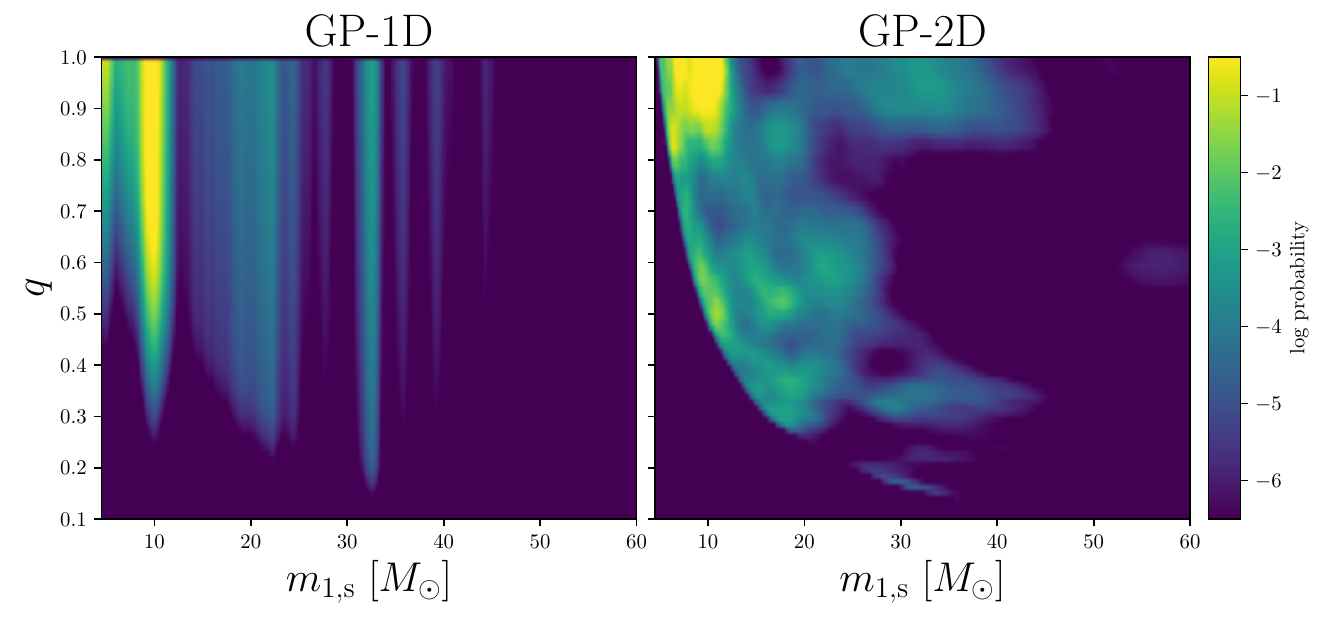}
    \caption{Reconstructed median (\textit{top}) and uncertainty (\textit{middle}, cf.~Eq.~\eqref{eq:def uncertainty measure}) for the GP-1D and GP-2D mass models for GWTC-5.
    In the lower panel we pick a mass-distribution posterior sample at random. This illustrates (similar to the prior draw of Fig.~\ref{fig:prior_draws}) the versatility and difference in structure of the mass models.
    }
    \label{fig: reconstructed mass distribution comparison gp1d vs gp2d gwtc5}
\end{figure}

\begin{figure}
    \centering
    \includegraphics[width=1.0\linewidth]{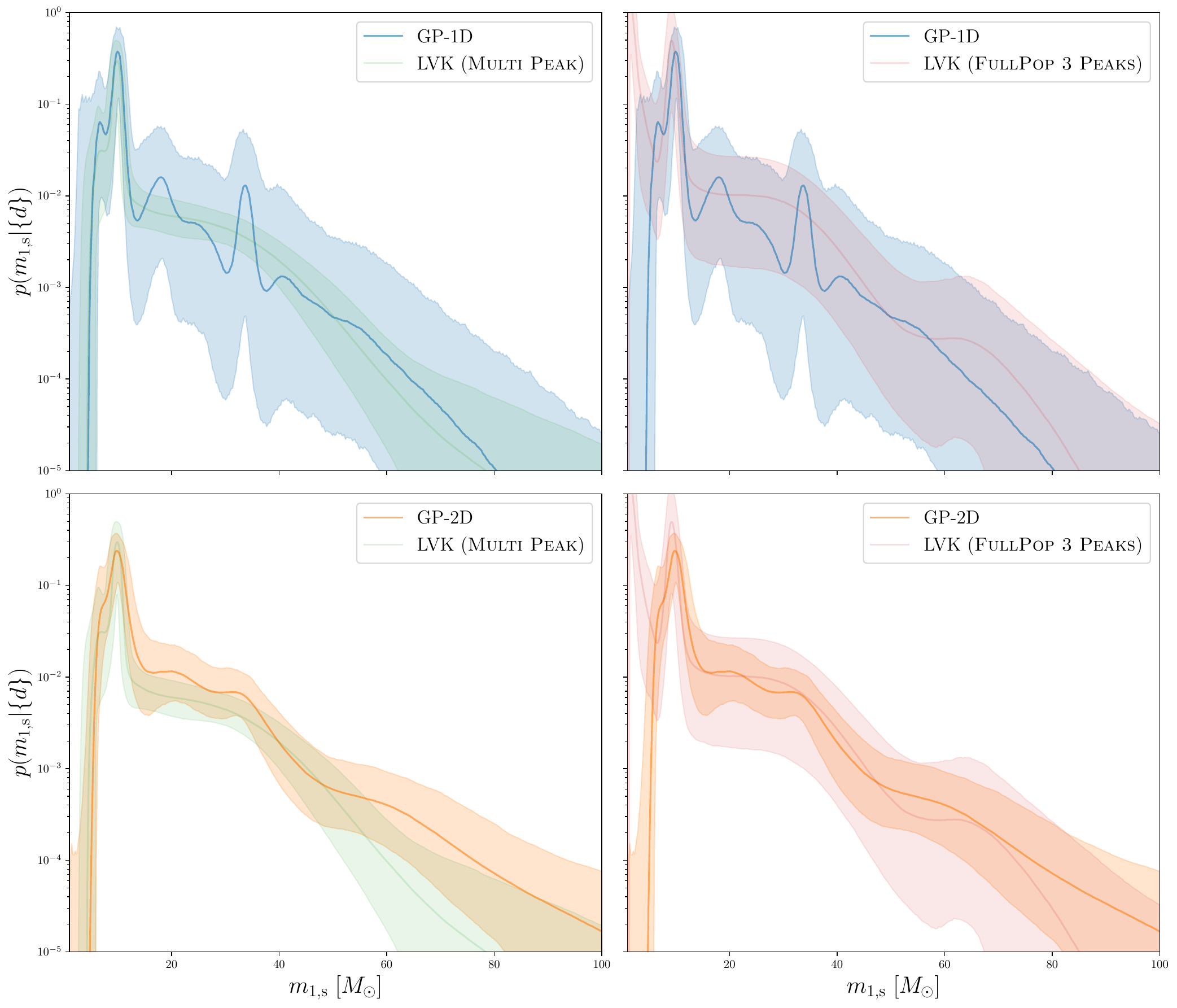}
    \caption{Reconstructed source-frame mass distribution for GWTC-5 (projection of the top panel of Fig.~\ref{fig: reconstructed mass distribution comparison gp1d vs gp2d gwtc5} before the median is computed).
    Indicated are the median (solid curve) with the central 95\% credible band.
    The two top panels show the GP-1D mass model compared to the two LVK mass models (left column: \mltp, right column: \fullpopthreepeak).
    Analogously, the bottom panels show the same comparisons for the GP-2D mass model.
    Since the \fullpopthreepeak model describes all binaries (including neutron stars), it extends to lower masses. Therefore, we have adjusted the normalization scale to facilitate visual comparison.
    }
    \label{fig:comparison gp1d vs gp2d gwtc5 marginal m1s}
\end{figure}

We analyze the 231 \glspl{bbh} of GWTC-5 using the GP-1D and GP-2D mass models. 
The power-spectrum parameters are marginalized over; their priors are listed in Tab.~\ref{tab:settings_diffs prior on power spectrum parameters}, and App.~\ref{app: meta choices} discusses variations of these parameters.

The Hubble constant is inferred to be $\gwtcfivehgponed$ (GP-1D) and $\gwtcfivehgptwod$ (GP-2D), where the quoted uncertainties represent the 16th and 84th percentiles around the median, a convention adopted throughout this work.
The corresponding $h$ posterior distributions are shown in Fig.~\ref{fig: hubble constant comparison gwtc5}.
While the $h$ uncertainty is similar for both models, the GP-1D mass model recovers a larger $h$ median value.
As shown below, this is tied to the GP-1D reconstructed mass distribution being skewed towards lower masses when compared to the GP-2D model.
The $h$ posteriors can be compared to the most recent LVK publication \cite{LIGOScientific:2026uyd}, which finds\footnote{In this work we assume a uniform prior of $p(h) = \mathcal{U}(0.1, 1.2)$. To adjust the LVK results with a prior of $p(h) = \mathcal{U}(0.1, 2.0)$ to the results presented here, we reduce the LVK posteriors (\mltp and \fullpopthreepeak) to this prior range. } $\gwtcfivehlvkmltp$ (\mltp mass model \cite{KAGRA:2021duu}) and $\gwtcfivehlvkfullpop$ (\fullpopthreepeak mass model \cite{Pierra:2026ffj}).
Note that the \fullpopthreepeak model uses five additional \glspl{cbc}, as it also models systems that contain \glspl{ns}.
While the GP-1D mass model infers $h$ with an almost identical median as the \mltp, the GP-2D $h$ posterior is shifted to smaller values with a similar uncertainty.
This similarity of the GP-1D $h$ posterior to the LVK \mltp posterior might be related to the shared structure of $p(\massratio | \msone, \hyper)$, which both the parametric LVK mass distributions and the GP-1D model describe as a power law.
Even though both GP mass models show similar $h$ uncertainty, this cannot be directly tied to the uncertainty in the reconstructed $p(\msone | \hyper)$, which differs between the two models, as shown below.
Overall, the results remain statistically consistent, with discrepancies below the $1\sigma$ level.
Both GP-based mass models have slightly increased $h$ uncertainty, as expected from a mass model that introduces more degrees of freedom. 

Fig.~\ref{fig: reconstructed mass distribution comparison gp1d vs gp2d gwtc5} plots the two-dimensional mass distribution $p(\msone, \massratio)$.
The correlation structure for the GP-2D process carries over to the reconstructed two-dimensional mass distribution: some features are extended along lines of constant $\logMtot$ (cf.~Eq.~\eqref{eq: def logMtot and minuslogq}), which corresponds to features at $\msone = e^{\logMtot} / (1 + \massratio)$.
Similar structures are already visible in the prior draws of Fig.~\ref{fig:prior_draws}.
To facilitate comparison, we plot the marginal $p(\msone | \hyper)$ distribution in Fig.~\ref{fig:comparison gp1d vs gp2d gwtc5 marginal m1s}.
The top panels compare the GP-1D mass model against the LVK \mltp and \fullpopthreepeak mass model (left and right columns, respectively), whereas the bottom panels show the same comparison for the GP-2D model.
While the median of the GP-1D model exhibits more pronounced peaks\footnote{The more strongly varying GP-1D median is directly related to the amplitude of the power spectrum.
For a smaller value of the amplitude of the power spectrum one can enforce a more smoothly varying median, see App.~\ref{app: meta choices}. }, the 95\% credible band encompasses the strongly parametric models for almost the full mass range.
The GP-1D model finds peaks at approximately $\gwtcfivegponedpeakone\msun$, $\gwtcfivegponedpeaktwo\msun$, and $\gwtcfivegponedpeakthree\msun$.
%
The median of the GP-2D mass model also approximately recovers the $\gwtcfivegponedpeakone\msun$ peak of the GP-1D distribution, while the two remaining peaks (for the GP-1D model at $\gwtcfivegponedpeaktwo\msun$ and $\gwtcfivegponedpeakthree\msun$) are less apparent and resemble more breaks than peaks.
The $\gwtcfivegponedpeakone\msun$ peak and the break at $\gwtcfivegponedpeakthree\msun$ are compatible with (but not identical to) the LVK population analysis result \cite{LIGOScientific:2026ctl} (with the latter analysis fixing $h$).
The GP-2D model also finds a break at $\sim63\msun$, corroborating the feature found with GWTC-4 in \cite{Pierra:2026ffj}.  
 
The inferred power-law slope parameters are found to be $\alphazerogwtcfivegponed$ (GP-1D) and $\alphazerogwtcfivegptwod$ (GP-2D).
The GP-2D model favors a steeper overall power-law slope parameter.
However, due to the inferred Gaussian field, the GP-2D distribution still has a longer tail to large masses, which is also visible in Fig.~\ref{fig:comparison gp1d vs gp2d gwtc5 marginal m1s}.
As mentioned previously, the GP-1D finds more mass probability at low masses than the GP-2D model, which instead has a longer tail to high masses; this leads to a lower inferred $h$ value for the GP-2D model.

Let us stress that the median is just one possible summary statistic of the $p(\msone | \hyper)$ distribution, and the emerging structure (e.g.~the aforementioned peaks) should not be overinterpreted.
For example, for different choices of the power-spectrum parameter $\powerspectrumcutoff$, the $\gwtcfivegponedpeaktwo\msun$ break can resemble a peak for settings~[3] (cf.~Tab.~\ref{tab:settings_diffs prior on power spectrum parameters}) with the GP-2D model.

For the $\gwtcfivegptwodpeakone\msun$ peak, the GP-2D model finds a mass-ratio distribution peaking at unity. 
The mass-ratio distribution at $\gwtcfivegptwodpeaktwo\msun$ is roughly flat, which is difficult to accommodate within the GP-1D model, since its conditional mass-ratio distribution is restricted to a running power-law form.
The reconstructed mass distribution with the GP-2D model of Fig.~\ref{fig: reconstructed mass distribution comparison gp1d vs gp2d gwtc5} suggests that the underlying joint mass distribution might deviate from the simplified conditional structure $p(\massratio|\hyper,\msone)$ imposed by the GP-1D model.
This interpretation is also consistent with several studies that infer the source-frame mass distribution while fixing $h$, see \cite{Li:2022jge, Farah:2023swu, Godfrey:2023oxb, Li:2024jzi, Roy:2025ktr, Banagiri:2025dmy, Sridhar:2025kvi, Flanagan:2026ayy, LIGOScientific:2025pvj, Tong:2025wpz, LIGOScientific:2026ctl, Ray:2026uur, Mould:2026sww, Godfrey:2026pbc, Guttman:2026cnv}.

We define the relative uncertainty (plotted in the middle row of Fig.~\ref{fig: reconstructed mass distribution comparison gp1d vs gp2d gwtc5}) as
\begin{equation}
\label{eq:def uncertainty measure}
    \text{uncertainty measure}
    \deffrom
    \log
    \left[
        \frac{p_{84\%}(\ms) - p_{16\%}(\ms)}{2\,p_{50\%}(\ms)}
    \right]
    \,,
\end{equation}
where $p_{X\%}(\ms)$ is the $X$th percentile for the source-frame mass distribution at mass $\ms$.
This uncertainty is expected to be lower in regions with more detected events, a property of any Poisson process. 

Indeed, for both distributions in Fig.~\ref{fig: reconstructed mass distribution comparison gp1d vs gp2d gwtc5}, a region of low uncertainty is the $\gwtcfivegponedpeakone\msun$ peak, $\massratio\approx 1$.
While the uncertainty significantly increases for unequal mass ratios at this peak for the GP-2D model, the GP-1D extrapolates low uncertainties down to mass ratios as low as $0.6$.
Another region of low uncertainty for both models is the $\gwtcfivegponedpeakthree\msun$ peak. The GP-2D model infers a large region around this peak to have low uncertainty which is tied to the large inferred correlation lengthscale that we discuss below. 
The GP-1D reconstructs the peak itself with low uncertainties, but the probability of the neighboring masses shows large uncertainty.  
Across the entire parameter space, both models show a similar uncertainty range.
However, for $p(\msone | \hyper)$  the uncertainty of the GP-2D model  is noticeably smaller than for the GP-1D model (Fig.~\ref{fig:comparison gp1d vs gp2d gwtc5 marginal m1s}). 
This is partially driven by the GP-2D model averaging over more fluctuations along the mass-ratio direction. Since the two-dimensional model imposes isotropy along both dimensions (via the power spectrum), marginalizing over a large enough volume yields a small variance.

The power-spectrum parameters (cf.~Eq.~\eqref{eq: power spectrum 1d} and Eq.~\eqref{eq: power spectrum 2d}) are inferred for both mass models. 
For the power-spectrum amplitude, $\powerspectrumamplitude$, the models recover $\powerspectrumamplitudegwtcfivegponed$ (GP-1D) and $\powerspectrumamplitudegwtcfivegptwod$ (GP-2D), while
the measured cutoff frequency of the correlations differs strongly: the GP-1D mass model finds $\powerspectrumcutoffgwtcfivegponed$, whereas the GP-2D model favors $\powerspectrumcutoffgwtcfivegptwod$.
The two numbers should not be compared directly, since $\powerspectrumcutoff$ is conjugate to different coordinates in the two models (cf.~App.~\ref{app:gaussian process} and Tab.~\ref{tab:binning_details}): the linear primary mass $\msone$ for the GP-1D model, and the logarithmic coordinates $\logMtot$ and $\minuslogq$ for the GP-2D model, each rescaled to unit length.
The correlation length of the GP-1D model is therefore a fixed absolute mass scale, whereas for the GP-2D model it is a fixed fractional one.
This preference of fewer fluctuations for the GP-2D model manifests itself in Fig.~\ref{fig: reconstructed mass distribution comparison gp1d vs gp2d gwtc5}, where the mass distribution varies more smoothly than in the GP-1D case. 
If the true correlation lengthscale is different for the two dimensions ($\logMtot$ and $\minuslogq$, cf.~Eq.~\eqref{eq: def logMtot and minuslogq}), this could bias the final result, since the inferred correlation lengthscale would be a weighted average of the two: a smooth $\minuslogq$ distribution could enforce a smoothed-out $\logMtot$ distribution, or vice versa.

In principle, this could suggest that a model inferring a lower $\powerspectrumcutoff$ has lower inferred uncertainties (since a given mass bin is informed by more sources). However, upon varying $\powerspectrumcutoff$ we find no clear trend in the uncertainty of $p(\msone | \hyper)$, see App.~\ref{app: meta choices}.
App.~\ref{app: meta choices} also shows that the $h$ posterior is robust against the particular choice of the power-spectrum parameters. For other parameters, such as the baseline power-law trend parameter, $\alpharef$, the GP-2D model varies more strongly under a variation of the power-spectrum parameters than the GP-1D model. 
We therefore recommend marginalizing over the correlation lengthscale when inferring a two-dimensional mass distribution.

\subsection{Simulated O5-like data}
\label{sec:results_simulated}

\begin{figure}
    \centering
    \includegraphics[width=0.8\linewidth]{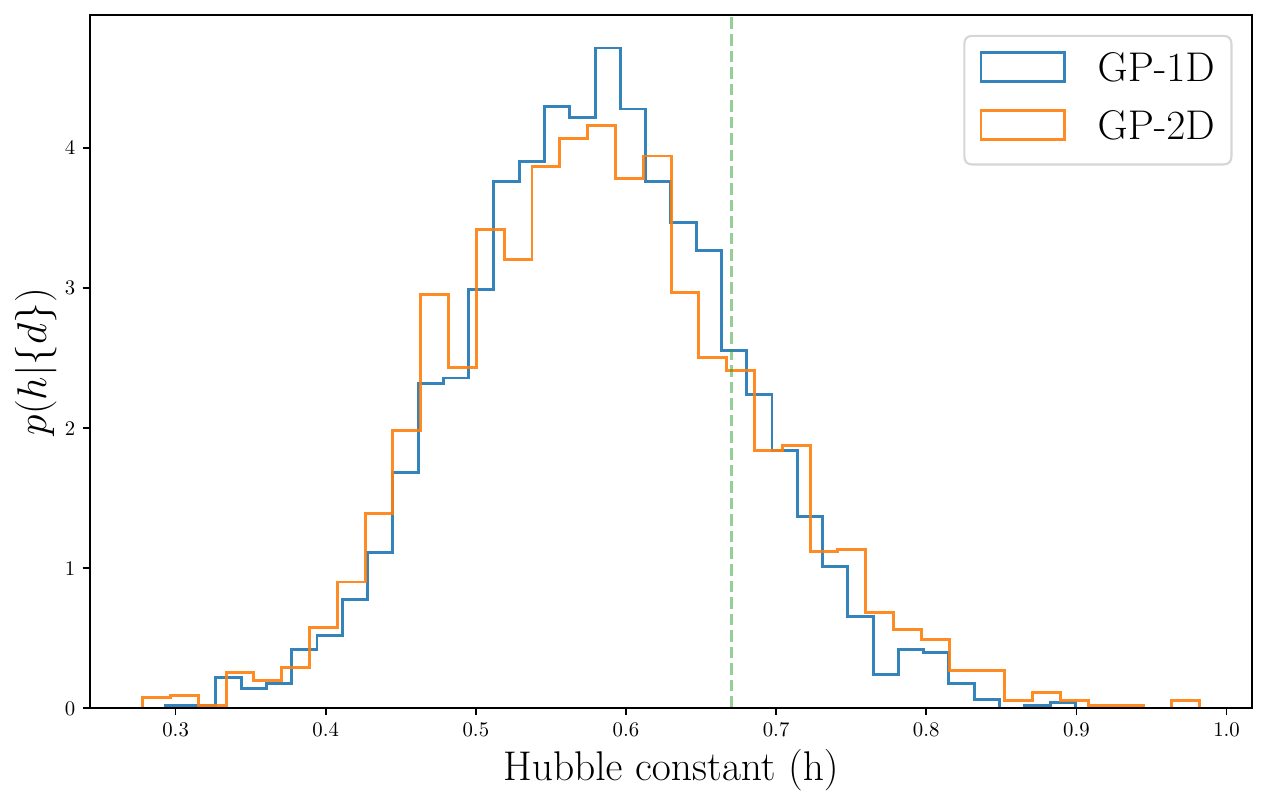}
    \caption{Hubble constant posterior using $\numsim$ O5-like simulated GW events with the two different GP mass models.
    The corresponding reconstructed mass distributions are shown in Fig.~\ref{fig: reconstructed mass distribution comparison sim} and Fig.~\ref{fig: reconstructed marginal mass distribution comparison sim}.
    The simulated $h=0.67$ value is marked with a dashed green line. 
    }
    \label{fig: hubble constant comparison sim}
\end{figure}

\begin{figure}
    \centering
    \includegraphics[width=\textwidth]{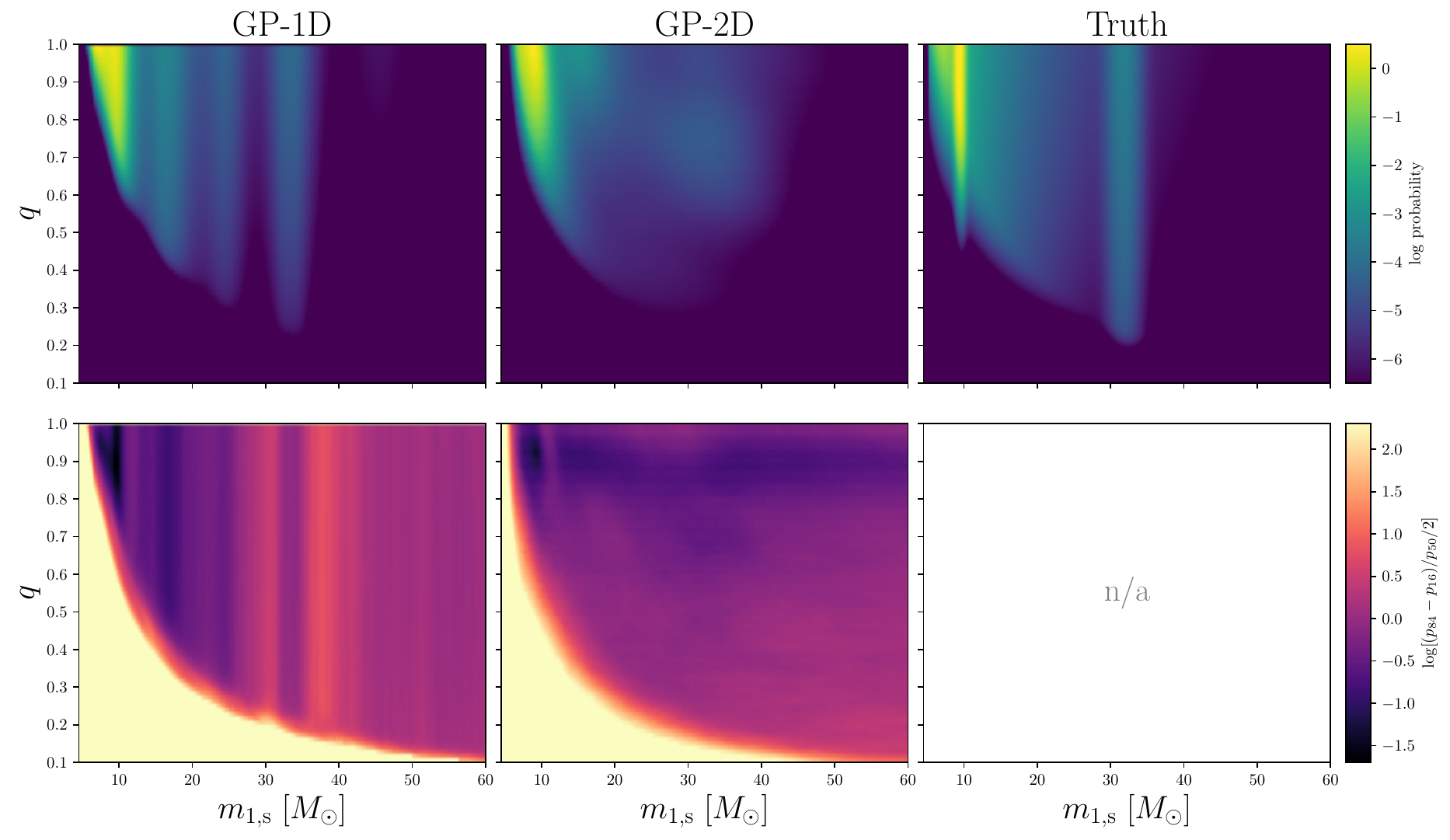}
    \includegraphics[width=\textwidth]{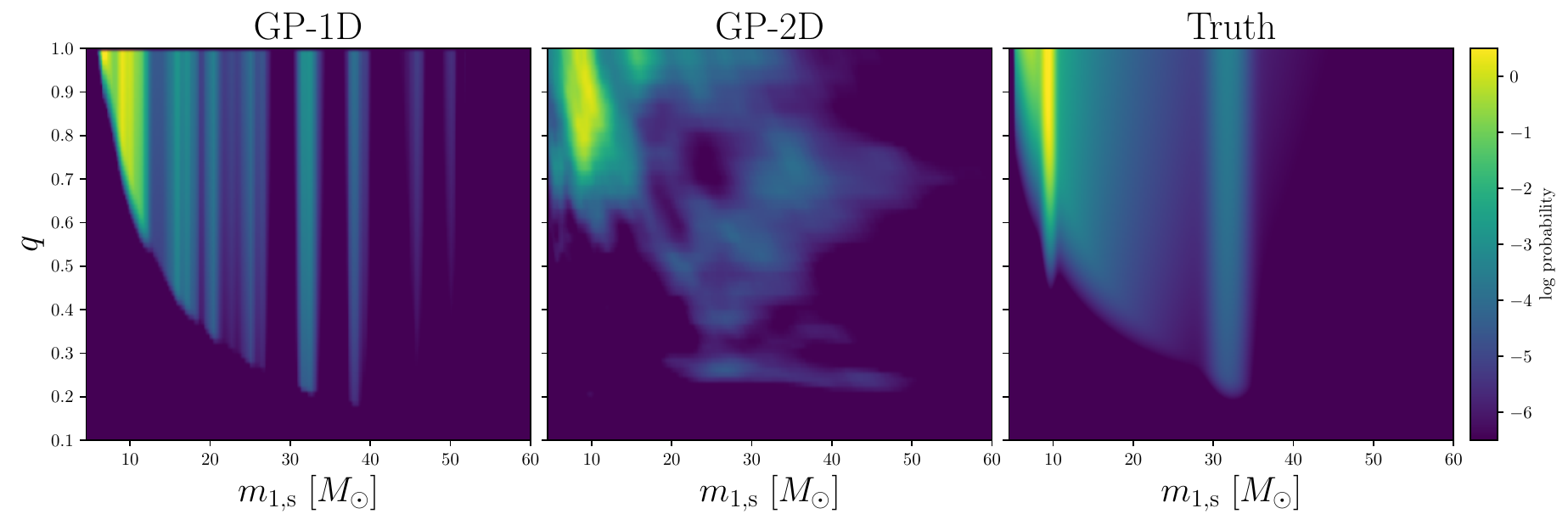}
    \caption{
    (\textit{Top row}) Reconstructed median of $p(\msone, \massratio)$ for the GW catalog of $\numsim$ O5-like events, with the true distribution on the right column.
    (\textit{Middle row}) Uncertainty (cf.~Eq.~\eqref{eq:def uncertainty measure}) for the reconstructed mass distribution.
    (\textit{Bottom row}) One randomly chosen posterior sample to illustrate the difference in structure of the mass distribution.
    }
    \label{fig: reconstructed mass distribution comparison sim}
\end{figure}

\begin{figure}
    \centering
    \includegraphics[width=0.8\textwidth]{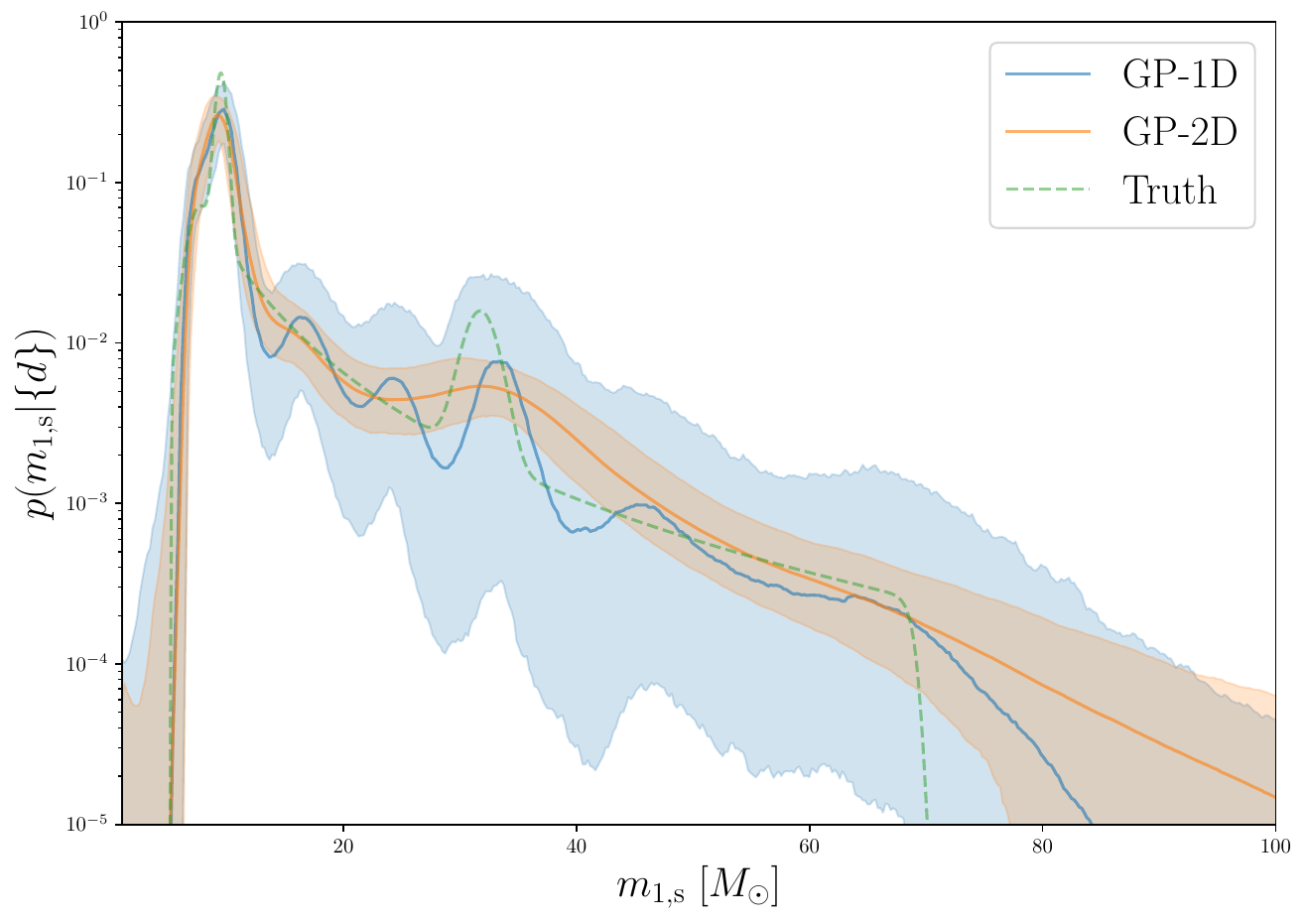}
    \caption{
    Reconstructed median (solid line) and 95\% credible regions (shaded regions) for two mass distributions, with the simulated distribution (dashed line).
    The median of the GP-1D model recovers the peak at $32\msun$ more accurately, but also introduces a spurious underdensity at masses above this peak.
    }
    \label{fig: reconstructed marginal mass distribution comparison sim}
\end{figure}

This section mirrors the preceding one, but uses simulated data to assess the faithfulness of the reconstruction in a controlled setting with known mass and redshift distributions and a known cosmological model.
We analyze the simulated catalog of $\numsim$ events with the two mass models: GP-1D and GP-2D.
For a simulated $h=0.67$, the inferred $h$ values are $\simhgponed$ (GP-1D) and $\simhgptwod$ (GP-2D), respectively, see Fig.~\ref{fig: hubble constant comparison sim}.
The inferred $h$ posterior is robust to the choice of flexible mass model (GP-1D versus GP-2D), despite noticeable differences in the reconstructed mass distributions, plotted in Fig.~\ref{fig: reconstructed mass distribution comparison sim}.
The discrepancy between the two models is negligible, in contrast to the results with GWTC-5. 
For both mass models, the true value is recovered at the $1\sigma$ level, with the median shifted to lower $h$; this is partially caused by a tail of the reconstructed mass distribution to higher masses, see the reconstructed mass distribution below.
We have verified that this downward $h$ shift is not present when analyzing a catalog with a different random seed.

The GP-1D mass distribution reconstructs the true distribution more faithfully than the GP-2D model (cf.~Fig.~\ref{fig: reconstructed mass distribution comparison sim}) because it imposes a more rigid structure on the two-dimensional mass distribution.
In this particular example, the modeled correlation structure matches the simulated correlation structure.
However, this is not generally true, which motivates the use of the GP-2D model.
See App.~\ref{app: wrong mass ratio distribution gp1d} for the $h$ inference when replacing the $p(\massratio| \hyper , \msone)$ power law by a truncated normal distribution.

While both mass models broadly agree with the simulated mass distribution plotted on the rightmost panel of Fig.~\ref{fig: reconstructed mass distribution comparison sim}, the GP-2D model naturally exhibits the previously discussed $\logMtot = \log(\msone + \mstwo)$ correlation.
The GP-2D mass model also reconstructs an overdensity at the $32\msun$ peak, albeit with lower mass ratios when compared to the true population that peaks at $\massratio = 1$.
At this catalog size, however, this shift in the reconstructed mass-ratio distribution is not by itself responsible for the offset in the $h$ measurement.

The two-dimensional uncertainty (cf.~Fig.~\ref{fig: reconstructed mass distribution comparison sim} and Eq.~\eqref{eq:def uncertainty measure}) of the reconstructed mass distribution is similar for both models, and the regions of lowest uncertainty are the $10\msun$ and $32\msun$ peaks, as seen previously on GWTC-5.

Fig.~\ref{fig: reconstructed marginal mass distribution comparison sim} shows the marginal $p(\msone | \hyper)$ distribution.
Again as for GWTC-5, the GP-1D mass model exhibits significantly larger uncertainties. 
While both distributions recover the shape of the mass distribution reasonably well, the GP-1D model recovers the $32\msun$ peak more faithfully than the GP-2D model that produces a smoother reconstruction.
This is also visible at the low-mass end of the distribution.
At masses near, but above the $32\msun$ peak, the median of the GP-1D distribution ``undershoots'' the true power law.
This further illustrates that the median should not be overinterpreted as an indicator of the true mass distribution.
The GP-1D reconstruction also reproduces the high-mass cutoff more accurately, whereas the GP-2D model yields a smoother transition.
We emphasize that the GP-1D has the (unjustified) advantage that the modeled mass-ratio distribution matches the simulated one, which is unlikely for real observations.
Overall, the GP-2D model favors a smoother mass distribution. This could be partially driven by the shared correlation lengthscale in $\logMtot$ and $\minuslogq$ as elaborated in Sec.~\ref{sec:results_gwtc5}. 

In the present case, the inferred power-spectrum parameters differ between the two models, with $\simpowerspectrumamplitudegponed$ (GP-1D) and $\simpowerspectrumamplitudegptwod$ (GP-2D), the inferred power-spectrum cutoff with $\simpowerspectrumcutoffgponed$ (GP-1D), and $\simpowerspectrumcutoffgptwod$ (GP-2D).
Analogously to the GWTC-5 results, the GP-2D model finds a higher amplitude and a longer correlation length when compared to the GP-1D mass model.

\section{Conclusions}
\label{sec:conclusions}

We have presented the GW hierarchical inference code \cosmopyro.
It is implemented in \jax, and therefore benefits from automatic differentiation, GPU acceleration, and just-in-time compilation.
Through gradient-based sampling, \cosmopyro allows us to constrain Gaussian-process-based source-frame mass distributions for two cases: a one-dimensional model (GP-1D), with a mass-ratio distribution that is a running power law, and a two-dimensional distribution (GP-2D).
We have intentionally formulated the correlations on the GP-2D mass model in the space of logarithmic source-frame total mass and logarithmic mass ratio, in order to avoid the standard correlation structure of $\msone$ and $\mstwo$ as encountered in many mass models.

For the 231 confidently identified BBH systems of GWTC-5 \cite{LIGOScientific:2026sit}, the two models yield $\gwtcfivehgponed$ (GP-1D) and $\gwtcfivehgptwod$ (GP-2D).
Compared to the spectral siren analysis by the LVK \cite{LIGOScientific:2026uyd} these error bars are slightly increased, but the $h$ medians are compatible within the $1\sigma$ interval.
With the GP models, we find similar structures to the LVK parametric models, with features at $\sim \gwtcfivegponedpeakone\msun$, $\sim \gwtcfivegponedpeaktwo\msun$ and $\sim \gwtcfivegponedpeakthree\msun$, although the strength of these features varies with the assumed power-spectrum parameters. The reconstructed two-dimensional source-frame mass distribution, see Fig.~\ref{fig: reconstructed mass distribution comparison gp1d vs gp2d gwtc5}, hints that the underlying mass distribution of GWTC-5 has a more complicated correlation structure in the mass ratio than the \mltp model.
This is consistent with the literature studying GW populations (i.e.~fixing $h$) \cite{Li:2022jge, Farah:2023swu, Godfrey:2023oxb, Li:2024jzi, Roy:2025ktr, Banagiri:2025dmy, Sridhar:2025kvi, Flanagan:2026ayy, LIGOScientific:2025pvj, Tong:2025wpz, LIGOScientific:2026ctl, Ray:2026uur, Mould:2026sww, Godfrey:2026pbc}.
The two mass models infer different power-spectrum parameters, with GP-1D (GP-2D) preferring small-scale (large-scale) fluctuations.
However, since the correlation lengthscales for the GP-2D mass model are shared among the two coordinates, this might introduce a bias where the relatively smooth distribution in the mass ratio pushes the inferred correlation lengthscale to larger values.
The agnostic nature of the GP models allows us to fit a large variety of possible mass models that could be realized in nature, and makes this particularly well suited for GW cosmology with future datasets.

We then simulated an O5-like catalog of $\numsim$ sources using the \mltp mass model, compatible with the lower bound of expected BBH detections during the fifth observing run (O5).
In this scenario, we recover $\simhgponed$ (GP-1D) and $\simhgptwod$ (GP-2D) for a simulated value of $h =0.67$.
The inferred $h$ values and uncertainties agree at the $1\sigma$ level, showing that one can recover $h$ robustly when the family of mass distributions is not known.
The measured $h$ values with GWTC-5 do not agree as well between the GP-1D and GP-2D mass models, further indicating that the underlying mass distribution of merging \glspl{bbh} has a more complex correlation structure than assumed in the \mltp and GP-1D mass distributions.

However, further development for future precision cosmology is needed. 
For example, the \cosmopyro pipeline assumes that the source-frame mass distribution is redshift independent, and, depending on the strength of the redshift evolution, this might warrant further exploration of a GP model that accounts for mass-redshift correlations; see \cite{Mukherjee:2021rtw, Pierra:2023deu} for studies of the impact of this misspecification.
Let us note here that \cosmopyro also allows for other non-parametric tests of fundamental physics, such as GW modified gravity propagation tests on cosmological scales, see \cite{mg_elena_cosmopyro} for applications on current and future data.
Additionally, although mass-spin correlations are known to have a weak impact on the $h$ measurement at the current size of the GW catalog \cite{LIGOScientific:2026uyd}, it could be increasingly important to model this for larger catalogs, as this systematic is not well explored.
Beyond the scope of this work, it would be desirable to compute evidences for Bayes-factor comparisons between the GP-1D and GP-2D models, something we leave for future work as this is not an immediate product from NUTS sampling.

\acknowledgments

We thank M.~Mancarella, A.~Papadopoulos, W.~Enzi, S.~Mastrogiovanni, J.~Heinzel, S.~Roy, W.~Farr, M.~Isi, T.~Baker, T.~Wagg, M.~Mould and M.~Williams for helpful discussions.
For the internal LVK review of the paper draft, we would like to thank M.~Tagliazucchi.
E.C. is supported by ERC Starting Grant SHADE (grant no.\,StG 949572).
Numerical computations were (in part) carried out on the \texttt{Sciama} High Performance Computing (HPC) cluster, which is supported by the Institute of Cosmology and Gravitation (ICG), the South-East Physics Network (SEPNet) and the University of Portsmouth.
The computations reported in this paper were (in part) performed using resources made available by the Flatiron Institute.
The Center for Computational Astrophysics at the Flatiron Institute is supported by the Simons Foundation.

This research has made use of data or software obtained from the Gravitational Wave Open Science Center (gwosc.org), a service of the LIGO Scientific Collaboration, the Virgo Collaboration, and KAGRA. This material is based upon work supported by NSF's LIGO Laboratory which is a major facility fully funded by the National Science Foundation, as well as the Science and Technology Facilities Council (STFC) of the United Kingdom, the Max-Planck-Society (MPS), and the State of Niedersachsen/Germany for support of the construction of Advanced LIGO and construction and operation of the GEO600 detector. Additional support for Advanced LIGO was provided by the Australian Research Council. Virgo is funded, through the European Gravitational Observatory (EGO), by the French Centre National de Recherche Scientifique (CNRS), the Italian Istituto Nazionale di Fisica Nucleare (INFN) and the Dutch Nikhef, with contributions by institutions from Belgium, Germany, Greece, Hungary, Ireland, Japan, Monaco, Poland, Portugal, Spain. KAGRA is supported by Ministry of Education, Culture, Sports, Science and Technology (MEXT), Japan Society for the Promotion of Science (JSPS) in Japan; National Research Foundation (NRF) and Ministry of Science and ICT (MSIT) in Korea; Academia Sinica (AS) and National Science and Technology Council (NSTC) in Taiwan.

\appendix

\section{\texorpdfstring{\mltp}{Multi Peak} mass model}
\label{app:mltp}

The parametric \mltp model as implemented in \cosmopyro is closely related to the \mltp model used in \gls{lvk} population analyses~\cite{Talbot:2018cva, LIGOScientific:2021aug}, with a slight modification to the low-mass smoothing.

The primary mass distribution is a mixture of a truncated power law and two Gaussian peaks:
\begin{align}
    \label{eq:mltp_m1}
    p(\msone \mid \hypermass) &\propto
    \Bigl[
        (1 - \lambdag)\,\mathcal{P}(\msone)
        + \lambdag\,\lambdaglow\,\mathcal{G}(\msone; \muglow, \sigmaglow)
        \nonumber
        \\
        &
        \quad
        \quad
        + \lambdag\,(1 - \lambdaglow)\,\mathcal{G}(\msone; \mughigh, \sigmaghigh)
    \Bigr] \Ssmooth(\msone)\,,
\end{align}
where
\begin{itemize}
    \item $\mathcal{P}(\msone) \propto \msone^{-\alphapl}$ is a power law truncated to $[\mmin, \mmax]$ by the smooth window function $w(\msone)$ of Eq.~\eqref{eq: 1D window function};
    \item $\mathcal{G}(\msone; \mu, \sigma)$ is a Gaussian with mean $\mu$ and standard deviation $\sigma$;
    \item $\lambdag \in [0,1]$ is the total Gaussian weight, and $\lambdaglow \in [0,1]$ allocates the fraction to the lower peak;
    \item $\Ssmooth(\msone)$ is a low-mass smoothing function (see below).
\end{itemize}
Instead of the standard $\Ssmooth$ function of~\cite{Talbot:2018cva}, we use an approximation
\begin{equation}
    \label{eq:smoothing}
    \log \Ssmooth(\msone; \mmin, \deltam)
    =
        -c\left[
        \log\!\left(
        1+\exp\!\left(
        \frac{m_{\min}+\deltam+a-\msone}{b\,\deltam}
        \right)
        \right)
        \right]^2
    \,,
\end{equation}
where we calibrate $a=-0.1626$, $b=0.0786$ and $c=0.0250$ to match the $\deltam$ in the above expression against the LVK smoothing window.

The mass-ratio distribution of the \mltp model follows Eq.~\eqref{eq:mltp_q} with a constant power-law exponent, i.e.~$\betaone = 0$ and $\betaq(\msone) = \betazero$ (denoted simply by $\betaq$ in Tab.~\ref{tab:mass_model_parameters}), and with $\deltamtwo = \deltam$.

\section{Priors}
\label{app: priors}

For the GP-1D and GP-2D models, we list the priors for all parameters except the Gaussian whitened field components in Tab.~\ref{tab:priors_gp1d_gp2d_combined} for GWTC-5 and for the O5-like simulated data.
With a few exceptions, the priors for the real and simulated datasets are almost identical.
The whitened Gaussian-field components $\gaussianfieldwhitened$ follow a zero-mean unit-variance Normal prior in all cases.
Tab.~\ref{tab:settings_diffs prior on power spectrum parameters} summarizes the choices of power-spectrum parameters that govern the strength of amplitude fluctuations and the correlation length.

\begin{table}[t]
\centering

\textbf{GWTC-5}

\vspace{0.4em}
\begin{minipage}[t]{0.48\textwidth}
\vspace{0pt}
\centering
\begin{tabular}{lll}
\toprule
\multicolumn{3}{c}{\textbf{GP-1D priors}} \\
\midrule
Parameter & Units & Prior \\
\midrule
\multicolumn{3}{l}{\textbf{Cosmology}}\\
$h$ & -- & $\mathcal{U}(0.1,\,1.2)$ \\
$\Omega_{\rm m}$ & -- & $\delta(0.3)$ \\[0.5ex]

\multicolumn{3}{l}{\textbf{Redshift evolution}}\\
$\gamma$ & -- & $\mathcal{U}(-2,\,5)$ \\
$\kappa$ & -- & $\mathcal{U}(0,\,6)$ \\
$z_{\rm p}$ & -- & $\mathcal{U}(0,\,4)$ \\[0.5ex]

\multicolumn{3}{l}{\textbf{Primary-mass model}}\\
$\mmin$ & $\msun$ & $\mathcal{U}(2,\,7)$ \\
$\mmax$ & $\msun$ & $\mathcal{U}(40,\,220)$ \\
$\sigmalowfrac$ & -- & $\mathcal{LU}(10^{-4},\,0.1)$ \\
$\sigmahighfrac$ & -- & $\mathcal{LU}(0.003,\,0.3)$ \\
$\alpharef$ & -- & $\mathcal{U}(-6.0,\,-1.5)$ \\
$\powerspectrumamplitude$ & -- & varying, see Tab.~\ref{tab:settings_diffs prior on power spectrum parameters} \\
$\powerspectrumcutoff$ & -- & varying, see Tab.~\ref{tab:settings_diffs prior on power spectrum parameters} \\[0.5ex]

\multicolumn{3}{l}{\textbf{Mass-ratio model}}\\
$\beta_{0}$ & -- & $\mathcal{U}(-2,\,4)$ \\
$\beta_{1}$ & -- & $\mathcal{U}(-0.1,\,0.1)$ \\
$\deltamtwo$ & $\msun$ & $\mathcal{U}(0.1,\,10)$ \\
$\msref$ & $\msun$ & $\delta(10)$ \\
\bottomrule
\end{tabular}
\end{minipage}
\hfill
\begin{minipage}[t]{0.48\textwidth}
\vspace{0pt}
\centering
\begin{tabular}{lll}
\toprule
\multicolumn{3}{c}{\textbf{GP-2D priors}} \\
\midrule
Parameter & Units & Prior \\
\midrule
\multicolumn{3}{l}{\textbf{Cosmology}}\\
$h$ & -- & $\mathcal{U}(0.1,\,1.2)$ \\
$\Omega_{\rm m}$ & -- & $\delta(0.3)$ \\[0.5ex]

\multicolumn{3}{l}{\textbf{Redshift evolution}}\\
$\gamma$ & -- & $\mathcal{U}(-2,\,5)$ \\
$\kappa$ & -- & $\mathcal{U}(0,\,6)$ \\
$z_{\rm p}$ & -- & $\mathcal{U}(0,\,4)$ \\[0.5ex]

\multicolumn{3}{l}{\textbf{Mass model}}\\
$\mmin$ & $\msun$ & $\mathcal{U}(3.5,\,10.0)$ \\
$\mmax$ & $\msun$ & $\mathcal{U}(40,\,220)$ \\
$\sigmalowfrac$ & -- & $\mathcal{LU}(10^{-4},\,0.1)$ \\
$\sigmahighfrac$ & -- & $\mathcal{LU}(0.003,\,0.3)$ \\
$\alpharef$ & -- & $\mathcal{U}(-6.0,\,-1.5)$ \\
$\betaref$ & -- & $\delta(1.0)$ \\
$\powerspectrumamplitude$ & -- & varying, see Tab.~\ref{tab:settings_diffs prior on power spectrum parameters} \\
$\powerspectrumcutoff$ & -- & varying, see Tab.~\ref{tab:settings_diffs prior on power spectrum parameters} \\
\bottomrule
\end{tabular}
\end{minipage}

\vspace{1.2em}
\textbf{O5-like simulated data}

\vspace{0.4em}
\begin{minipage}[t]{0.48\textwidth}
\vspace{0pt}
\centering
\begin{tabular}{lll}
\toprule
\multicolumn{3}{c}{\textbf{GP-1D priors (differences only)}} \\
\midrule
Parameter & Units & Prior \\
\midrule
\multicolumn{3}{l}{\textbf{Cosmology}}\\
$h$ & -- & $\mathcal{U}(0.1,\,1.0)$ \\

\multicolumn{3}{l}{\textbf{Primary-mass model}}\\
$\mmin$ & $\msun$ & $\mathcal{U}(2,\,10)$ \\
$\sigmalowfrac$ & -- & $\mathcal{LU}(0.01,\,0.1)$ \\[0.5ex]

\multicolumn{3}{l}{\textbf{Mass-ratio model}}\\
$\deltamtwo$ & $\msun$ & $\mathcal{U}(0.1,\,5)$ \\
\bottomrule
\end{tabular}
\end{minipage}
\hfill
\begin{minipage}[t]{0.48\textwidth}
\vspace{0pt}
\centering
\begin{tabular}{lll}
\toprule
\multicolumn{3}{c}{\textbf{GP-2D priors (differences only)}} \\
\midrule
Parameter & Units & Prior \\
\midrule
\multicolumn{3}{l}{\textbf{Cosmology}}\\
$h$ & -- & $\mathcal{U}(0.1,\,1.0)$ \\

\multicolumn{3}{l}{\textbf{Mass model}}\\
$\sigmalowfrac$ & -- & $\mathcal{LU}(0.01,\,0.1)$ \\
\bottomrule
\end{tabular}
\end{minipage}

\caption{Priors used for GWTC-5 (top) and O5-like simulated data (bottom) in the one- and two-dimensional Gaussian-process mass models. For the simulated data, only differences relative to GWTC-5 are listed. Uniform priors are denoted by $\mathcal{U}$, log-uniform priors by $\mathcal{LU}$, and fixed parameters by $\delta$.}
\label{tab:priors_gp1d_gp2d_combined}
\end{table}

\begin{table}[htbp]
\centering
\small

\textbf{GWTC-5}

\vspace{0.4em}
\textbf{GP-1D}

\vspace{0.4em}
\begin{tabular}{lcccccc}
\toprule
Parameter &
Marginalization &
\textcolor[HTML]{1F77B4}{\rule{1.5ex}{1.5ex}}~[1] &
\textcolor[HTML]{FF7F0E}{\rule{1.5ex}{1.5ex}}~[2] &
\textcolor[HTML]{2CA02C}{\rule{1.5ex}{1.5ex}}~[3] &
\textcolor[HTML]{D62728}{\rule{1.5ex}{1.5ex}}~[4] &
\textcolor[HTML]{9467BD}{\rule{1.5ex}{1.5ex}}~[5] \\
\midrule
$\powerspectrumamplitude$ & $\mathcal{LU}(0.1,100)$ & 15.0 & 7.5 & 30.0 & 15.0 & 15.0 \\
$\powerspectrumcutoff$    & $\mathcal{LU}(10,1000)$ & 500.0 & 500.0 & 500.0 & 250.0 & 1000.0 \\
\bottomrule
\end{tabular}

\vspace{1.0em}
\textbf{GP-2D}

\vspace{0.4em}
\begin{tabular}{lcccccc}
\toprule
Parameter &
Marginalization &
\textcolor[HTML]{1F77B4}{\rule{1.5ex}{1.5ex}}~[1] &
\textcolor[HTML]{FF7F0E}{\rule{1.5ex}{1.5ex}}~[2] &
\textcolor[HTML]{2CA02C}{\rule{1.5ex}{1.5ex}}~[3] &
\textcolor[HTML]{D62728}{\rule{1.5ex}{1.5ex}}~[4] &
\textcolor[HTML]{9467BD}{\rule{1.5ex}{1.5ex}}~[5] \\
\midrule
$\powerspectrumamplitude$ & $\mathcal{LU}(1,300)$ & 60.0 & 30.0 & 120.0 & 60.0 & 60.0 \\
$\powerspectrumcutoff$    & $\mathcal{LU}(10,300)$ & 40.0 & 40.0 & 40.0 & 20.0 & 80.0 \\
\bottomrule
\end{tabular}

\caption{Comparison of the Gaussian-process power-spectrum hyperparameter settings used for the GP-1D and GP-2D analyses of GWTC-5. The fiducial analysis choices in Sec.~\ref{sec:results} marginalize over the log-uniform prior ($\mathcal{LU}$) indicated.
The colors correspond to panels of Fig.~\ref{fig: reconstructed mass distribution metachoices gp1d gwtc5} (GP-1D) and Fig.~\ref{fig: reconstructed mass distribution metachoices gp2d gwtc5} (GP-2D).
Note that the first settings~[1] represent the reference settings, from which we then vary either the power-spectrum amplitude or the cutoff separately.
}
\label{tab:settings_diffs prior on power spectrum parameters}
\end{table}

\begin{table}[htp]
\centering

\begin{minipage}[t]{0.48\textwidth}
\vspace{0pt}
\centering
\begin{tabular}{lcc}
\toprule
\multicolumn{3}{c}{\textbf{GP-1D}}\\
\midrule
Parameter & Range & Num. bins \\
\midrule
$\msone~[\msun]$ & $[0.9,\,500.0]$ & 1500 \\
$\massratio$ & $[0.01,\,1.0]$ & 1000 \\
$z$ & $[0.0,\,7.5]$ & 5000 \\
\bottomrule
\end{tabular}
\end{minipage}
\hfill
\begin{minipage}[t]{0.48\textwidth}
\vspace{0pt}
\centering
\begin{tabular}{lcc}
\toprule
\multicolumn{3}{c}{\textbf{GP-2D}}\\
\midrule
Parameter & Range & Num. bins \\
\midrule
$\logMtot$ & $[0.6,\,7.0]$ & 250 \\
$\minuslogq$ & $[0,\,4.2]$ & 100 \\
$z$ & $[0,\,8]$ & 3000 \\
\bottomrule
\end{tabular}
\end{minipage}

\caption{The binning adopted for the GP-1D and GP-2D population models.
For the GP-1D (GP-2D) mass model, the number of inferred Gaussian-field components is equal to the number of bins in $\msone$ (the product of the numbers of bins in $\logMtot$ and $\minuslogq$).
All settings apply to GWTC-5 and O5-like simulated data alike.
}
\label{tab:binning_details}

\end{table}

\section{Gaussian process}
\label{app:gaussian process}

In the following, we detail the generation process for draws of the Gaussian-process mass distribution.
We infer the mass distribution on a grid and interpolate with a cubic scheme in the logarithmic probability.
To this end, we first draw white-noise values on the grid from a zero-mean unit-variance Normal distribution.
We then Fourier transform them, to obtain $\gaussianfieldwhitened(k)$, where $k$ is the wavenumber conjugate to the relevant mass coordinate, after rescaling the grid range (cf.~Tab.~\ref{tab:binning_details}) to unit length; $\powerspectrumcutoff$ is therefore dimensionless.
In the one-dimensional case, the relevant coordinate is $\msone$, in the two-dimensional case, these are $\logMtot$ and $\minuslogq$ (cf.~Eq.~\eqref{eq: def logMtot and minuslogq}).

The correlated field is obtained by multiplying the white-noise coefficients by $\sqrt{P(k)}$ in Fourier space and transforming back to the mass domain via inverse \gls{fft}.
The Gaussian field (cf.~Eq.~\eqref{eq: link gaussian process mass distribution 1d} and Eq.~\eqref{eq: link gaussian process mass distribution 2d}) is then
\begin{equation}
    \gaussianfield = \text{FT}^{-1}\left[\sqrt{P(k)} \gaussianfieldwhitened(k)
    \right]
    \,,
\end{equation}
where $\text{FT}^{-1}$ is the inverse Fourier transform, and we define the relevant power spectra, $P(k)$, below.

For the one-dimensional power spectrum, we take
\begin{equation}
\label{eq: power spectrum 1d}
P_{\rm 1D}(k)
=
\frac{\powerspectrumamplitude}{\powerspectrumcutoff}
\frac{\left(k/\powerspectrumcutoff\right)^2}
{1+\left(k/\powerspectrumcutoff\right)^6}
\,,
\end{equation}
and for the two-dimensional case we take
\begin{equation}
\label{eq: power spectrum 2d}
P_{\rm 2D}(\mathbf{k})
=
\frac{\powerspectrumamplitude}{\powerspectrumcutoff^2}
\frac{|\mathbf{k}|/\powerspectrumcutoff}
{\left(
1+\dfrac{|\mathbf{k}|^2}{\powerspectrumcutoff^2}
\right)^{3}},
\qquad
|\mathbf{k}| \equiv \sqrt{k_1^2+k_2^2}
\,.
\end{equation}
While $\powerspectrumamplitude$ controls the amplitude of the fluctuations, $\powerspectrumcutoff$ determines (the inverse of) the smallest lengthscale of correlations.
In both cases the numerator suppresses the largest-scale modes, which would otherwise only be degenerate with the overall normalization of the mass distribution, and the prefactors $1/\powerspectrumcutoff$ and $1/\powerspectrumcutoff^2$ are chosen such that the variance of the Gaussian field is set by $\powerspectrumamplitude$ alone, independently of $\powerspectrumcutoff$.
We explore the impact of the choice for these power-spectrum parameters in App.~\ref{app: meta choices}.

\section{Implementation details and likelihood variance penalty function}
\label{app: implementation details}

To generate prior draws of the Gaussian random field, we pre-compute the source-frame mass distribution on a grid and then evaluate each sample via cubic interpolation. 
Tab.~\ref{tab:binning_details} lists the resolution for the various runs. 
We have verified that the binning resolution does not affect the resulting posterior distributions appreciably.

A common concern when approximating the integral in the likelihood of Eq.~\eqref{eq:hyperlikelihood} is the variance due to a finite number of samples. 
We follow the literature \cite{Farr:2019rap} and introduce a penalty factor in the likelihood that disfavors points in $\hyper$ parameter space that yield large variance of the estimated likelihood. 
Say $\{w_j\}_{j=1}^N$ are the weights of the samples in the integral of Eq.~\eqref{eq:numerator} (for simplicity we consider here only event $i$, and combine the variance over all events below).
Then the estimated mean of numerator of the likelihood is given by
\begin{equation}
    \mu_{\mathcal{L}} \deffrom \frac{1}{N}\sum_{j=1}^N w_j \,.
\end{equation}
The estimated variance of the mean estimate is 
\begin{equation}
    \text{Var}(\mu_{\mathcal{L}}) = \frac{1}{N^2}\left[\sum_{j=1}^N w_j^2 \right]- \frac{\mu_{\mathcal{L}}^2}{N} \,.
\end{equation}
The relative variance is then defined by
\begin{equation}
    \mathcal{V} \deffrom 
    \frac{\text{Var}(\mu_{\mathcal{L}}) }{\mu_{\mathcal{L}}^2}
    \,.
\end{equation}
We denote by $\mathcal{V}_{{\rm num}, i}$ the relative variance of GW event $i$, and by $\mathcal{V}_{{\rm den}}$ the relative variance of the denominator of Eq.~\eqref{eq:hyperlikelihood}. The total relative variance of the population likelihood is then given by
\begin{equation}
    \mathcal{V}_{\hyper} = \sum_{i=1}^{\nobs} \mathcal{V}_{{\rm num}, i} + \nobs ^ 2 \mathcal{V}_{{\rm den}} \,.
\end{equation}
Note that the number of observations enters quadratically in the denominator (i.e.~selection effect) variance rather than linearly like the relative variance due to each event.
We then introduce the following log-likelihood penalty function that down-weights all $\hyper$ choices with $\mathcal{V}_{\hyper} > 1$ with 
\begin{equation}
    \text{penalty} = - s \cdot (\mathcal{V}_{\hyper} - 1)^2 \cdot H_{0.01}(\mathcal{V}_{\hyper} - 1)
    \,,
\end{equation}
where the Heaviside function is smoothly approximated by 
\begin{equation}
    H_\sigma(x) = \frac{1}{2} \left[ 1 + \operatorname{erf}\left( \frac{x}{\sqrt{2}\sigma} \right) \right]\,.
\end{equation}
We fix $s = 100$ throughout, and find no noticeable change when instead assuming $s=1000$.

\section{Simplified single-event inference}
\label{app: mock pe}

\begin{figure}[t]
    \centering
    \includegraphics[width=0.75\textwidth]{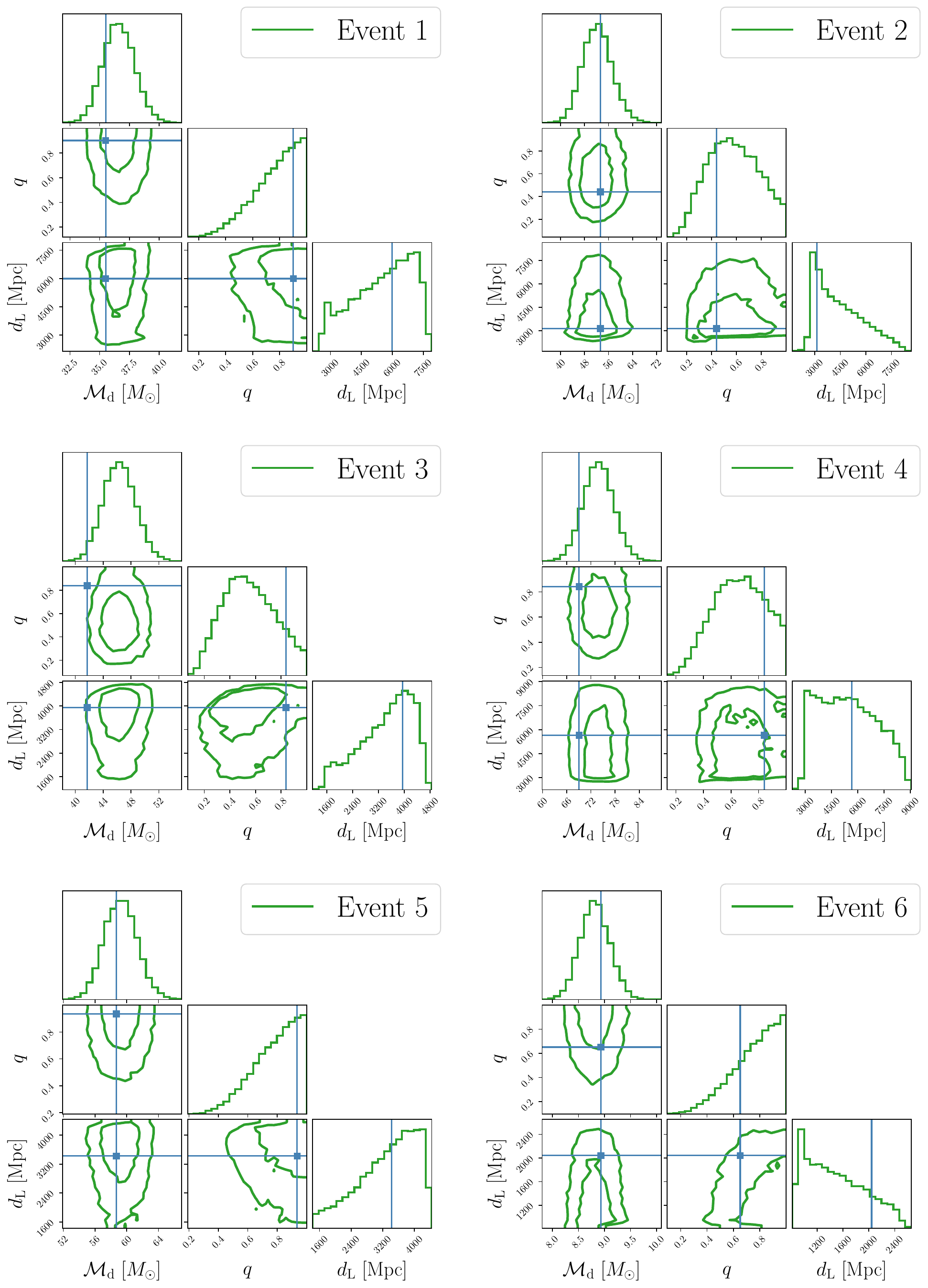}
    \caption{%
    Single-event posteriors in detector-frame chirp mass, mass ratio and luminosity distance obtained with the simplified parameter estimation described in this appendix, for six randomly selected events of the simulated O5-like catalog.
    The true parameter values are indicated in each panel (blue), and we draw the $0.5$ and $0.9$ levels. 
    The distance and mass-ratio posteriors are substantially broader than the chirp-mass posterior.
    }
    \label{fig: mock pe corner}
\end{figure}

The simulated catalog of Sec.~\ref{sec: data simulated_data} is not produced with a full gravitational-wave parameter-estimation pipeline.
Instead, we adopt a simplified model of the detection and measurement process, following the spirit of \cite{Finn:1992xs, 2018ApJ...863L..41F}, in which the \gls{snr} is available in closed form.
This is sufficient for our purpose: the model reproduces the features that matter for the reconstruction of the mass distribution and the inference of $h$, namely that measurement uncertainties are much broader in luminosity distance and mass ratio than in chirp mass, and that the horizon depends on mass and inclination.
It does not aim to reproduce realistic single-event posteriors and does not account for parameters such as coalescence time, sky position or spins.

For a binary with detector-frame chirp mass $\mathcal{M}_{\rm d}$, mass ratio $\massratio$, luminosity distance $d_L$ and inclination $\iota$, we take the optimal \gls{snr} to be
\begin{equation}
\label{eq: snr approximation}
    \rho_{\rm opt}
    =
    \mathcal{A}\,
    \frac{\mathcal{M}_{\rm d}^{5/6}}{d_L}\,
    \Theta(\iota)\,
    b_{\rm bw}(M_{\rm tot, d})
    \,,
\end{equation}
where the $\mathcal{M}_{\rm d}^{5/6}/d_L$ scaling is the leading-order inspiral amplitude, and
\begin{equation}
    \Theta(\iota) = \sqrt{\frac{\left(1+\cos^2\iota\right)^2}{4} + \cos^2\iota}
    \,,
    \qquad
    b_{\rm bw}(M_{\rm tot, d}) = \left[1 + \left(\frac{M_{\rm tot, d}}{M_{\rm cut, d}}\right)^{2}\right]^{-1}
    \,.
\end{equation}
The factor $\Theta$ encodes the antenna response of a single detector to the two polarizations at fixed sky location, and varies by a factor of $2\sqrt{2}$ between an edge-on and a face-on orientation.
The factor $b_{\rm bw}$ is a bandwidth penalty that suppresses the \gls{snr} of high-mass systems, which merge at increasingly low frequencies; we fix $M_{\rm cut, d} = 200\,\msun$, with $M_{\rm tot, d}$ the detector-frame total mass.

The overall amplitude $\mathcal{A}$ sets the sensitivity of the network, and we fix $\mathcal{A} = 4\times 10^{3}\,\Mpc\,\msun^{-5/6}$ to approximate an O5 sensitivity configuration \cite{KAGRA:2013rdx, Kiendrebeogo:2023hzf}.
With the SNR threshold quoted below, this corresponds to a horizon of $\sim 3.2\,$Gpc for an optimally oriented equal-mass binary with $\mathcal{M}_{\rm d} = 10\,\msun$, and $\sim 10\,$Gpc at $\mathcal{M}_{\rm d} = 60\,\msun$. 

The matched-filter \gls{snr} of a detected event differs from $\rho_{\rm opt}$ because of the noise realization.
We mimic this by scattering the optimal SNR by a random variable, $s$, drawn from a $\chi^2$ distribution with two degrees of freedom,
\begin{equation}
\label{eq: scattered snr}
    \rho_{\rm obs} = \sqrt{\rho_{\rm opt}^2 + s}
    \,,
    \qquad
    p(s) = \chi^2_{2}
    \,,
\end{equation}
which approximates the non-central $\chi^2$ distribution expected for $\rho_{\rm obs}^2$.
An event is included in the catalog if the observed (i.e.~not the optimal) \gls{snr} passes the threshold, $\rho_{\rm obs} \geq 12$.
The injection set used to estimate the selection function in Eq.~\eqref{eq:hyperlikelihood} is generated according to the same procedure.

For each detected event we generate mock data and posterior samples for the three parameters that enter the population likelihood, $x \in \{\mathcal{M}_{\rm d}, \massratio, d_L\}$.
All three are positive, and we take the measurement of each of them to be described by a Gamma likelihood whose rate parameter is the true value,
\begin{equation}
\label{eq: gamma likelihood}
    p(\hat{x}\mid x)
    =
    \frac{x^{k_x}}{\Gamma(k_x)}\,
    \hat{x}^{\,k_x - 1}
    e^{-x \hat{x}}
    \,,
\end{equation}
where $\hat{x}$ denotes the mock data and $k_x$ is a concentration parameter that controls the measurement precision.
This choice is convenient because it is self-conjugate under the exchange of data and parameter: for a prior flat in $x$, the posterior is again a Gamma distribution, $p(x\,|\,\hat{x}) \propto x^{k_x} e^{-\hat{x} x}$. 
Note that this posterior has a relative width of $1/\sqrt{k_x + 1}$.
The measurement is thus generated by drawing one $\hat{x}$ per event from Eq.~\eqref{eq: gamma likelihood} given the true value of $x$. 
Similarly, posterior samples can be obtained by sampling from a Gamma distribution with concentration parameter $k_x + 1$, instead of $k_x$ as before, at fixed $\hat{x}$.
In addition, the mass-ratio posterior is restricted to the physical range $\massratio \in (0, 1]$.
The concentration values will be specified below.

An observation contains more information than the three terms in Eq.~\eqref{eq: gamma likelihood}.
The fact that the event was detected with a given $\rho_{\rm obs}$ is itself a measurement, and constrains the combination of $(\mathcal{M}_{\rm d}, \massratio, d_L, \iota)$ appearing in Eq.~\eqref{eq: snr approximation}.
We therefore take the single-event likelihood to be
\begin{equation}
\label{eq: mock pe likelihood}
    p(\datagwsub{i}\mid \mathcal{M}_{\rm d}, \massratio, d_L, \iota)
    =
    p(\hat{\mathcal{M}}_{\rm d}\,|\,\mathcal{M}_{\rm d})\,
    p(\hat{\massratio}\,|\,\massratio)\,
    p(\hat{d}_L\,|\,d_L)\,
    p(\rho_{\rm obs}\,|\,\mathcal{M}_{\rm d}, \massratio, d_L, \iota)
    \,,
\end{equation}
where the last factor follows from Eq.~\eqref{eq: scattered snr}, and we assume a prior that is flat in $(\mathcal{M}_{\rm d}, \massratio, d_L)$ and isotropic in $\cos\iota$.
In principle, posterior samples could be constructed by importance sampling: proposals would be drawn from the individual Gamma posteriors above (in $\mathcal{M}_{\rm d}$, $d_L$ and $\massratio$, according to the first three terms), the remaining factor would be absorbed into an importance weight, and the weighted proposals would be resampled to the number of posterior samples given in Sec.~\ref{sec: data simulated_data}.

However, the \gls{snr} factor in Eq.~\eqref{eq: mock pe likelihood} is much sharper than the distance measurement, since at fixed inclination it determines $d_L$ to within a few percent; proposing the distance from its Gamma posterior and reweighting is therefore inefficient.
Instead we exploit that the last term of Eq.~\eqref{eq: mock pe likelihood} can be rewritten as 
\begin{equation}
    p(\rho_{\rm obs}\mid\mathcal{M}_{\rm d}, \massratio, d_L, \iota)
    =
    p(s\mid\mathcal{M}_{\rm d}, \massratio, d_L, \iota)
    =p(s)
    \,.
\end{equation}
Then we can invert Eq.~\eqref{eq: snr approximation}: writing $\rho_{\rm opt} = \mathcal{C}(\mathcal{M}_{\rm d}, \massratio, \iota)/d_L$, drawing the residual $s$ from its $\chi^2_2$ distribution and solving
\begin{equation}
    d_L = \frac{\mathcal{C}(\mathcal{M}_{\rm d}, \massratio, \iota)}{\sqrt{\rho^2_{\rm obs} - s}}
\end{equation}
samples the \gls{snr} factor exactly, so that it cancels against the proposal density.
The importance weight then reduces to the broad distance likelihood times the Jacobian $|\mathrm{d}s/\mathrm{d}d_L| = 2\mathcal{C}^2/d_L^3$,
\begin{equation}
    \log w = \log p(\hat{d}_L\mid d_L) + 3\log d_L - \log\left(2\mathcal{C}^2\right) + \log w_{\massratio} + \log |\partial(\mathcal{M}_{\rm d}, \massratio)/\partial(\mdone, \mdtwo)|
    \,,
\end{equation}
where $\log w_{\massratio}$ enforces the truncation of the mass ratio to $(0,1]$.
This targets the same posterior as the scheme outlined above while increasing the effective sample size by two to three orders of magnitude.
We monitor the effective sample size of the reweighting for every event and require it to exceed the number of requested posterior samples, so that no event is represented by a small number of repeated proposals.
The mass Jacobian is given by $|\partial(\mathcal{M}_{\rm d}, \massratio)/\partial(\mdone, \mdtwo)| = \mathcal{M}_{\rm d}/\mdone^2$ and ensures that the above posterior samples follow a prior that is uniform in detector-frame component masses.

We draw the concentration for each event according to $k_{\mathcal{M}_{\rm d}} \sim \mathcal{U}(10, 1000)$, $k_{\massratio} \sim \mathcal{U}(3, 7)$ and $k_{d_L} \sim \mathcal{U}(1, 5)$, for $\mathcal{U}$ the uniform distribution. These values are chosen such that we match the resulting relative uncertainties to expected uncertainties of an O5-like catalog \cite{KAGRA:2013rdx, Kiendrebeogo:2023hzf}. For our catalog we find a mean relative uncertainty of $6\%$ (detector-frame chirp mass), $27\%$ (mass ratio) and $26\%$ (luminosity distance).
Fig.~\ref{fig: mock pe corner} shows the resulting posteriors for six randomly selected events of the simulated catalog, together with their true parameter values.

\section{Validation against LVK \mltp GWTC-5 result}
\label{app: validation against lvk gwtc-5 result}

\begin{figure}[t]
    \centering
    \includegraphics[width=\textwidth]{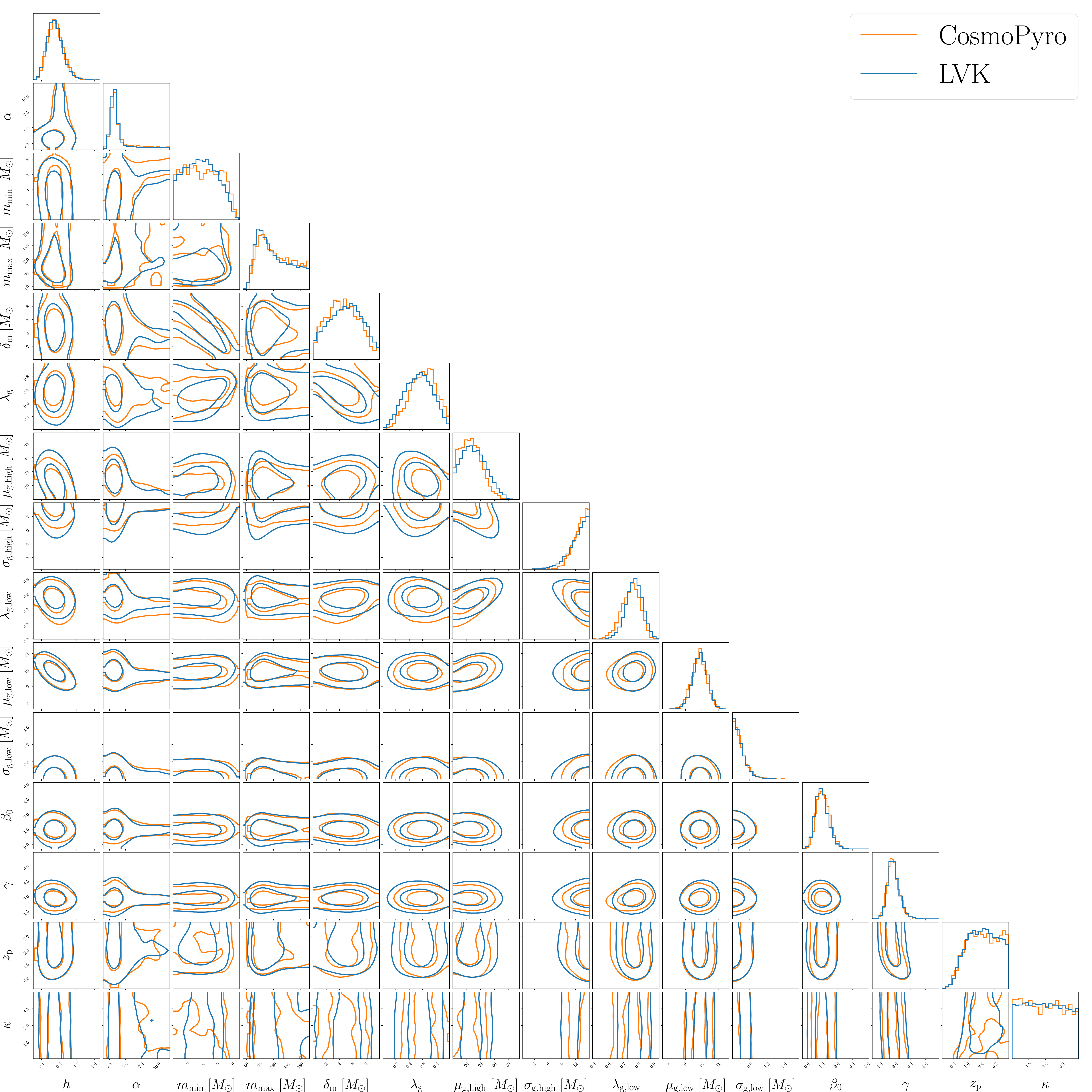}
    \caption{%
        Corner plot of the posterior distribution of $\Hubble$ and mass hyperparameters from the GWTC-5 analysis, using the \mltp mass model.
        Indicated are the $0.5$ and $0.9$ levels. 
        We compare a mixture of equal weight of the \gwcosmo \cite{Gray:2023wgj, Papadopoulos:2026puy} and \icarogw \cite{Mastrogiovanni:2021wsd, Mastrogiovanni:2023zbw} codes (LVK) against the \cosmopyro result.
        Differences in the parameter $\deltam$ arise from slight differences in the definitions of the smoothing function at low masses, see App.~\ref{app:mltp}.
        We find good agreement between the LVK analyses and the \cosmopyro result.
    }
    \label{fig:gwtc5_corner}
\end{figure}

To validate the \cosmopyro codebase against the LVK-reviewed \gwcosmo \cite{Gray:2019ksv, Gray:2023wgj, Papadopoulos:2026puy} and \icarogw \cite{Mastrogiovanni:2021wsd, Mastrogiovanni:2023zbw, Mastrogiovanni:2023emh} packages, we perform an additional comparison.
For this purpose, Fig.~\ref{fig:gwtc5_corner} compares the two-dimensional marginal distribution of the hyperparameter posteriors with the \mltp mass model, using the GWTC-5 catalog \cite{LIGOScientific:2026uyd}.
We use identical priors to the ones used in \cite{LIGOScientific:2026uyd}. In particular we enforce the wide prior $p(h) = \mathcal{U}(0.1, 2.0)$, which results in slightly different values than the ones given in Sec.~\ref{sec:results_gwtc5}.

The two posteriors show a few differences, most pronounced for $\deltam$ and $\mmin$. 
This arises from the aforementioned difference between the smoothing function definition of the LVK, and the implementation of \cosmopyro (cf.~Eq.~\eqref{eq:smoothing}).
Overall, we find good agreement between the two results, with $\hmeasurelvkmltpgwtcfivewide$ (LVK) and $\hmeasurecosmopyromltpgwtcfivewide$ (\cosmopyro) \cite{LIGOScientific:2026uyd}.

\section{Assessment of meta-choices on GWTC-5}
\label{app: meta choices}

The choice of the power-spectrum parameters is not unique and can potentially influence the inferred $h$ value.
While the main text marginalizes over these parameters, in this appendix we repeat the GWTC-5 analysis of Sec.~\ref{sec:results_gwtc5}, instead fixing $\powerspectrumamplitude$ and $\powerspectrumcutoff$ (cf.~Eq.~\eqref{eq: power spectrum 1d} and Eq.~\eqref{eq: power spectrum 2d}) to a range of different values.
Tab.~\ref{tab:settings_diffs prior on power spectrum parameters} lists the adopted values for both mass models.

\begin{figure}
    \centering
    \includegraphics[width=0.98\linewidth]{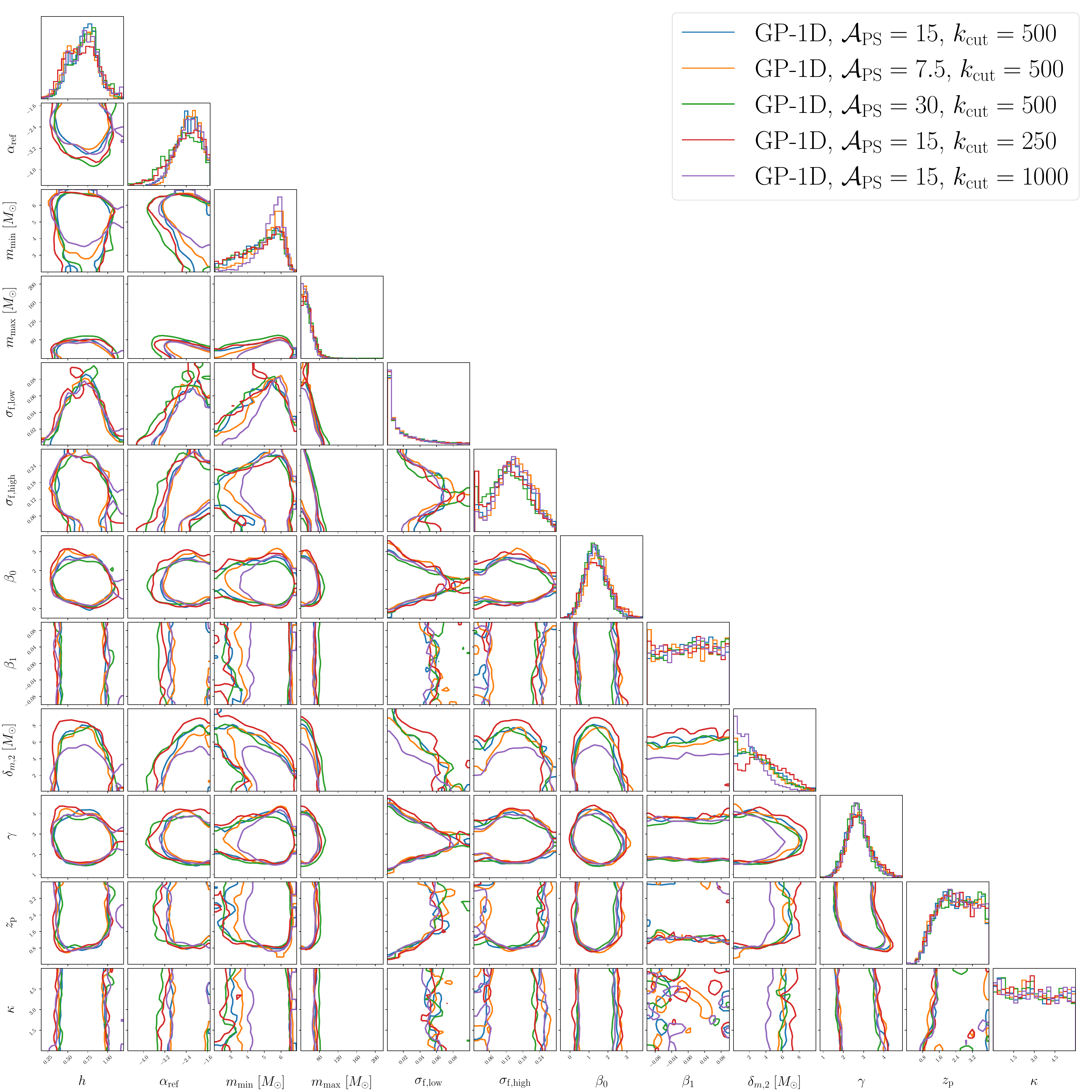}
    \caption{Two-dimensional marginal distribution for GWTC-5 and the GP-1D mass model.
    The contours correspond to the 90\% level. }
    \label{fig: corner gp1d gwtc5 meta choices}
\end{figure}

\begin{figure}
    \centering
    \includegraphics[width=0.98\linewidth]{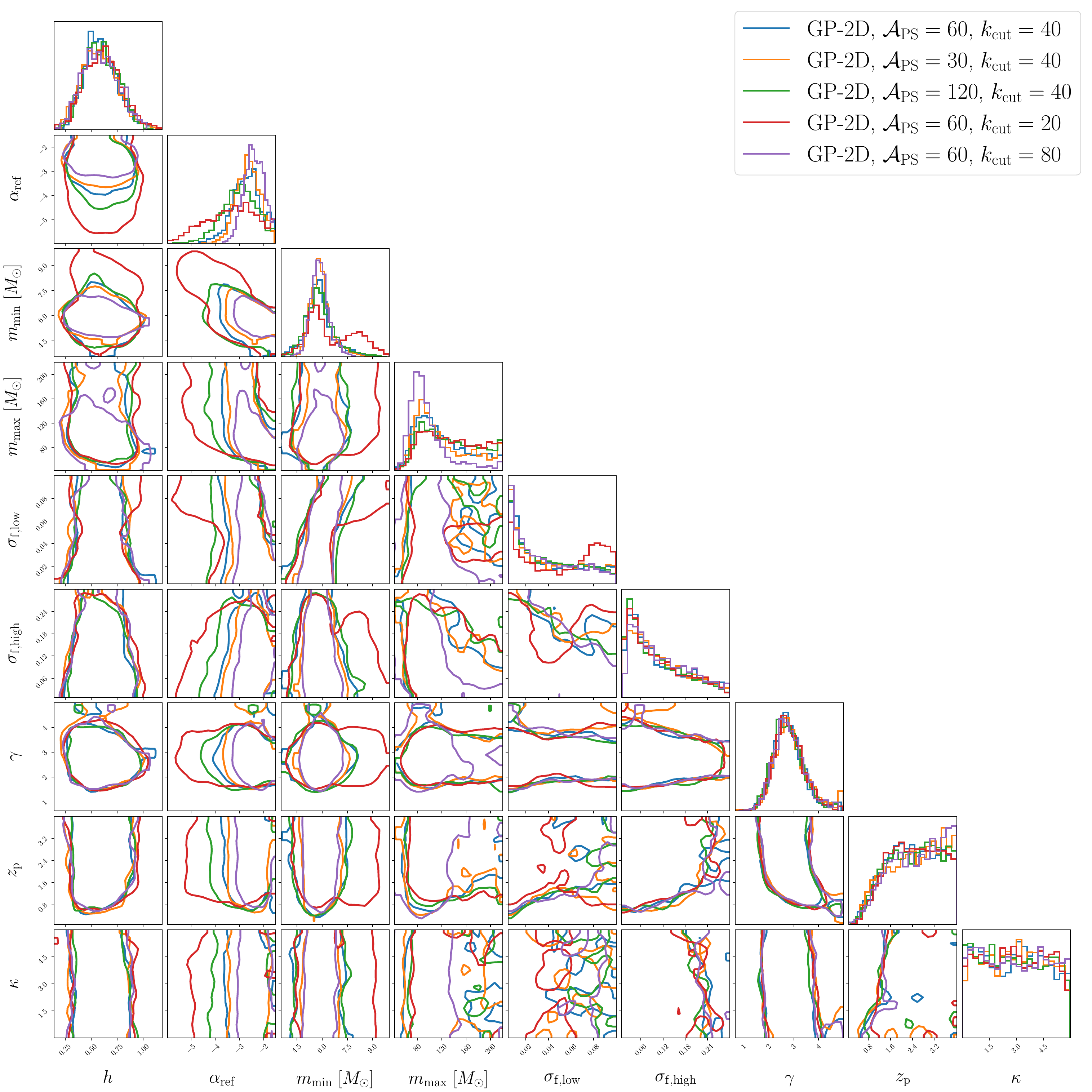}
    \caption{Two-dimensional marginal distribution for GWTC-5 and the GP-2D mass model. The contours correspond to the 90\% level. }
    \label{fig: corner gp2d gwtc5 meta choices}
\end{figure}

\begin{landscape}
\begin{figure}[p]
    \centering
    \includegraphics[width=22.5cm]{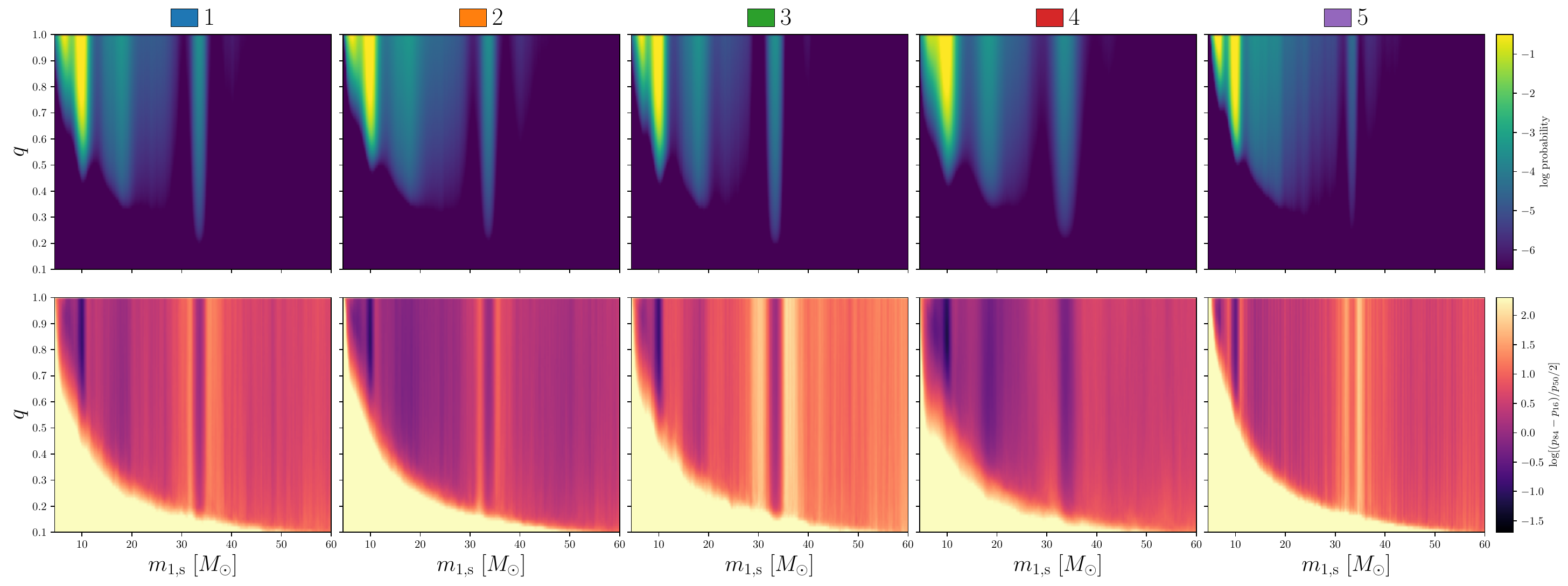}
    \vspace{0.3cm}
    \includegraphics[width=22.5cm]{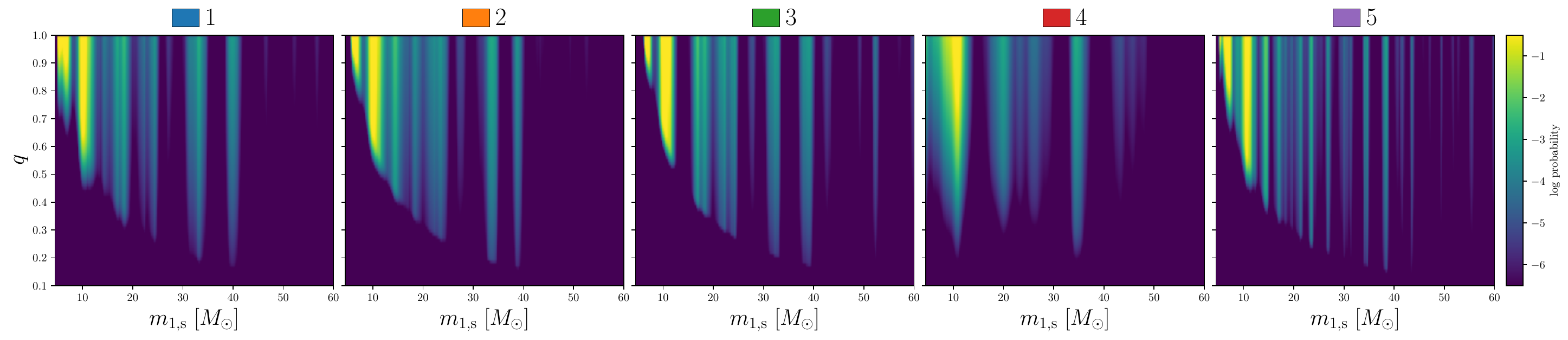}
    \caption{
    (\textit{Top}) Median of the reconstructed mass distribution in $\msone$ and $q$ for varying power-spectrum meta-choices analyzing GWTC-5. 
    The choices are summarized in Tab.~\ref{tab:settings_diffs prior on power spectrum parameters}.
    (\textit{Middle}) Relative uncertainty defined in Eq.~\eqref{eq:def uncertainty measure}.
    The peak of the mass distribution shows a low relative uncertainty.
    (\textit{Bottom}) One chosen posterior sample (at random) to illustrate varying power-spectrum parameters.
    The colors correspond to Fig.~\ref{fig: corner gp1d gwtc5 meta choices}.
    }
    \label{fig: reconstructed mass distribution metachoices gp1d gwtc5}
\end{figure}
\end{landscape}

\begin{landscape}
\begin{figure}[p]
    \centering
    \includegraphics[width=22.5cm]{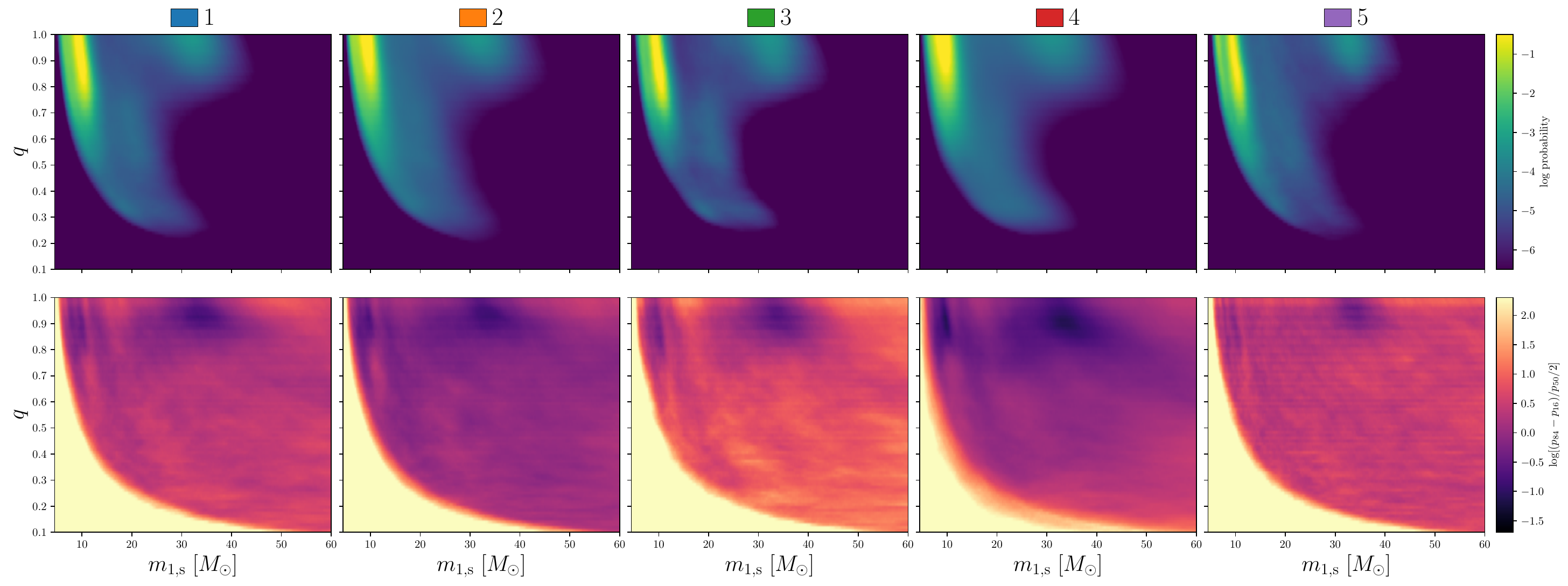}
    \vspace{0.3cm}
    \includegraphics[width=22.5cm]{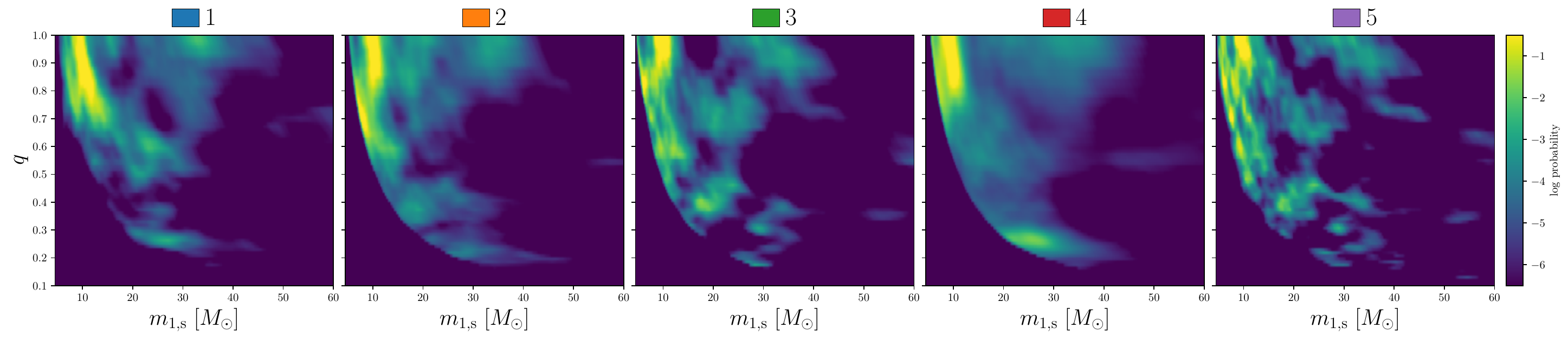}
    \caption{
    Reconstructed mass distribution for GWTC-5 for varying power-spectrum meta-choices.
    The panel structure follows exactly the same conventions as described in the caption of Fig.~\ref{fig: reconstructed mass distribution metachoices gp1d gwtc5}.
    See Tab.~\ref{tab:settings_diffs prior on power spectrum parameters} for the choices of power-spectrum parameters.
    }
    \label{fig: reconstructed mass distribution metachoices gp2d gwtc5}
\end{figure}
\end{landscape}

Fig.~\ref{fig: corner gp1d gwtc5 meta choices} and Fig.~\ref{fig: corner gp2d gwtc5 meta choices} compare the one- and two-dimensional marginal distributions of all parameters, excluding the whitened Gaussian-field components, for the GP-1D and GP-2D mass models, respectively.
Similarly, Fig.~\ref{fig: reconstructed mass distribution metachoices gp1d gwtc5} and Fig.~\ref{fig: reconstructed mass distribution metachoices gp2d gwtc5} show the reconstructed median mass distribution and uncertainty (cf.~Eq.~\eqref{eq:def uncertainty measure}), again for both mass models.
We now discuss the impact of the choice of power-spectrum parameters for each model.

We caveat that the following conclusions only hold in the explored region of the power-spectrum parameters. 
For more extreme values of $\powerspectrumamplitude$ and $\powerspectrumcutoff$, the $h$ posterior can be more affected, particularly in the case of the GP-2D mass model. 

\modelheading{GP-1D}
Although the fluctuations allowed by the GP-1D model (controlled by the $\powerspectrumamplitude$ and $\powerspectrumcutoff$ parameters) vary widely, the resulting corner plots agree well across the different meta-choices.
The parameters with the largest discrepancies are $\deltamtwo$ and $\mmin$, which control the low-mass end of the mass distribution, as well as the baseline power-law trend parameter $\alpharef$.
However, the inferred $h$ values are similar.
All reconstructed mass distributions for the GP-1D model (Fig.~\ref{fig: reconstructed mass distribution metachoices gp1d gwtc5}) recover four visible peaks, with the exception of settings~[4] (red), where the low value of $\powerspectrumcutoff$ enforces a mass distribution that varies more smoothly.

\modelheading{GP-2D}
Similarly to the GP-1D model, the inferred $h$ posterior is robust against the choice of power-spectrum parameters. 
However, unlike for the GP-1D model, the GP-2D posterior distributions show a larger scatter (Fig.~\ref{fig: corner gp2d gwtc5 meta choices}), with $\alpharef$, $\mmin$, $\mmax$ and $\sigmalowfrac$ the parameters with most variation. 
This is a result of the increased flexibility of the two-dimensional model: for example, settings~[2] (orange) have the lowest value for $\powerspectrumamplitude$, recovering a smoother mass distribution (cf.~Fig.~\ref{fig: reconstructed mass distribution metachoices gp2d gwtc5}).
No parameter is strongly correlated with $h$.

We do not find a clear trend between the power-spectrum parameters and the median of the power-law trend parameter $\alpharef$ (cf.~Eq.~\eqref{eq:prior_2D}).
Its width, however, does respond: for the GP-2D mass model, the smoother the distribution (low $\powerspectrumcutoff$), the broader the $\alpharef$ posterior.

\section{Misspecification for the mass-ratio distribution}
\label{app: wrong mass ratio distribution gp1d}

\begin{figure}
    \centering
    \includegraphics[width=0.8\linewidth]{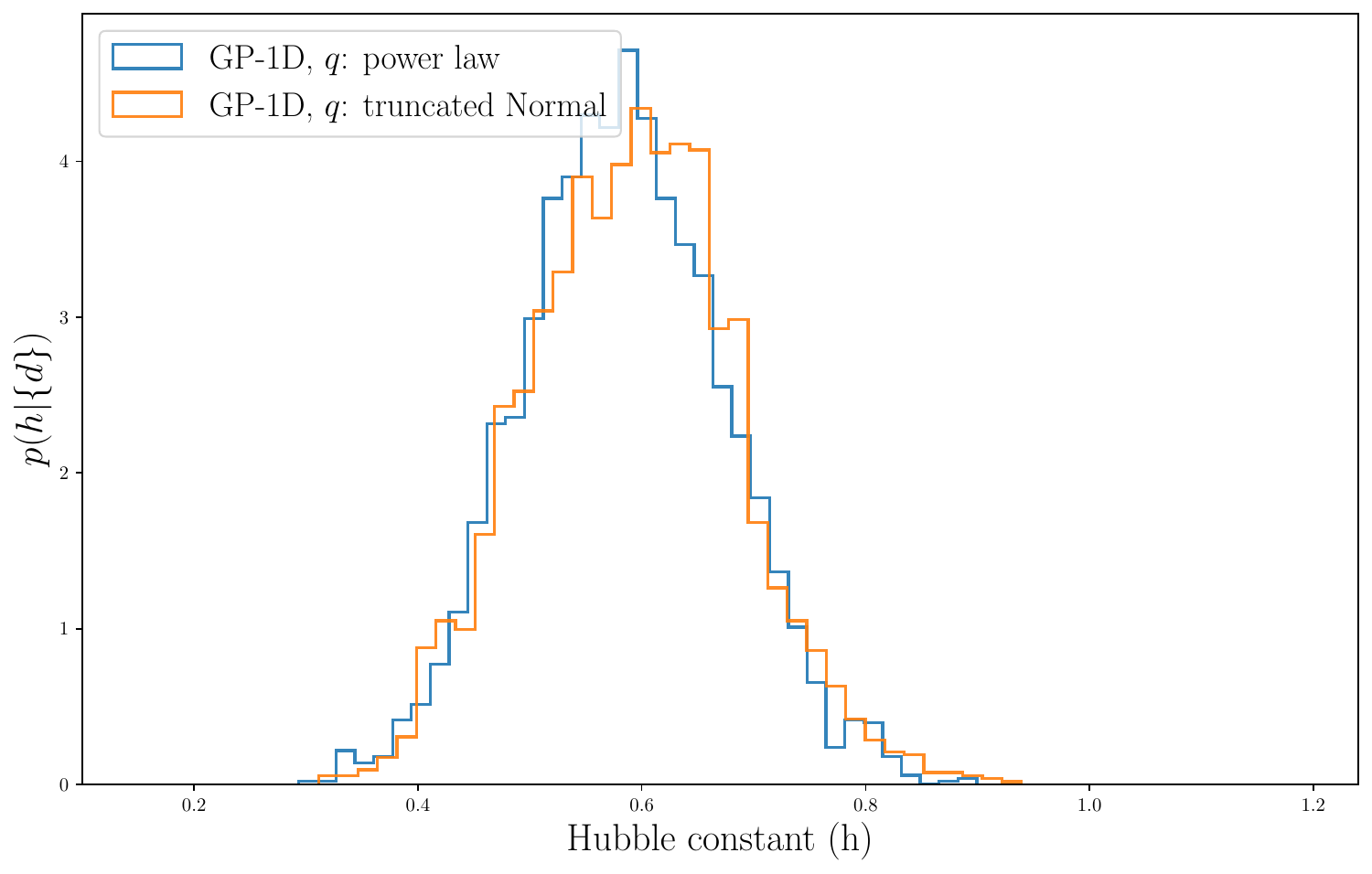}
    \caption{Inferred $h$ posterior under two assumptions on the mass-ratio distribution.
    The power law (cf.~Eq.~\eqref{eq:mltp_q}, blue) and the truncated normal (cf.~Eq.~\eqref{eq:truncated normal q}, orange).
    }
    \label{fig: hubble constant comparison sim pl vs truncated gaussian mass ratio}
\end{figure}

In Sec.~\ref{sec:results_simulated}, when applying \cosmopyro to the simulated catalog of $\numsim$ GW events, we compared the GP-1D with the GP-2D mass model.
In contrast to the flexible GP-2D model, the GP-1D model uses a power law to model the mass-ratio distribution, with $p(\massratio | \hyper, \msone) \propto \massratio ^ {\betaq(\msone)}$.
The latter power-law behavior matches the simulated population. This appendix studies the change in the $h$ posterior when we assume a different mass-ratio distribution that does not match the simulated one.

For this purpose, we define a truncated normal distribution, namely
\begin{equation}
\label{eq:truncated normal q}
    p(\massratio | \hyper, \msone)
    \propto
    \exp\left(- \frac{(\massratio - \mumassratio) ^ 2}{2 \sigmamassratio ^ 2}\right)
    \,,
    \qquad
    \massratio \in (0, 1] \,,
\end{equation}
where $\mumassratio$ and $\sigmamassratio$ are two hyperparameters that are jointly inferred along with the other mass parameters.
The proportionality can be converted into an equality via normalization.
We marginalize over $\powerspectrumamplitude$ and $\powerspectrumcutoff$, and assume priors identical to Tab.~\ref{tab:priors_gp1d_gp2d_combined}, and $p(\mumassratio) = \mathcal{U}(0.4, 2)$ and $p(\sigmamassratio) = \mathcal{LU}(0.05, 1)$.

The simulated data are then analyzed twice with the GP-1D mass model. 
For the mass ratio, the first analysis assumes the running power law of Eq.~\eqref{eq:mltp_q}, the second analysis the aforementioned truncated Normal distribution. 
Fig.~\ref{fig: hubble constant comparison sim pl vs truncated gaussian mass ratio} shows the inferred $h$ posterior comparing the baseline posterior of Sec.~\ref{sec:results_simulated} against the $h$ posterior given the truncated normal mass-ratio distribution.
Although introducing a slight shift, $h$ is robust against mismodeling the mass-ratio distribution for this setup.
Note that this might no longer hold for GW catalogs with larger sizes, and for mass distributions that have a more complicated structure than the \mltp mass model.

\bibliographystyle{JHEP}
\bibliography{references}

@article{LIGOScientific:2025jau,
    author = "LIGO Scientific, VIRGO, KAGRA Collaborations",
    collaboration = "LIGO Scientific, VIRGO, KAGRA",
    title = "{GWTC-4.0: Constraints on the Cosmic Expansion Rate and Modified Gravitational-wave Propagation}",
    eprint = "2509.04348",
    archivePrefix = "arXiv",
    primaryClass = "astro-ph.CO",
    reportNumber = "LIGO-P2400152",
    month = "9",
    year = "2025",
    journal = "arXiv:2509.04348"
}

@article{LIGOScientific:2025pvj,
    author = "Abac, A. G. and others",
    collaboration = "LIGO Scientific, VIRGO, KAGRA",
    title = "{GWTC-4.0: Population Properties of Merging Compact Binaries}",
    eprint = "2508.18083",
    archivePrefix = "arXiv",
    primaryClass = "astro-ph.HE",
    reportNumber = "LIGO-P2400004",
    month = "8",
    year = "2025",
    journal = "arXiv:2508.18083"
}

@article{LIGOScientific:2025slb,
    author = "Abac, A. G. and others",
    collaboration = "LIGO Scientific, VIRGO, KAGRA",
    title = "{GWTC-4.0: Updating the Gravitational-Wave Transient Catalog with Observations from the First Part of the Fourth LIGO-Virgo-KAGRA Observing Run}",
    eprint = "2508.18082",
    archivePrefix = "arXiv",
    primaryClass = "gr-qc",
    reportNumber = "LIGO-P2400386",
    month = "8",
    year = "2025",
    journal = "arXiv:2508.18082"
}

@article{bingham2019pyro,
  author    = {Eli Bingham and
               Jonathan P. Chen and
               Martin Jankowiak and
               Fritz Obermeyer and
               Neeraj Pradhan and
               Theofanis Karaletsos and
               Rohit Singh and
               Paul A. Szerlip and
               Paul Horsfall and
               Noah D. Goodman},
  title     = {Pyro: Deep Universal Probabilistic Programming},
  journal   = {J. Mach. Learn. Res.},
  volume    = {20},
  pages     = {28:1--28:6},
  year      = {2019},
  url       = {http://jmlr.org/papers/v20/18-403.html}
}

@article{Mastrogiovanni:2021wsd,
    author = "Mastrogiovanni, S. and Leyde, K. and Karathanasis, C. and Chassande-Mottin, E. and Steer, D. A. and Gair, J. and Ghosh, A. and Gray, R. and Mukherjee, S. and Rinaldi, S.",
    title = "{On the importance of source population models for gravitational-wave cosmology}",
    eprint = "2103.14663",
    archivePrefix = "arXiv",
    primaryClass = "gr-qc",
    doi = "10.1103/PhysRevD.104.062009",
    journal = "Phys. Rev. D",
    volume = "104",
    number = "6",
    pages = "062009",
    year = "2021"
}

@article{Pratten:2020ceb,
    author = "Pratten, Geraint and others",
    title = "{Computationally efficient models for the dominant and subdominant harmonic modes of precessing binary black holes}",
    eprint = "2004.06503",
    archivePrefix = "arXiv",
    primaryClass = "gr-qc",
    doi = "10.1103/PhysRevD.103.104056",
    journal = "Phys. Rev. D",
    volume = "103",
    number = "10",
    pages = "104056",
    year = "2021"
}

@article{Varma:2019csw,
    author = "Varma, Vijay and Field, Scott E. and Scheel, Mark A. and Blackman, Jonathan and Gerosa, Davide and Stein, Leo C. and Kidder, Lawrence E. and Pfeiffer, Harald P.",
    title = "{Surrogate models for precessing binary black hole simulations with unequal masses}",
    eprint = "1905.09300",
    archivePrefix = "arXiv",
    primaryClass = "gr-qc",
    doi = "10.1103/PhysRevResearch.1.033015",
    journal = "Phys. Rev. Research.",
    volume = "1",
    pages = "033015",
    year = "2019"
}

@article{LIGOScientific:2021djp,
    author = "Abbott, R. and others",
    collaboration = "KAGRA, VIRGO, LIGO Scientific",
    title = "{GWTC-3: Compact Binary Coalescences Observed by LIGO and Virgo during the Second Part of the Third Observing Run}",
    eprint = "2111.03606",
    archivePrefix = "arXiv",
    primaryClass = "gr-qc",
    reportNumber = "LIGO-P2000318",
    doi = "10.1103/PhysRevX.13.041039",
    journal = "Phys. Rev. X",
    volume = "13",
    number = "4",
    pages = "041039",
    year = "2023"
}

@article{Schutz:1986gp,
    author = "Schutz, Bernard F.",
    title = "{Determining the Hubble Constant from Gravitational Wave Observations}",
    doi = "10.1038/323310a0",
    journal = "Nature",
    volume = "323",
    pages = "310--311",
    year = "1986"
}

@article{KAGRA:2013rdx,
    author = "Abbott, B. P. and others",
    collaboration = "KAGRA, LIGO Scientific, Virgo, VIRGO",
    title = "{Prospects for observing and localizing gravitational-wave transients with Advanced LIGO, Advanced Virgo and KAGRA}",
    eprint = "1304.0670",
    archivePrefix = "arXiv",
    primaryClass = "gr-qc",
    reportNumber = "LIGO-P1200087, VIR-0288A-12",
    doi = "10.1007/s41114-020-00026-9",
    journal = "Living Rev. Rel.",
    volume = "21",
    number = "1",
    pages = "3",
    year = "2018"
}

@article{Mukherjee:2021rtw,
    author = "Mukherjee, Suvodip",
    title = "{The redshift dependence of black hole mass distribution: is it reliable for standard sirens cosmology?}",
    eprint = "2112.10256",
    archivePrefix = "arXiv",
    primaryClass = "astro-ph.CO",
    doi = "10.1093/mnras/stac2152",
    journal = "Mon. Not. Roy. Astron. Soc.",
    volume = "515",
    number = "4",
    pages = "5495--5505",
    year = "2022"
}

@article{det1-aligo2015,
	doi = {10.1088/0264-9381/32/7/074001},
	url = {https://doi.org/10.1088/0264-9381/32/7/074001},
	year = 2015,
	month = {mar},
	publisher = {{IOP} Publishing},
	volume = {32},
	number = {7},
	pages = {074001},
	author = {Aasi, J. and others},
	title = {Advanced {LIGO}},
	journal = {Classical and Quantum Gravity}
}

@article{det2-aLIGO:2020wna,
    author = "Buikema, Aaron and others",
    collaboration = "aLIGO",
    title = "{Sensitivity and performance of the Advanced LIGO detectors in the third observing run}",
    eprint = "2008.01301",
    archivePrefix = "arXiv",
    primaryClass = "astro-ph.IM",
    doi = "10.1103/PhysRevD.102.062003",
    journal = "Phys. Rev. D",
    volume = "102",
    number = "6",
    pages = "062003",
    year = "2020"
}

@article{det3-Tse:2019wcy,
    author = "Tse, M. and others",
    title = "{Quantum-Enhanced Advanced LIGO Detectors in the Era of Gravitational-Wave Astronomy}",
    doi = "10.1103/PhysRevLett.123.231107",
    journal = "Phys. Rev. Lett.",
    volume = "123",
    number = "23",
    pages = "231107",
    year = "2019"
}

@article{det4-VIRGO:2014yos,
    author = "Acernese, F. and others",
    collaboration = "VIRGO",
    title = "{Advanced Virgo: a second-generation interferometric gravitational wave detector}",
    eprint = "1408.3978",
    archivePrefix = "arXiv",
    primaryClass = "gr-qc",
    doi = "10.1088/0264-9381/32/2/024001",
    journal = "Class. Quant. Grav.",
    volume = "32",
    number = "2",
    pages = "024001",
    year = "2015"
}

@article{det5-Virgo:2019juy,
    author = "Acernese, F. and others",
    collaboration = "Virgo",
    title = "{Increasing the Astrophysical Reach of the Advanced Virgo Detector via the Application of Squeezed Vacuum States of Light}",
    doi = "10.1103/PhysRevLett.123.231108",
    journal = "Phys. Rev. Lett.",
    volume = "123",
    number = "23",
    pages = "231108",
    year = "2019"
}

@article{DelPozzo:2011vcw,
    author = "Del Pozzo, Walter",
    title = "{Inference of the cosmological parameters from gravitational waves: application to second generation interferometers}",
    eprint = "1108.1317",
    archivePrefix = "arXiv",
    primaryClass = "astro-ph.CO",
    doi = "10.1103/PhysRevD.86.043011",
    journal = "Phys. Rev. D",
    volume = "86",
    pages = "043011",
    year = "2012"
}

@article{DES:2020nay,
    author = "Palmese, A. and others",
    collaboration = "DES",
    title = "{A statistical standard siren measurement of the Hubble constant from the LIGO/Virgo gravitational wave compact object merger GW190814 and Dark Energy Survey galaxies}",
    eprint = "2006.14961",
    archivePrefix = "arXiv",
    primaryClass = "astro-ph.CO",
    reportNumber = "FERMILAB-PUB-20-216-AE, DES-2020-0548",
    doi = "10.3847/2041-8213/abaeff",
    journal = "Astrophys. J. Lett.",
    volume = "900",
    number = "2",
    pages = "L33",
    year = "2020"
}

@article{Ezquiaga:2022zkx,
    author = "Ezquiaga, Jose Mar\'\i{}a and Holz, Daniel E.",
    title = "{Spectral Sirens: Cosmology from the Full Mass Distribution of Compact Binaries}",
    eprint = "2202.08240",
    archivePrefix = "arXiv",
    primaryClass = "astro-ph.CO",
    doi = "10.1103/PhysRevLett.129.061102",
    journal = "Phys. Rev. Lett.",
    volume = "129",
    number = "6",
    pages = "061102",
    year = "2022"
}

@article{Madau:2014bja,
    author = "Madau, Piero and Dickinson, Mark",
    title = "{Cosmic Star Formation History}",
    eprint = "1403.0007",
    archivePrefix = "arXiv",
    primaryClass = "astro-ph.CO",
    doi = "10.1146/annurev-astro-081811-125615",
    journal = "Ann. Rev. Astron. Astrophys.",
    volume = "52",
    pages = "415--486",
    year = "2014"
}

@article{Finke:2021aom,
    author = "Finke, Andreas and Foffa, Stefano and Iacovelli, Francesco and Maggiore, Michele and Mancarella, Michele",
    title = "{Cosmology with LIGO/Virgo dark sirens: Hubble parameter and modified gravitational wave propagation}",
    eprint = "2101.12660",
    archivePrefix = "arXiv",
    primaryClass = "astro-ph.CO",
    doi = "10.1088/1475-7516/2021/08/026",
    journal = "JCAP",
    volume = "08",
    pages = "026",
    year = "2021"
}

@article{Gray:2019ksv,
    author = "Gray, Rachel and others",
    title = "{Cosmological inference using gravitational wave standard sirens: A mock data analysis}",
    eprint = "1908.06050",
    archivePrefix = "arXiv",
    primaryClass = "gr-qc",
    reportNumber = "LIGO-P1900017",
    doi = "10.1103/PhysRevD.101.122001",
    journal = "Phys. Rev. D",
    volume = "101",
    number = "12",
    pages = "122001",
    year = "2020"
}

@article{Mandel:2018mve,
    author = "Mandel, Ilya and Farr, Will M. and Gair, Jonathan R.",
    title = "{Extracting distribution parameters from multiple uncertain observations with selection biases}",
    eprint = "1809.02063",
    archivePrefix = "arXiv",
    primaryClass = "physics.data-an",
    doi = "10.1093/mnras/stz896",
    journal = "Mon. Not. Roy. Astron. Soc.",
    volume = "486",
    number = "1",
    pages = "1086--1093",
    year = "2019"
}

@article{Taylor:2012db,
    author = "Taylor, Stephen R. and Gair, Jonathan R.",
    title = "{Cosmology with the lights off: standard sirens in the Einstein Telescope era}",
    eprint = "1204.6739",
    archivePrefix = "arXiv",
    primaryClass = "astro-ph.CO",
    doi = "10.1103/PhysRevD.86.023502",
    journal = "Phys. Rev. D",
    volume = "86",
    pages = "023502",
    year = "2012"
}

@article{Taylor:2011fs,
    author = "Taylor, Stephen R. and Gair, Jonathan R. and Mandel, Ilya",
    title = "{Hubble without the Hubble: Cosmology using advanced gravitational-wave detectors alone}",
    eprint = "1108.5161",
    archivePrefix = "arXiv",
    primaryClass = "gr-qc",
    doi = "10.1103/PhysRevD.85.023535",
    journal = "Phys. Rev. D",
    volume = "85",
    pages = "023535",
    year = "2012"
}

@article{You:2020wju,
    author = "You, Zhi-Qiang and Zhu, Xing-Jiang and Ashton, Gregory and Thrane, Eric and Zhu, Zong-Hong",
    title = "{Standard-siren cosmology using gravitational waves from binary black holes}",
    eprint = "2004.00036",
    archivePrefix = "arXiv",
    primaryClass = "astro-ph.CO",
    doi = "10.3847/1538-4357/abd4d4",
    journal = "Astrophys. J.",
    volume = "908",
    number = "2",
    pages = "215",
    year = "2021"
}

@article{Farr:2019twy,
    author = "Farr, Will M. and Fishbach, Maya and Ye, Jiani and Holz, Daniel",
    title = "{A Future Percent-Level Measurement of the Hubble Expansion at Redshift 0.8 With Advanced LIGO}",
    eprint = "1908.09084",
    archivePrefix = "arXiv",
    primaryClass = "astro-ph.CO",
    reportNumber = "LIGO P1900252",
    doi = "10.3847/2041-8213/ab4284",
    journal = "Astrophys. J. Lett.",
    volume = "883",
    number = "2",
    pages = "L42",
    year = "2019"
}

@article{Ashton:2018jfp,
    author = "Ashton, Gregory and others",
    title = "{BILBY: A user-friendly Bayesian inference library for gravitational-wave astronomy}",
    eprint = "1811.02042",
    archivePrefix = "arXiv",
    primaryClass = "astro-ph.IM",
    doi = "10.3847/1538-4365/ab06fc",
    journal = "Astrophys. J. Suppl.",
    volume = "241",
    number = "2",
    pages = "27",
    year = "2019"
}

@ARTICLE{2018ApJ...863L..41F,
       author = {{Fishbach}, Maya and {Holz}, Daniel E. and {Farr}, Will M.},
        title = "{Does the Black Hole Merger Rate Evolve with Redshift?}",
      journal = {\apjl},
         year = 2018,
        month = aug,
       volume = {863},
       number = {2},
          eid = {L41},
        pages = {L41},
          doi = {10.3847/2041-8213/aad800},
archivePrefix = {arXiv},
       eprint = {1805.10270},
 primaryClass = {astro-ph.HE},
       adsurl = {https://ui.adsabs.harvard.edu/abs/2018ApJ...863L..41F}
}

@article{Vitale:2020aaz,
    author = "Vitale, Salvatore and Gerosa, Davide and Farr, Will M. and Taylor, Stephen R.",
    title = "{Inferring the properties of a population of compact binaries in presence of selection effects}",
    eprint = "2007.05579",
    archivePrefix = "arXiv",
    primaryClass = "astro-ph.IM",
    doi = "10.1007/978-981-15-4702-7_45-1",
    month = "7",
    year = "2020",
    journal = "arXiv:2007.05579"
}

@article{Talbot:2018cva,
    author = "Talbot, Colm and Thrane, Eric",
    title = "{Measuring the binary black hole mass spectrum with an astrophysically motivated parameterization}",
    eprint = "1801.02699",
    archivePrefix = "arXiv",
    primaryClass = "astro-ph.HE",
    doi = "10.3847/1538-4357/aab34c",
    journal = "Astrophys. J.",
    volume = "856",
    number = "2",
    pages = "173",
    year = "2018"
}

@article{Talbot:2020oeu,
    author = "Talbot, Colm and Thrane, Eric",
    title = "{Flexible and Accurate Evaluation of Gravitational-wave Malmquist Bias with Machine Learning}",
    eprint = "2012.01317",
    archivePrefix = "arXiv",
    primaryClass = "gr-qc",
    doi = "10.3847/1538-4357/ac4bc0",
    journal = "Astrophys. J.",
    volume = "927",
    number = "1",
    pages = "76",
    year = "2022"
}

@article{Thrane:2018qnx,
    author = "Thrane, Eric and Talbot, Colm",
    title = "{An introduction to Bayesian inference in gravitational-wave astronomy: parameter estimation, model selection, and hierarchical models}",
    eprint = "1809.02293",
    archivePrefix = "arXiv",
    primaryClass = "astro-ph.IM",
    doi = "10.1017/pasa.2019.2",
    journal = "Publ. Astron. Soc. Austral.",
    volume = "36",
    pages = "e010",
    year = "2019",
    note = "[Erratum: Publ.Astron.Soc.Austral. 37, e036 (2020)]"
}

@article{Usman:2015kfa,
    author = "Usman, Samantha A. and others",
    title = "{The PyCBC search for gravitational waves from compact binary coalescence}",
    eprint = "1508.02357",
    archivePrefix = "arXiv",
    primaryClass = "gr-qc",
    reportNumber = "LIGO-P1500086",
    doi = "10.1088/0264-9381/33/21/215004",
    journal = "Class. Quant. Grav.",
    volume = "33",
    number = "21",
    pages = "215004",
    year = "2016"
}

@article{Sachdev:2019vvd,
    author = "Sachdev, Surabhi and others",
    title = "{The GstLAL Search Analysis Methods for Compact Binary Mergers in Advanced LIGO's Second and Advanced Virgo's First Observing Runs}",
    eprint = "1901.08580",
    archivePrefix = "arXiv",
    primaryClass = "gr-qc",
    month = "1",
    year = "2019",
    journal = "arXiv:1901.08580"
}

@article{Aubin:2020goo,
    author = "Aubin, F. and others",
    title = "{The MBTA pipeline for detecting compact binary coalescences in the third LIGO\textendash{}Virgo observing run}",
    eprint = "2012.11512",
    archivePrefix = "arXiv",
    primaryClass = "gr-qc",
    doi = "10.1088/1361-6382/abe913",
    journal = "Class. Quant. Grav.",
    volume = "38",
    number = "9",
    pages = "095004",
    year = "2021"
}

@article{Farr:2019rap,
    author = "Farr, Will M.",
    title = "{Accuracy Requirements for Empirically-Measured Selection Functions}",
    eprint = "1904.10879",
    archivePrefix = "arXiv",
    primaryClass = "astro-ph.IM",
    doi = "10.3847/2515-5172/ab1d5f",
    journal = "Research Notes of the AAS",
    volume = "3",
    number = "5",
    pages = "66",
    year = "2019"
}

@article{Gair:2022zsa,
    author = "Gair, Jonathan R. and others",
    title = "{The Hitchhiker{\textquoteright}s Guide to the Galaxy Catalog Approach for Dark Siren Gravitational-wave Cosmology}",
    eprint = "2212.08694",
    archivePrefix = "arXiv",
    primaryClass = "gr-qc",
    doi = "10.3847/1538-3881/acca78",
    journal = "Astron. J.",
    volume = "166",
    number = "1",
    pages = "22",
    year = "2023"
}

@article{KAGRA:2021duu,
    author = "Abbott, R. and others",
    collaboration = "KAGRA, VIRGO, LIGO Scientific",
    title = "{Population of Merging Compact Binaries Inferred Using Gravitational Waves through GWTC-3}",
    eprint = "2111.03634",
    archivePrefix = "arXiv",
    primaryClass = "astro-ph.HE",
    reportNumber = "LIGO-P2100239 ; Data release: https://zenodo.org/record/5655785, LIGO-P2100239",
    doi = "10.1103/PhysRevX.13.011048",
    journal = "Phys. Rev. X",
    volume = "13",
    number = "1",
    pages = "011048",
    year = "2023"
}

@article{LIGOScientific:2021aug,
    author = "Abbott, R. and others",
    collaboration = "LIGO Scientific, Virgo,, KAGRA, VIRGO",
    title = "{Constraints on the Cosmic Expansion History from GWTC\textendash{}3}",
    eprint = "2111.03604",
    archivePrefix = "arXiv",
    primaryClass = "astro-ph.CO",
    reportNumber = "LIGO-P2100185-v6, LIGO-P2100185-v5",
    doi = "10.3847/1538-4357/ac74bb",
    journal = "Astrophys. J.",
    volume = "949",
    number = "2",
    pages = "76",
    year = "2023"
}

@article{Mastrogiovanni:2023emh,
    author = "Mastrogiovanni, Simone and Laghi, Danny and Gray, Rachel and Santoro, Giada Caneva and Ghosh, Archisman and Karathanasis, Christos and Leyde, Konstantin and Steer, Daniele A. and Perries, Stephane and Pierra, Gregoire",
    title = "{Joint population and cosmological properties inference with gravitational waves standard sirens and galaxy surveys}",
    eprint = "2305.10488",
    archivePrefix = "arXiv",
    primaryClass = "astro-ph.CO",
    doi = "10.1103/PhysRevD.108.042002",
    journal = "Phys. Rev. D",
    volume = "108",
    number = "4",
    pages = "042002",
    year = "2023"
}

@article{Turski:2023lxq,
    author = "Turski, Cezary and Bilicki, Maciej and D\'alya, Gergely and Gray, Rachel and Ghosh, Archisman",
    title = "{Impact of modelling galaxy redshift uncertainties on the gravitational-wave dark standard siren measurement of the Hubble constant}",
    eprint = "2302.12037",
    archivePrefix = "arXiv",
    primaryClass = "gr-qc",
    doi = "10.1093/mnras/stad3110",
    journal = "Mon. Not. Roy. Astron. Soc.",
    volume = "526",
    number = "4",
    pages = "6224--6233",
    year = "2023"
}

@article{Gray:2021sew,
    author = "Gray, Rachel and Messenger, Chris and Veitch, John",
    title = "{A pixelated approach to galaxy catalogue incompleteness: improving the dark siren measurement of the Hubble constant}",
    eprint = "2111.04629",
    archivePrefix = "arXiv",
    primaryClass = "astro-ph.CO",
    doi = "10.1093/mnras/stac366",
    journal = "Mon. Not. Roy. Astron. Soc.",
    volume = "512",
    number = "1",
    pages = "1127--1140",
    year = "2022"
}

@article{Gray:2023wgj,
    author = "Gray, Rachel and others",
    title = "{Joint cosmological and gravitational-wave population inference using dark sirens and galaxy catalogues}",
    eprint = "2308.02281",
    archivePrefix = "arXiv",
    primaryClass = "astro-ph.CO",
    doi = "10.1088/1475-7516/2023/12/023",
    journal = "JCAP",
    volume = "12",
    pages = "023",
    year = "2023"
}

@article{Edelman:2021zkw,
    author = "Edelman, Bruce and Doctor, Zoheyr and Godfrey, Jaxen and Farr, Ben",
    title = "{Ain\textquoteright{}t No Mountain High Enough: Semiparametric Modeling of LIGO\textendash{}Virgo\textquoteright{}s Binary Black Hole Mass Distribution}",
    eprint = "2109.06137",
    archivePrefix = "arXiv",
    primaryClass = "astro-ph.HE",
    reportNumber = "LIGO-P2100300",
    doi = "10.3847/1538-4357/ac3667",
    journal = "Astrophys. J.",
    volume = "924",
    number = "2",
    pages = "101",
    year = "2022"
}

@article{Edelman:2022ydv,
    author = "Edelman, Bruce and Farr, Ben and Doctor, Zoheyr",
    title = "{Cover Your Basis: Comprehensive Data-driven Characterization of the Binary Black Hole Population}",
    eprint = "2210.12834",
    archivePrefix = "arXiv",
    primaryClass = "astro-ph.HE",
    reportNumber = "LIGO-P2200312",
    doi = "10.3847/1538-4357/acb5ed",
    journal = "Astrophys. J.",
    volume = "946",
    number = "1",
    pages = "16",
    year = "2023"
}

@article{Farah:2024xub,
    author = "Farah, Amanda M. and Callister, Thomas A. and Ezquiaga, Jose Mar{\'\i}a and Zevin, Michael and Holz, Daniel E.",
    title = "{No Need to Know: Toward Astrophysics-free Gravitational-wave Cosmology}",
    eprint = "2404.02210",
    archivePrefix = "arXiv",
    primaryClass = "astro-ph.CO",
    doi = "10.3847/1538-4357/ad9253",
    journal = "Astrophys. J.",
    volume = "978",
    number = "2",
    pages = "153",
    year = "2025"
}

@inproceedings{Phan:2019elc,
    author = "Phan, Du and Pradhan, Neeraj and Jankowiak, Martin",
    title = "{Composable Effects for Flexible and Accelerated Probabilistic Programming in NumPyro}",
    eprint = "1912.11554",
    archivePrefix = "arXiv",
    primaryClass = "stat.ML",
    month = "12",
    year = "2019"
}

@book{Neal:2011mrf,
    author = "Neal, Radford M.",
    title = "{Handbook of Markov Chain Monte Carlo}",
    eprint = "1206.1901",
    archivePrefix = "arXiv",
    primaryClass = "stat.CO",
    doi = "10.1201/b10905",
    month = "5",
    year = "2011"
}

@article{Hoffman:2011ukg,
    author = "Hoffman, Matthew D. and Gelman, Andrew",
    title = "{The No-U-Turn Sampler: Adaptively Setting Path Lengths in Hamiltonian Monte Carlo}",
    eprint = "1111.4246",
    archivePrefix = "arXiv",
    primaryClass = "stat.CO",
    month = "11",
    year = "2011"
}

@article{Betancourt:2017ebh,
    author = "Betancourt, Michael",
    title = "{A Conceptual Introduction to Hamiltonian Monte Carlo}",
    eprint = "1701.02434",
    archivePrefix = "arXiv",
    primaryClass = "stat.ME",
    month = "1",
    year = "2017"
}

@article{Pierra:2023deu,
    author = "Pierra, Gr\'egoire and Mastrogiovanni, Simone and Perri\`es, St\'ephane and Mapelli, Michela",
    title = "{Study of systematics on the cosmological inference of the Hubble constant from gravitational wave standard sirens}",
    eprint = "2312.11627",
    archivePrefix = "arXiv",
    primaryClass = "astro-ph.CO",
    reportNumber = "LIGO-P2300442",
    doi = "10.1103/PhysRevD.109.083504",
    journal = "Phys. Rev. D",
    volume = "109",
    number = "8",
    pages = "083504",
    year = "2024"
}

@software{jax2018github,
  author = {James Bradbury and Roy Frostig and Peter Hawkins and Matthew James Johnson and Chris Leary and Dougal Maclaurin and George Necula and Adam Paszke and Jake Vander{P}las and Skye Wanderman-{M}ilne and Qiao Zhang},
  title = {{JAX}: composable transformations of {P}ython+{N}um{P}y programs},
  url = {http://github.com/google/jax},
  version = {0.3.13},
  year = {2018},
}

@article{Mali:2024wpq,
    author = "Mali, Utkarsh and Essick, Reed",
    title = "{Striking a Chord with Spectral Sirens: Multiple Features in the Compact Binary Population Correlate with H$_{0}$}",
    eprint = "2410.07416",
    archivePrefix = "arXiv",
    primaryClass = "astro-ph.HE",
    doi = "10.3847/1538-4357/ad9de7",
    journal = "Astrophys. J.",
    volume = "980",
    number = "1",
    pages = "85",
    year = "2025"
}

@article{Mastrogiovanni:2023zbw,
    author = "Mastrogiovanni, Simone and Pierra, Gr\'egoire and Perri\`es, St\'ephane and Laghi, Danny and Caneva Santoro, Giada and Ghosh, Archisman and Gray, Rachel and Karathanasis, Christos and Leyde, Konstantin",
    title = "{ICAROGW: A python package for inference of astrophysical population properties of noisy, heterogeneous, and incomplete observations}",
    eprint = "2305.17973",
    archivePrefix = "arXiv",
    primaryClass = "astro-ph.CO",
    doi = "10.1051/0004-6361/202347007",
    journal = "Astron. Astrophys.",
    volume = "682",
    pages = "A167",
    year = "2024"
}

@article{Borghi:2023opd,
    author = "Borghi, Nicola and Mancarella, Michele and Moresco, Michele and Tagliazucchi, Matteo and Iacovelli, Francesco and Cimatti, Andrea and Maggiore, Michele",
    title = "{Cosmology and Astrophysics with Standard Sirens and Galaxy Catalogs in View of Future Gravitational Wave Observations}",
    eprint = "2312.05302",
    archivePrefix = "arXiv",
    primaryClass = "astro-ph.CO",
    doi = "10.3847/1538-4357/ad20eb",
    journal = "Astrophys. J.",
    volume = "964",
    number = "2",
    pages = "191",
    year = "2024"
}

@article{Essick:2023toz,
    author = "Essick, Reed",
    title = "{Semianalytic sensitivity estimates for catalogs of gravitational-wave transients}",
    eprint = "2307.02765",
    archivePrefix = "arXiv",
    primaryClass = "gr-qc",
    doi = "10.1103/PhysRevD.108.043011",
    journal = "Phys. Rev. D",
    volume = "108",
    number = "4",
    pages = "043011",
    year = "2023"
}

@article{Mancarella:2025uat,
    author = "Mancarella, Michele and Gerosa, Davide",
    title = "{Sampling the full hierarchical population posterior distribution in gravitational-wave astronomy}",
    eprint = "2502.12156",
    archivePrefix = "arXiv",
    primaryClass = "gr-qc",
    doi = "10.1103/PhysRevD.111.103012",
    journal = "Phys. Rev. D",
    volume = "111",
    number = "10",
    pages = "103012",
    year = "2025"
}

@article{Naveed:2025kgk,
    author = "Naveed, Khuzaifa and Turski, Cezary and Ghosh, Archisman",
    title = "{Dark standard siren cosmology with bright galaxy subsets}",
    eprint = "2505.11268",
    archivePrefix = "arXiv",
    primaryClass = "astro-ph.CO",
    reportNumber = "LIGO-P2500266",
    month = "5",
    year = "2025",
    journal = "arXiv:2505.11268"
}

@article{Tiwari:2020vym,
    author = "Tiwari, Vaibhav",
    title = "{VAMANA: modeling binary black hole population with minimal assumptions}",
    eprint = "2006.15047",
    archivePrefix = "arXiv",
    primaryClass = "astro-ph.HE",
    doi = "10.1088/1361-6382/ac0b54",
    journal = "Class. Quant. Grav.",
    volume = "38",
    number = "15",
    pages = "155007",
    year = "2021"
}

@article{Sadiq:2023zee,
    author = "Sadiq, Jam and Dent, Thomas and Gieles, Mark",
    title = "{Binary Vision: The Mass Distribution of Merging Binary Black Holes via Iterative Density Estimation}",
    eprint = "2307.12092",
    archivePrefix = "arXiv",
    primaryClass = "astro-ph.HE",
    doi = "10.3847/1538-4357/ad0ce6",
    journal = "Astrophys. J.",
    volume = "960",
    number = "1",
    pages = "65",
    year = "2024"
}

@article{Callister:2023tgi,
    author = "Callister, Thomas A. and Farr, Will M.",
    title = "{Parameter-Free Tour of the Binary Black Hole Population}",
    eprint = "2302.07289",
    archivePrefix = "arXiv",
    primaryClass = "astro-ph.HE",
    doi = "10.1103/PhysRevX.14.021005",
    journal = "Phys. Rev. X",
    volume = "14",
    number = "2",
    pages = "021005",
    year = "2024"
}

@article{Ray:2023upk,
    author = "Ray, Anarya and Maga{\~n}a Hernandez, Ignacio and Mohite, Siddharth and Creighton, Jolien and Kapadia, Shasvath",
    title = "{Nonparametric Inference of the Population of Compact Binaries from Gravitational-wave Observations Using Binned Gaussian Processes}",
    eprint = "2304.08046",
    archivePrefix = "arXiv",
    primaryClass = "gr-qc",
    reportNumber = "LIGO-P2300098",
    doi = "10.3847/1538-4357/acf452",
    journal = "Astrophys. J.",
    volume = "957",
    number = "1",
    pages = "37",
    year = "2023"
}

@article{Ray:2024hos,
    author = "Ray, Anarya and Maga{\~n}a Hernandez, Ignacio and Breivik, Katelyn and Creighton, Jolien",
    title = "{Searching for Binary Black Hole Subpopulations in Gravitational-wave Data Using Binned Gaussian Processes}",
    eprint = "2404.03166",
    archivePrefix = "arXiv",
    primaryClass = "astro-ph.HE",
    reportNumber = "LIGO-P2400115",
    doi = "10.3847/1538-4357/adf22a",
    journal = "Astrophys. J.",
    volume = "991",
    number = "1",
    pages = "17",
    year = "2025"
}

@article{MaganaHernandez:2024uty,
    author = "Maga{\~n}a Hernandez, Ignacio and Ray, Anarya",
    title = "{Beyond Gaps and Bumps: Spectral Siren Cosmology with Non-Parametric Population Models}",
    eprint = "2404.02522",
    archivePrefix = "arXiv",
    primaryClass = "astro-ph.CO",
    month = "4",
    year = "2024"
}

@article{Tagliazucchi:2026gxn,
    author = "Tagliazucchi, Matteo and Moresco, Michele and Borghi, Nicola and Ciapetti, Chiara",
    title = "{Mind the peak: improving cosmological constraints from GWTC-4.0 spectral sirens using semiparametric mass models}",
    eprint = "2601.03347",
    archivePrefix = "arXiv",
    primaryClass = "astro-ph.CO",
    month = "1",
    year = "2026"
}

@article{Heinzel:2024jlc,
    author = "Heinzel, Jack and Mould, Matthew and {\'A}lvarez-L{\'o}pez, Sof{\'\i}a and Vitale, Salvatore",
    title = "{High resolution nonparametric inference of gravitational-wave populations in multiple dimensions}",
    eprint = "2406.16813",
    archivePrefix = "arXiv",
    primaryClass = "astro-ph.HE",
    doi = "10.1103/PhysRevD.111.063043",
    journal = "Phys. Rev. D",
    volume = "111",
    number = "6",
    pages = "063043",
    year = "2025"
}

@article{Heinzel:2023hlb,
    author = "Heinzel, Jack and Biscoveanu, Sylvia and Vitale, Salvatore",
    title = "{Probing correlations in the binary black hole population with flexible models}",
    eprint = "2312.00993",
    archivePrefix = "arXiv",
    primaryClass = "astro-ph.HE",
    doi = "10.1103/PhysRevD.109.103006",
    journal = "Phys. Rev. D",
    volume = "109",
    number = "10",
    pages = "103006",
    year = "2024"
}

@article{Alvarez-Lopez:2025ltt,
    author = "Alvarez-Lopez, Sofia and Heinzel, Jack and Mould, Matthew and Vitale, Salvatore",
    title = "{Nowhere left to hide: revealing realistic gravitational-wave populations in high dimensions and high resolution with PixelPop}",
    eprint = "2506.20731",
    archivePrefix = "arXiv",
    primaryClass = "astro-ph.HE",
    month = "6",
    year = "2025"
}

@article{Heinzel:2024hva,
    author = "Heinzel, Jack and Mould, Matthew and Vitale, Salvatore",
    title = "{Nonparametric analysis of correlations in the binary black hole population with LIGO-Virgo-KAGRA data}",
    eprint = "2406.16844",
    archivePrefix = "arXiv",
    primaryClass = "astro-ph.HE",
    doi = "10.1103/PhysRevD.111.L061305",
    journal = "Phys. Rev. D",
    volume = "111",
    number = "6",
    pages = "L061305",
    year = "2025"
}

@article{Pierra:2025hoc,
    author = "Pierra, Gr{\'e}goire and Colombo, Alberto and Mastrogiovanni, Simone",
    title = "{Non-Parametric Reconstruction of the Hubble Parameter from the Fourth Gravitational Wave Transient Catalog and DESI Baryonic Acoustic Oscillations}",
    eprint = "2511.11795",
    archivePrefix = "arXiv",
    primaryClass = "astro-ph.CO",
    month = "11",
    year = "2025"
}

@article{Dupletsa:2026uqs,
    author = "Dupletsa, Ulyana and others",
    title = "{Radio sirens: inferring $H_0$ with binary black holes and neutral hydrogen in the era of the Einstein Telescope and the SKA Observatory}",
    eprint = "2605.12606",
    archivePrefix = "arXiv",
    primaryClass = "astro-ph.CO",
    month = "5",
    year = "2026"
}

@article{LIGOScientific:2026sit,
    author = "Abac, None and others",
    collaboration = "LIGO Scientific, VIRGO, KAGRA",
    title = "{GWTC-5.0: An Introduction to Version 5.0 of the Gravitational-Wave Transient Catalog}",
    eprint = "2605.27223",
    archivePrefix = "arXiv",
    primaryClass = "gr-qc",
    reportNumber = "LIGO-P2500701",
    month = "5",
    year = "2026"
}

@article{LIGOScientific:2026wfs,
    collaboration = "LIGO Scientific, VIRGO, KAGRA",
    title = "{GWTC-5.0: Observations from the Second Part of the Fourth LIGO-Virgo-KAGRA Observing Run and Updates to the Gravitational-Wave Transient Catalog}",
    eprint = "2605.27225",
    archivePrefix = "arXiv",
    primaryClass = "gr-qc",
    reportNumber = "LIGO-P2600152",
    month = "5",
    year = "2026"
}

@article{LIGOScientific:2026ctl,
    collaboration = "LIGO Scientific, VIRGO, KAGRA",
    title = "{GWTC-5.0: Population Properties of Merging Compact Binaries}",
    eprint = "2605.27226",
    archivePrefix = "arXiv",
    primaryClass = "astro-ph.HE",
    reportNumber = "LIGO-P2600045",
    month = "5",
    year = "2026"
}

@article{LIGOScientific:2026uyd,
    collaboration = "LIGO Scientific, VIRGO, KAGRA",
    title = "{GWTC-5.0: Constraints on the Cosmic Expansion Rate and Modified Gravitational-wave Propagation}",
    eprint = "2605.27227",
    archivePrefix = "arXiv",
    primaryClass = "astro-ph.CO",
    reportNumber = "LIGO-P2600018",
    month = "5",
    year = "2026"
}

@article{LIGOScientific:2026ifv,
    collaboration = "LIGO Scientific, VIRGO, KAGRA",
    title = "{GWTC-5.0: Methods for Identifying and Characterizing Gravitational-wave Transients}",
    eprint = "2605.27224",
    archivePrefix = "arXiv",
    primaryClass = "gr-qc",
    reportNumber = "LIGO-P2600166",
    month = "5",
    year = "2026"
}

@article{Essick:2025zed,
    author = "Essick, Reed and others",
    title = "{Compact binary coalescence sensitivity estimates with injection campaigns during the LIGO-Virgo-KAGRA Collaborations{\textquoteright} fourth observing run}",
    eprint = "2508.10638",
    archivePrefix = "arXiv",
    primaryClass = "gr-qc",
    doi = "10.1103/44x3-hv3y",
    journal = "Phys. Rev. D",
    volume = "112",
    number = "10",
    pages = "102001",
    year = "2025"
}

@article{Tagliazucchi:2025ofb,
    author = "Tagliazucchi, Matteo and Moresco, Michele and Borghi, Nicola and Fiebig, Manfred",
    title = "{Accelerating the standard siren method: Improved constraints on modified gravitational-wave propagation with future data}",
    eprint = "2504.02034",
    archivePrefix = "arXiv",
    primaryClass = "astro-ph.CO",
    doi = "10.1051/0004-6361/202554827",
    journal = "Astron. Astrophys.",
    volume = "702",
    pages = "A244",
    year = "2025"
}

@article{Borghi:2025pav,
    author = "Borghi, Nicola and Moresco, Michele and Tagliazucchi, Matteo and Cuomo, Giulia",
    title = "{Echoes from the dark: Galaxy catalog incompleteness in standard siren cosmology}",
    eprint = "2509.18243",
    archivePrefix = "arXiv",
    primaryClass = "astro-ph.CO",
    doi = "10.1051/0004-6361/202557354",
    journal = "Astron. Astrophys.",
    volume = "706",
    pages = "A199",
    year = "2026"
}

@article{DESI:2023fij,
    author = "Ballard, W. and others",
    collaboration = "DESI",
    title = "{A Dark Siren Measurement of the Hubble Constant with the LIGO/Virgo Gravitational Wave Event GW190412 and DESI Galaxies}",
    eprint = "2311.13062",
    archivePrefix = "arXiv",
    primaryClass = "astro-ph.CO",
    doi = "10.3847/2515-5172/ad0eda",
    journal = "Res. Notes AAS",
    volume = "7",
    number = "11",
    pages = "250",
    year = "2023"
}

@article{Alfradique:2023giv,
    author = "Alfradique, V. and others",
    title = "{A dark siren measurement of the Hubble constant using gravitational wave events from the first three LIGO/Virgo observing runs and DELVE}",
    eprint = "2310.13695",
    archivePrefix = "arXiv",
    primaryClass = "astro-ph.CO",
    reportNumber = "FERMILAB-PUB-23-550-LDRD-PPD",
    doi = "10.1093/mnras/stae086",
    journal = "Mon. Not. Roy. Astron. Soc.",
    volume = "528",
    number = "2",
    pages = "3249--3259",
    year = "2024"
}

@article{Palmese:2021mjm,
    author = "Palmese, Antonella and Bom, Clecio R. and Mucesh, Sunil and Hartley, William G.",
    title = "{A Standard Siren Measurement of the Hubble Constant Using Gravitational-wave Events from the First Three LIGO/Virgo Observing Runs and the DESI Legacy Survey}",
    eprint = "2111.06445",
    archivePrefix = "arXiv",
    primaryClass = "astro-ph.CO",
    doi = "10.3847/1538-4357/aca6e3",
    journal = "Astrophys. J.",
    volume = "943",
    number = "1",
    pages = "56",
    year = "2023"
}

@article{LIGOScientific:2018gmd,
    author = "Fishbach, M. and others",
    collaboration = "LIGO Scientific, Virgo",
    title = "{A Standard Siren Measurement of the Hubble Constant from GW170817 without the Electromagnetic Counterpart}",
    eprint = "1807.05667",
    archivePrefix = "arXiv",
    primaryClass = "astro-ph.CO",
    reportNumber = "LIGO-P1800192",
    doi = "10.3847/2041-8213/aaf96e",
    journal = "Astrophys. J. Lett.",
    volume = "871",
    number = "1",
    pages = "L13",
    year = "2019"
}

@article{Beirnaert:2025wcx,
    author = "Beirnaert, Freija and D{\'a}lya, Gergely and Ghosh, Archisman",
    title = "{A Hubble constant estimation with dark standard sirens and galaxy cluster catalogues}",
    eprint = "2505.14077",
    archivePrefix = "arXiv",
    primaryClass = "astro-ph.CO",
    doi = "10.1093/mnras/staf1432",
    journal = "Mon. Not. Roy. Astron. Soc.",
    volume = "542",
    number = "4",
    pages = "3346--3353",
    year = "2025"
}

@article{Cross-Parkin:2025xwf,
    author = "Cross-Parkin, Madeline L. and Howlett, Cullan and Davis, Tamara M. and Khetan, Nandita",
    title = "{Dark sirens and the impact of redshift precision}",
    eprint = "2502.17747",
    archivePrefix = "arXiv",
    primaryClass = "astro-ph.CO",
    month = "2",
    year = "2025"
}

@article{McMahon:2026nhi,
    author = "McMahon, Isaac and others",
    title = "{Measurement of the Hubble constant using the Dark Energy Survey Year 6 Gold galaxy catalogue and the fourth Gravitational-Wave Transient Catalogue}",
    eprint = "2602.04766",
    archivePrefix = "arXiv",
    primaryClass = "astro-ph.CO",
    doi = "10.1093/mnras/stag1167",
    month = "2",
    year = "2026"
}

@article{Li:2025hrh,
    author = "Li, Zhuotao and Gray, Rachel and Heng, Ik Siong",
    title = "{Using Gravitational-wave Dark Sirens to Choose between Host Galaxy Weighting Models}",
    eprint = "2508.15574",
    archivePrefix = "arXiv",
    primaryClass = "astro-ph.CO",
    doi = "10.3847/1538-4357/ae6100",
    journal = "Astrophys. J.",
    volume = "1003",
    number = "2",
    pages = "176",
    year = "2026"
}

@article{Andrade-Oliveira:2026jjm,
    author = "Andrade-Oliveira, Felipe and Sanchez-Cid, David and Laghi, Danny and Soares-Santos, Marcelle",
    title = "{First measurement of the Hubble constant from a combined weak-lensing and gravitational-wave standard siren analysis}",
    eprint = "2601.04774",
    archivePrefix = "arXiv",
    primaryClass = "astro-ph.CO",
    doi = "10.1051/0004-6361/202659044",
    journal = "Astron. Astrophys.",
    volume = "709",
    pages = "L14",
    year = "2026"
}

@article{Turski:2025flk,
    author = "Turski, Cezary and Brozzetti, Maria Lisa and D{\'a}lya, Gergely and Punturo, Michele and Ghosh, Archisman",
    title = "{The Luminosity of the Darkness: Schechter function in dark sirens}",
    eprint = "2505.13568",
    archivePrefix = "arXiv",
    primaryClass = "astro-ph.CO",
    doi = "10.1093/mnras/stag218",
    journal = "Mon. Not. Roy. Astron. Soc.",
    volume = "546",
    number = "4",
    pages = "stag218",
    year = "2026"
}

@article{Dalang:2024gfk,
    author = "Dalang, Charles and Fiorini, Bartolomeo and Baker, Tessa",
    title = "{Large scale structure prior knowledge in the dark siren method}",
    eprint = "2410.03275",
    archivePrefix = "arXiv",
    primaryClass = "astro-ph.CO",
    doi = "10.1088/1475-7516/2026/01/034",
    journal = "JCAP",
    volume = "01",
    pages = "034",
    year = "2026"
}

@article{Tagliazucchi:2026dpr,
    author = "Tagliazucchi, Matteo and Moresco, Michele and Agapito, Alessandro and Mancarella, Michele and Ferraiuolo, Sarah and Mastrogiovanni, Simone and Borghi, Nicola and Pannarale, Francesco and Bonacorsi, Daniele",
    title = "{Pushing spectral siren cosmology into the third-generation era: a blinded mock data challenge}",
    eprint = "2602.17756",
    archivePrefix = "arXiv",
    primaryClass = "astro-ph.CO",
    month = "2",
    year = "2026"
}

@article{Flanagan:2026ayy,
    author = "Flanagan, Elizabeth and Antonini, Fabio and Callister, Thomas and Chattopadhyay, Debatri and Dosopoulou, Fani and Romero-Shaw, Isobel and Stegmann, Jakob",
    title = "{Transitions in the Mass-ratio and Spin Properties of Binary Black Holes in GWTC-5}",
    eprint = "2606.14472",
    archivePrefix = "arXiv",
    primaryClass = "astro-ph.HE",
    month = "6",
    year = "2026"
}

@article{Farah:2023swu,
    author = "Farah, Amanda M. and Fishbach, Maya and Holz, Daniel E.",
    title = "{Two of a Kind: Comparing Big and Small Black Holes in Binaries with Gravitational Waves}",
    eprint = "2308.05102",
    archivePrefix = "arXiv",
    primaryClass = "astro-ph.HE",
    doi = "10.3847/1538-4357/ad0558",
    journal = "Astrophys. J.",
    volume = "962",
    number = "1",
    pages = "69",
    year = "2024"
}

@article{Banagiri:2025dmy,
    author = "Banagiri, Sharan and Thrane, Eric and Lasky, Paul D.",
    title = "{Evidence for Three Subpopulations of Merging Binary Black Holes at Different Primary Masses}",
    eprint = "2509.15646",
    archivePrefix = "arXiv",
    primaryClass = "astro-ph.HE",
    doi = "10.1103/blyb-lqv6",
    journal = "Phys. Rev. Lett.",
    volume = "137",
    number = "2",
    pages = "021403",
    year = "2026"
}

@article{Li:2024jzi,
    author = "Li, Yin-Jie and Tang, Shao-Peng and Gao, Shi-Jie and Wu, Dao-Cheng and Wang, Yuan-Zhu",
    title = "{Exploring Field-evolution and Dynamical-capture Coalescing Binary Black Holes in GWTC-3}",
    eprint = "2404.09668",
    archivePrefix = "arXiv",
    primaryClass = "astro-ph.HE",
    doi = "10.3847/1538-4357/ad83b5",
    journal = "Astrophys. J.",
    volume = "977",
    number = "1",
    pages = "67",
    year = "2024"
}

@article{Li:2022jge,
    author = "Li, Yin-Jie and Wang, Yuan-Zhu and Tang, Shao-Peng and Yuan, Qiang and Fan, Yi-Zhong and Wei, Da-Ming",
    title = "{Divergence in Mass Ratio Distributions between Low-mass and High-mass Coalescing Binary Black Holes}",
    eprint = "2201.01905",
    archivePrefix = "arXiv",
    primaryClass = "astro-ph.HE",
    doi = "10.3847/2041-8213/ac78dd",
    journal = "Astrophys. J. Lett.",
    volume = "933",
    number = "1",
    pages = "L14",
    year = "2022"
}

@article{Roy:2025ktr,
    author = "Roy, Soumendra Kishore and van Son, Lieke A. C. and Farr, Will M.",
    title = "{A mid-thirties crisis: dissecting the properties of gravitational wave sources near the 35 solar mass peak}",
    eprint = "2507.01086",
    archivePrefix = "arXiv",
    primaryClass = "astro-ph.HE",
    reportNumber = "LIGO document number LIGO-P2500403",
    doi = "10.1088/1361-6382/ae1921",
    journal = "Class. Quant. Grav.",
    volume = "42",
    number = "22",
    pages = "225008",
    year = "2025"
}

@article{Ray:2026uur,
    author = "Ray, Anarya and Mukherjee, Shirsha and Zevin, Michael and Kalogera, Vicky",
    title = "{On the Astrophysical Origin of Binary Black Hole Subpopulations: A Tale of Three Channels?}",
    eprint = "2603.17987",
    archivePrefix = "arXiv",
    primaryClass = "astro-ph.HE",
    reportNumber = "LIGO-P2600074",
    doi = "10.3847/2041-8213/ae80cb",
    journal = "Astrophys. J. Lett.",
    volume = "1005",
    number = "2",
    pages = "L55",
    year = "2026"
}

@article{Sridhar:2025kvi,
    author = "Sridhar, Omkar and Ray, Anarya and Kalogera, Vicky",
    title = "{Characterizing Binary Black Hole Subpopulations in GWTC-4 with Binned Gaussian Processes: On the Origins of the 35 M$_{\odot}$ Peak}",
    eprint = "2511.22093",
    archivePrefix = "arXiv",
    primaryClass = "astro-ph.HE",
    reportNumber = "LIGO-P2500712",
    doi = "10.3847/2041-8213/ae8011",
    journal = "Astrophys. J. Lett.",
    volume = "1005",
    number = "2",
    pages = "L54",
    year = "2026"
}

@article{Tong:2025wpz,
    author = "Tong, Hui and others",
    title = "{Evidence of the pair-instability gap from black-hole masses}",
    eprint = "2509.04151",
    archivePrefix = "arXiv",
    primaryClass = "astro-ph.HE",
    doi = "10.1038/s41586-026-10359-0",
    journal = "Nature",
    volume = "652",
    number = "8111",
    pages = "874--877",
    year = "2026"
}

@article{Mould:2026sww,
    author = "Mould, Matthew and Heinzel, Jack and Alvarez-Lopez, Sofia and Plunkett, Cailin and Wolfe, Noah E. and Vitale, Salvatore",
    title = "{Measurement prospects for the pair-instability mass cutoff with gravitational waves}",
    eprint = "2602.11282",
    archivePrefix = "arXiv",
    primaryClass = "astro-ph.HE",
    doi = "10.1103/sqy9-nnqq",
    journal = "Phys. Rev. D",
    volume = "113",
    number = "10",
    pages = "103021",
    year = "2026"
}

@article{Godfrey:2026pbc,
    author = "Godfrey, Jaxen and van Son, Lieke and Farr, Ben",
    title = "{A Strongly Parametrized Mass Ratio Model for the Stable Mass Transfer Channel: a Case Study of the $10 \, \rm{M}_{\odot}$ Peak}",
    eprint = "2605.23083",
    archivePrefix = "arXiv",
    primaryClass = "astro-ph.HE",
    month = "5",
    year = "2026"
}

@article{Godfrey:2023oxb,
    author = "Godfrey, Jaxen and Edelman, Bruce and Farr, Ben",
    title = "{Cosmic Cousins: Identification of a Subpopulation of Binary Black Holes Consistent with Isolated Binary Evolution}",
    eprint = "2304.01288",
    archivePrefix = "arXiv",
    primaryClass = "astro-ph.HE",
    reportNumber = "LIGO-P2300073",
    month = "4",
    year = "2023"
}

@article{Pierra:2026ffj,
    author = "Pierra, Gr{\'e}goire and Papadopoulos, Alexander",
    title = "{Heavy Black-Holes Also Matter in Standard Siren Cosmology}",
    eprint = "2601.03257",
    archivePrefix = "arXiv",
    primaryClass = "astro-ph.CO",
    month = "1",
    year = "2026"
}

@article{Papadopoulos:2026puy,
    author = "Papadopoulos, Alexander and Chapman-Bird, Christian E. A. and Gray, Rachel and Messenger, Christopher and Bertheas, Tom",
    title = "{Scalable Dark Siren Cosmology with gwcosmo: GPU Acceleration, Validation and Systematics}",
    eprint = "2605.23538",
    archivePrefix = "arXiv",
    primaryClass = "astro-ph.CO",
    month = "5",
    year = "2026"
}

@article{mg_elena_cosmopyro,
    author  = "Colangeli, Elena and Leyde, Konstantin and Baker, Tessa and Chen, Anson",
    title   = "{No parametrisation, No Problem: A Weakly Modelled Framework to Constrain the Luminosity Distance--Redshift Relation Using Gravitational Wave Sirens}",
    year    = "2026",
    note    = "In preparation",
    journal = "In preparation."
}

@article{Kiendrebeogo:2023hzf,
    author = "Kiendrebeogo, R. Weizmann and others",
    title = "{Updated Observing Scenarios and Multimessenger Implications for the International Gravitational-wave Networks O4 and O5}",
    eprint = "2306.09234",
    archivePrefix = "arXiv",
    primaryClass = "astro-ph.HE",
    doi = "10.3847/1538-4357/acfcb1",
    journal = "Astrophys. J.",
    volume = "958",
    number = "2",
    pages = "158",
    year = "2023"
}

@article{Gerosa:2020pgy,
    author = "Gerosa, Davide and Pratten, Geraint and Vecchio, Alberto",
    title = "{Gravitational-wave selection effects using neural-network classifiers}",
    eprint = "2007.06585",
    archivePrefix = "arXiv",
    primaryClass = "astro-ph.HE",
    doi = "10.1103/PhysRevD.102.103020",
    journal = "Phys. Rev. D",
    volume = "102",
    number = "10",
    pages = "103020",
    year = "2020"
}

@article{Callister:2024qyq,
    author = "Callister, Thomas A. and Essick, Reed and Holz, Daniel E.",
    title = "{Neural network emulator of the Advanced LIGO and Advanced Virgo selection function}",
    eprint = "2408.16828",
    archivePrefix = "arXiv",
    primaryClass = "astro-ph.HE",
    doi = "10.1103/PhysRevD.110.123041",
    journal = "Phys. Rev. D",
    volume = "110",
    number = "12",
    pages = "123041",
    year = "2024"
}

@article{Lorenzo-Medina:2024opt,
    author = "Lorenzo-Medina, Ana and Dent, Thomas",
    title = "{A physically modelled selection function for compact binary mergers in the LIGO-Virgo O3 run and beyond}",
    eprint = "2408.13383",
    archivePrefix = "arXiv",
    primaryClass = "gr-qc",
    doi = "10.1088/1361-6382/ad9c0e",
    journal = "Class. Quant. Grav.",
    volume = "42",
    number = "4",
    pages = "045008",
    year = "2025"
}

@article{Hussain:2025llf,
    author = "Hussain, Asad and Isi, Maximiliano and Zimmerman, Aaron",
    title = "{Living on the edge: Testing for compact population features at the edges of parameter space}",
    eprint = "2510.20010",
    archivePrefix = "arXiv",
    primaryClass = "astro-ph.IM",
    month = "10",
    year = "2025"
}

@article{Leyde:2026hvm,
    author = "Leyde, Konstantin and Green, Stephen R. and Dax, Maximilian and Mould, Matthew and Fabbri, Cecilia Maria and Gair, Jonathan",
    title = "{End-to-End Population Inference from Gravitational-Wave Strain using Transformers}",
    eprint = "2605.11274",
    archivePrefix = "arXiv",
    primaryClass = "gr-qc",
    month = "5",
    year = "2026"
}

@misc{ligo_scientific_collaboration_and_virgo_2026_20348006,
  author       = {LIGO Scientific Collaboration and Virgo Collaboration and KAGRA Collaboration},
  title        = {GWTC-5.0: Parameter estimation data release (Part
                   2 of 2)
                  },
  month        = may,
  year         = 2026,
  publisher    = {Zenodo},
  doi          = {10.5281/zenodo.20348006},
  url          = {https://doi.org/10.5281/zenodo.20348006},
}

@misc{ligo_scientific_collaboration_and_virgo_2025_16053484,
  author       = {LIGO Scientific Collaboration and Virgo Collaboration and KAGRA Collaboration},
  title        = {GWTC-4.0: Parameter estimation data release},
  month        = aug,
  year         = 2025,
  publisher    = {Zenodo},
  doi          = {10.5281/zenodo.16053484},
  url          = {https://doi.org/10.5281/zenodo.16053484},
}

@misc{ligo_scientific_collaboration_and_virgo_2021_5546663,
  author       = {LIGO Scientific Collaboration and Virgo Collaboration and KAGRA Collaboration},
  title        = {GWTC-3: Compact Binary Coalescences Observed by
                   LIGO and Virgo During the Second Part of the Third
                   Observing Run — Parameter estimation data release
                  },
  month        = nov,
  year         = 2021,
  publisher    = {Zenodo},
  doi          = {10.5281/zenodo.5546663},
  url          = {https://doi.org/10.5281/zenodo.5546663},
}

@misc{ligo_scientific_collaboration_and_virgo_2022_6513631,
  author       = {LIGO Scientific Collaboration and  Virgo Collaboration},
  title        = {GWTC-2.1: Deep Extended Catalog of Compact Binary
                   Coalescences Observed by LIGO and Virgo During the
                   First Half of the Third Observing Run - Parameter
                   Estimation Data Release
                  },
  month        = may,
  year         = 2022,
  publisher    = {Zenodo},
  version      = {v2},
  doi          = {10.5281/zenodo.6513631},
  url          = {https://doi.org/10.5281/zenodo.6513631},
}

@article{Guttman:2026cnv,
    author = "Guttman, Nir and Lasky, Paul D. and Thrane, Eric",
    title = "{Revealing Four Subpopulations of Binary Black-Hole Mergers with the Fifth Gravitational-Wave Transient Catalog}",
    eprint = "2607.22011",
    archivePrefix = "arXiv",
    primaryClass = "astro-ph.HE",
    month = "7",
    year = "2026"
}

@article{Finn:1992xs,
    author = "Finn, Lee Samuel and Chernoff, David F.",
    title = "{Observing binary inspiral in gravitational radiation: One interferometer}",
    eprint = "gr-qc/9301003",
    archivePrefix = "arXiv",
    reportNumber = "PRINT-93-0138 (NORTHWESTERN)",
    doi = "10.1103/PhysRevD.47.2198",
    journal = "Phys. Rev. D",
    volume = "47",
    pages = "2198--2219",
    year = "1993"
}

@article{Wouters:2025zju,
    author = "Wouters, Thibeau and Pang, Peter T. H. and Koehn, Hauke and Rose, Henrik and Somasundaram, Rahul and Tews, Ingo and Dietrich, Tim and Van Den Broeck, Chris",
    title = "{Leveraging differentiable programming in the inverse problem of neutron stars}",
    eprint = "2504.15893",
    archivePrefix = "arXiv",
    primaryClass = "astro-ph.HE",
    reportNumber = "LA-UR-25-23486",
    doi = "10.1103/v2y8-kxvx",
    journal = "Phys. Rev. D",
    volume = "112",
    number = "4",
    pages = "043037",
    year = "2025"
}

@article{Edwards:2023sak,
    author = "Edwards, Thomas D. P. and Wong, Kaze W. K. and Lam, Kelvin K. H. and Coogan, Adam and Foreman-Mackey, Daniel and Isi, Maximiliano and Zimmerman, Aaron",
    title = "{Differentiable and hardware-accelerated waveforms for gravitational wave data analysis}",
    eprint = "2302.05329",
    archivePrefix = "arXiv",
    primaryClass = "astro-ph.IM",
    doi = "10.1103/PhysRevD.110.064028",
    journal = "Phys. Rev. D",
    volume = "110",
    number = "6",
    pages = "064028",
    year = "2024"
}

@article{Demasi:2026ltw,
    author = "Demasi, Gabriele and others",
    title = "{The Sequential Monte Carlo goes NUTS: boosting gravitational-wave inference}",
    eprint = "2601.02336",
    archivePrefix = "arXiv",
    primaryClass = "gr-qc",
    doi = "10.1140/epjc/s10052-026-15840-8",
    journal = "Eur. Phys. J. C",
    volume = "86",
    number = "6",
    pages = "686",
    year = "2026"
}

@article{Talbot:2024yqw,
    author = "Talbot, Colm and Farah, Amanda and Galaudage, Shanika and Golomb, Jacob and Tong, Hui",
    title = "{GWPopulation: Hardware agnostic population inference for compact binaries and beyond}",
    eprint = "2409.14143",
    archivePrefix = "arXiv",
    primaryClass = "astro-ph.IM",
    doi = "10.21105/joss.07753",
    journal = "J. Open Source Softw.",
    volume = "10",
    number = "109",
    pages = "7753",
    year = "2025"
}

\end{document}